\documentclass[11pt]{article}
\usepackage{amssymb, amsmath, amsthm, mathtools}
\usepackage{bm}
\usepackage{mathrsfs}
\usepackage{hyperref,bookmark}
\usepackage{graphicx}
\usepackage{epsfig,color}
\usepackage{mathrsfs}
\usepackage{verbatim}
\usepackage{url}
\usepackage{subcaption}
\usepackage{float}
\usepackage[greek,english]{babel}
\usepackage{dcolumn}
\usepackage{accents}

\usepackage[
    a4paper,
    left=2.5cm,
    right=2.5cm,
    top=3cm,
    bottom=3cm
]{geometry}

\usepackage{cases}

\newcommand{\tr}{\operatorname{tr}}
\newcommand{\dd}{\operatorname{d}\!}

\newcommand{\e}{\bm{e}}

\newcommand{\body}{\mathcal{B}}
\newcommand{\free}{\mathcal{F}}

\newcommand{\bvarphi}{\bm{\varphi}}
\newcommand{\n}{\bm{n}}

\newcommand{\bF}{\bm{\mathrm{F}}}
\newcommand{\Id}{\bm{\mathrm{I}}}
\newcommand{\bL}{\bm{\mathrm{L}}}
\newcommand{\Frel}{\bm{\mathrm{F}}_{\mathrm{rel}}}

\newcommand{\Fel}{\bm{\mathrm{F}}_{\mathrm{e}}}
\newcommand{\bG}{\bm{\mathrm{F}}_{\mathrm{g}}}

\newcommand{\transp}{\mathrm{T}}

\newcommand{\radius}{R_0}
\newcommand{\bodynorm}{\mathcal{\bar{B}}_0}

\theoremstyle{definition}

\begin{document}
\latintext

\title{Morphogenesis driven by nematic defects in\\ active biological networks}
\author{Silvia Paparini\footnote{Corresponding author.} $^\dagger$ \qquad
	Giulio G. Giusteri\footnote{Dipartimento di Matematica “Tullio Levi-Civita”, Università degli Studi di Padova, Padua, Italy, and Gruppo Nazionale per la Fisica Matematica, Istituto Nazionale di Alta Matematica “Francesco Severi”. Email: \texttt{silvia.paparini@unipd.it}, \texttt{giulio.giusteri@unipd.it}}\qquad
L. Angela Mihai\footnote{School of Mathematics, Cardiff University, Cardiff, UK, Email: \texttt{MihaiLA@cardiff.ac.uk}.}}
\date{}
\maketitle

\begin{abstract}\vskip 6pt
\hrule\vskip 12pt
Cellular morphogenesis, the process by which biological tissues acquire shape and structure, remains a fundamental challenge in understanding pattern formation and the coordinated remodeling of cellular assemblies. Under appropriate conditions, cytoskeletal filaments can organize into a nematic phase exhibiting partial orientational order. Topological defects within this nematic organization generate localized mechanical stresses that destabilize the tissue and promote deformation and structural rearrangements to relieve internal stresses. We develop a continuum framework that models living tissues as active biological networks represented as nematic polymer networks capable of heterogeneous growth and remodeling. The model captures macroscopic effects through spatial variations in the fiber order parameter which drives the system away from equilibrium. Morphogenesis is described as a sequence of quasi-static equilibrium states governed by the coupling between nematic order, elasticity, stress-driven growth, and adaptive relaxation. Finite element simulations illustrate Hydra regeneration and development when topological defects are prescribed according to the mature organism’s expected morphology. The results show that defect topology controls stress localization and shape evolution: $+1$ defects drive protrusion formation, while $-1/2$ defects act as structural stabilizers with minimal growth. By varying the initial defect configuration, we model diverse morphogenetic outcomes, including uniaxial regeneration, tentacle formation, and biaxial development. \\



\vskip 12pt
\hrule
\end{abstract}


\section{Introduction}

One of the main challenges in modeling morphogenesis---the biological process of shape formation---consists in the intrinsic complexity of the underlying biological processes. Tissue development is governed by tightly coupled interactions among biochemical signaling, mechanical deformation, and surface curvature. Yet the precise role of mechanics in transforming a disordered state into an ordered pattern remains largely unclear.

Biological systems composed of cells and cytoskeleton elements can form nematic phases in which elongated constituents align, producing partial orientational order \cite{blanch:turbulent,Guillamat:2026:etal,keber:topology,kumar:tunable,narayan:long,saw:topological}. This order is described by a macroscopic unit vector field, the nematic director $\n$, which reflects the average local orientation. 
Topological defects are singularities in this field. In a two-dimensional setting, defects are classified by their half-integer topological charge, defined as the winding number of $\n$ along a closed loop around the defect. 
Half-integer charges appear due to the head-to-tail symmetry of nematic liquid crystals (LCs): since $\n \to -\n$ leaves the physical state unchanged, the presence of lines of discontinuity across which the director undergoes $\pm \pi$-rotations does not constitute a material singularity.
In our setting, the director field is defined on a three-dimensional domain, but it is constrained to be a tangent field on the boundary, so that two-dimensional defects (measured on the domain boundary) are central to our analysis.
Director fields enclosing defects of charge $+1$ and $-1/2$ rotate by $2\pi$ and $-\pi$, respectively, along any anticlockwise loop around the defect \cite{rosso:parallel}. The sign reflects whether the rotation is coherent with the loop orientation. 

In morphogenesis, defects act as organizing centers that drive growth or material depletion to relieve mechanical stress \cite{Bailles:2025:etal,kawaguchi:topological,copenhagen:topological,keber:topology,maroudas:topological,napoli2026:nematicbubbles,saw:topological}. A model organism for studying this process is Hydra, a freshwater basal invertebrate. It possesses two orthogonal arrays of supracellular actin fibers in the ectoderm and endoderm \cite{maroudas:topological}. The ectodermal fibers dominate mechanically, generating active stresses that keep the system far from equilibrium. Their alignment defines a slowly varying nematic field that remains coherent across large spatial and temporal scales, even under significant deformation. 
Although individual actin filaments are polar, their organization into supracellular fibers behaves macroscopically as a nematic system: active contractile stresses generated by myosin lack net polarity, justifying the use of an apolar director $\n$ \cite{doostmohammadi2018active,marchetti2013hydrodynamics,maroudas:topological}. 

\begin{figure}[t] 
	\centering
	\includegraphics[width=0.9\linewidth]{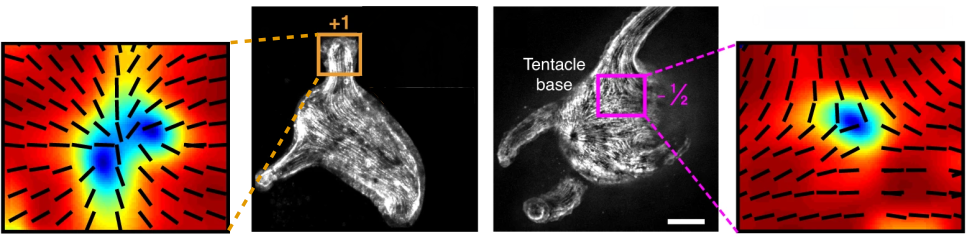}
\caption{Image of nematic actin fiber organization in the ectoderm of a small, mature Hydra (scale bar: $100\mu\mathrm{m}$). Adapted from \cite{maroudas:topological}.}
	\label{fig:hydra}
\end{figure}

Experiments show a strong correlation between defect positions and morphological features in Hydra. In uniaxial organisms, two $+1$ defects appear, one at the mouth and one at the foot \cite{maroudas:topological,maroudas2025:mechanical}. Tentacle tips also host $+1$ defects, while two $-1/2$ defects lie at each base (see Figure~\ref{fig:hydra}). This relationship is further illustrated in biaxial specimens, where two distinct heads correspond to two separate $+1$ defects.

\begin{figure}[t] 
	\centering
	\includegraphics[width=0.9\linewidth]{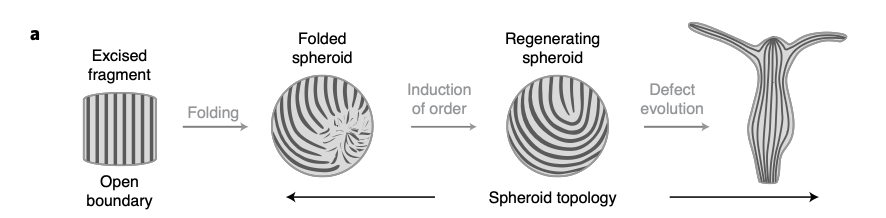}
\caption{Schematic of the nematic organization of actin fibers at successive stages of regeneration in an excised Hydra tissue fragment. Immediately after excision (left), the fragment exhibits a well-ordered array of aligned fibers. After approximately 2 hours (middle-left), the tissue folds into a closed spheroid, displaying partial nematic order alongside an extended disordered region. As ordering progresses over time, localized topological defects emerge (middle-right). These defects evolve over the course of several days while the tissue regenerates into a mature organism (right) \cite{maroudas:topological}.}
\label{fig:hydra_organization}
\end{figure}

Live-imaging studies \cite{maroudas:topological} reveal that, immediately after excision, tissue fragments retain an ordered nematic structure inherited from the parent. Within hours, the tissue folds into a closed spheroid, largely disrupting this order. Nematic alignment then gradually re-emerges as actin fibers tend to align tangentially to the surface. Assuming that Hydra is topologically equivalent to a sphere, a continuous tangent vector field cannot exist without singularities. Consequently, defect formation is unavoidable and constrained by the Poincar\'{e}-Hopf theorem \cite{Hopf1983:differential,poincare1886:courbes,stoker1989differential}, which requires the total topological charge to equal the Euler characteristic, $+2$ for a sphere. This re-established order encodes a memory of the body axis, guiding subsequent head and foot formation. While the mechanisms encoding body-axis polarity memory within the tissue remain elusive, this polarization is already evident during the earliest stages of regeneration. The underlying physical processes, driven by the feedback between biochemical and mechanical activity, are currently a subject of theoretical investigation. For instance, in \cite{wang:patterning}, axis formation is driven by the alignment of nematic fibers with a global morphogen concentration field, by analogy with nematic liquid crystals (LCs) under external fields. However, to reconcile this approach with empirical evidence showing that the signaling protein family Wnt is heavily localized to the head region, the authors in \cite{maroudas2025:mechanical} introduce a strain-regulated mechanism for morphogen expression: higher strain increases morphogen concentration, reinforcing fiber alignment along its gradient. Instead, in \cite{hernandez2026:mechanics}, the regenerating Hydra is modeled as a linearly elastic shell where the interplay between active nematic stresses and shell elasticity spontaneously identifies a body axis without a pre-existing morphogen gradient. This formulation aligns with a series of works on defect-mediated morphogenesis and tuneable defect-curvature coupling \cite{hoffmann2022:theory,hoffmann2023:tuneable,napoli2026:nematicbubbles} that employ linear elastic shell theory combined with LC defects. While these linear approximations succeed in describing the initial nucleation stages, their application is limited.

In this study, we introduce a continuum framework in which living tissues are modeled as \emph{active biological networks} (ABNs). We focus on the mechanical consequences of fiber alignment through a prescribed activation of the order parameter, without addressing its biochemical origins, and on the induced shape change. 
Within our framework, shape changes arise from quasi-static coupling between nematic order and mechanics (Section~\ref{sec:quasi-static}). 

Building on previous work on coupling mesogens disclination and polymeric chains in liquid crystal networks \cite{Paparini:2025:PGM}, we introduce our model for for ABNs. Inertial effects are neglected, and morphogenesis is described as a sequence of equilibrium states driven by elasticity and growth. Growth is not prescribed but guided by stress concentrations generated by distortions in the director field near defects. Because localized growth induces geometric incompatibilities, residual stresses develop; these relax over time through active remodeling, assumed to occur on timescales shorter than that of growth (Section~\ref{sec:energy}).

To validate our model, we apply it to Hydra morphogenesis (Section~\ref{sec:hydra}). After excision, tissue fragments fold into spherical bilayers and later break symmetry to form structures such as tentacles, hypostomes, and basal discs (see Figure~\ref{fig:hydra_organization}). 
Our finite element simulations, implemented in FEniCSx, show how defect topology determines stress localization, which in turn drives growth and shape formation (Section~\ref{sec:config}). By varying initial defect configurations, the new model captures uniaxial regeneration, tentacle formation, and biaxial development with two heads.

\section{A quasi-static continuum framework for active biological networks}\label{sec:quasi-static}

In our model, the morphogenetic process is represented as a sequence of \emph{quasi-static states}. At each state, the material configuration evolves through changes in both local volume and the scalar order parameter $s$, capturing tissue growth and progressive nematic ordering, respectively. This framework describes gradual tissue deformation while assuming mechanical equilibrium at every intermediate state. We first introduce the key ingredients of our continuum model for \emph{active biological networks} (ABNs), providing a mathematical analogue of deforming biological tissues, such as regenerating Hydra fragments.

\subsection{Conceptual framework and modeling assumptions}

We define ABNs as activable deformable materials composed of hydrophilic filament chains cross-linked by rod-like mesogenic segments that can form a nematic phase with orientational but no positional order. The local average orientation is described macroscopically by the unit director field $\n$, while the degree of orientation is quantified by the scalar order parameter $s\in[0,1]$, which vanishes in isotropic regions lacking any preferred direction.

The activation process in ABNs---how mechanics drive the emergence of order from disorder, and vice versa---is largely unknown. Rather than resolving the underlying biochemical signaling, our model focuses on its primary macroscopic manifestation: a \emph{change in the degree of orientation}, captured by $s$. Accordingly, $s$ acts as an activation parameter: its variations can drive the system out of equilibrium and induce macroscopic shape changes. Such deformations, arising from changes in $s$, are also characteristic of nematic liquid crystal elastomers (LCEs) and liquid crystal polymers (LCPs). In the isotropic phase, liquid crystal (LC) molecules have no preferred orientation and $s=0$. Upon activation, $s$ increases ($s>0$), and mesogens and cross-links align uniaxially along $\n$, while the filament network extends in the same direction. These local, reversible deformations generate macroscopic shape changes governed by the (possibly inhomogeneous) \emph{spatial distribution of the director field}. 

Similarly to LCPs \cite{Paparini:2025:PGM}, we assume tight cross-linking in ABNs, limiting director mobility relative to the filament network. The coupling between nematic order and the network is then strong enough to constrain the director field to follow macroscopic deformations. Experimental observations in \cite{maroudas:topological} support this hypothesis, showing that fiber orientation is preserved even under large tissue deformations, with topological defects (disclinations) aligning with morphological features. These singularities in nematic alignment, classified by topological charge, play a key role in the mechanical response of ABNs as strong stresses typically arise in their vicinity.

Finally, we assume that ABNs, as living tissues, undergo \emph{stress-driven growth} and \emph{remodeling}. During morphogenesis, tissue and organ formation involves not only cellular proliferation (mass increase through environmental uptake), but also substantial reorganization and differentiation of the material structure. While nematic order drives growth, accumulated elastic stress can hinder cellular activity. Because growth is slow, concurrent cellular flows promote microstructural reorganization that keeps stresses below a critical threshold. Since proliferation localizes near topological defects, driving, for example, elongation of the central body and extension of tentacles, our model treats local stress as a direct growth stimulus. These relaxation and remodeling processes are therefore tightly linked to the final functional shape of the organism.

\subsection{Kinematics of growth and relaxation}

Our continuum mechanical model aims at describing the behavior of ABNs, incorporating concepts from the theories of nematic LCs, LCEs, LCPs, elastic growth, and relaxation. We denote the undeformed body by $\body_0$ and the deformation map by $\bm \varphi:\body_0\to\mathbb{R}^3$. We define the
deformed body $\body= \bm \varphi(\body_0)$ and the deformation gradient $\bm{\mathrm{F}}=\nabla\bm\varphi$. We denote by $\bm X$ points in $\body_0$ and by $\bm x$ points in $\body$.

Active impulses generated by the contractile actin fibers in ABNs, which consume energy and do work on their surroundings, induce changes in the degree of orientation. Before activation, the tissue is assumed to be isotropic, corresponding to $s\equiv 0$. Based on experimental observations, we assume that, upon activation, the degree of orientation $s(\bm X)$ is prescribed in the reference configuration. The induction of order leads to a non-uniform nematic alignment $\n(\bm X)$ featuring localized topological defects, which correspond to regions of vanishing orientational order. Let $\n_0(\bm X)$ denote the director field in the reference configuration $\body_0$, prescribed in order to reproduce the experimentally observed defects that act as organizing centers of the developing body plan. Provided that the director field is convected by the macroscopic deformation, we have
\begin{equation}
\label{eq:n_deformed}
\n=\frac{\bm{\mathrm{F}}\n_0}{|\bm{\mathrm{F}}\n_0|}.
\end{equation}
We postulate that volume changes in ABNs occur solely as a result of growth. During this process, new material is added at the same density as the existing tissue, ensuring that mass and volume increase proportionally. Consequently, we treat the biological tissue as locally incompressible, implying a spatially uniform and constant mass density 
\begin{equation}
\rho(\bm{x}) = \bar{\rho},
\end{equation}
in the deformed configuration. For modeling purposes, it can be identified with the mass density of water to reflect the primary composition of living tissues. Accordingly, the reference mass density $\rho_0(\bm X)$, defined with respect to the reference configuration $\body_0$, must account for the occurrence of growth despite $\body_0$ remaining geometrically fixed. In this context, mass conservation dictates that
\begin{equation}
\label{eq:rho0}
\rho_0(\bm X) = \bar{\rho}\, J(\bm X),
\end{equation}
where $J(\bm X)$ is the local volume ratio associated with the deformation gradient $\bF$, defined as
\begin{equation}
J(\bm X) = \det \bF(\bm X).
\end{equation}

\subsection{Growth and relaxation}\label{sec:growth_relaxation}

As the strong mechanical stresses arising near topological defects impart instability and drive deformation, growth and structural rearrangements act to relax these internal stresses. Within this framework, growth and remodeling can produce a purposeful morphing of the cellular aggregate. In living organisms, growth mainly results from cell proliferation and enlargement, leading to a local volumetric expansion that is, to a first approximation, isotropic at the cellular scale. To mathematically describe this process, we introduce an isotropic growth tensor
\begin{equation}
\label{eq:Fg_isotropic}
\bm{\mathrm{F}_{\mathrm{g}}} = \sqrt[3]{J_{\mathrm{g}}}\, \bm{\mathrm{I}},
\end{equation}
where $J_{\mathrm{g}}(\bm X)$ governs the local amount of added mass. Because the biological tissue is modeled as locally incompressible, elastic deformations are volume-preserving. Consequently, the macroscopic volume change of the material stems entirely from this growth process, satisfying the kinematic constraint 
\begin{equation}
\label{eq:J_detG}
\det \bm{\mathrm{F}} = \det \bm{\mathrm{F}_{\mathrm{g}}} \equiv J_{\mathrm{g}}.
\end{equation}
To characterize the spatial heterogeneity of mass growth in ABNs, which is more pronounced in regions of elevated mechanical stress, the expansion factor $J_{\mathrm{g}}$ will be modeled as a volumetric expansion source governed by the local stress intensity. This will establish a bidirectional coupling between mechanical stresses (both generated and alleviated by growth) and stress-modulated growth.

To describe the remodeling process during material evolution, we adopt a framework akin to that of viscoelastic fluids with a comparatively long relaxation time. In such systems, applied deformation and growth generate stresses that relax slowly in the absence of motion. Within our quasi-static framework, this continuous stress-release mechanism is captured by a suitable relaxation tensor $\bm{\mathrm{F}_{\mathrm{rel}}}$ that enables the recursive updating of the elastically relaxed material configuration. This will become a central component of our model.

\section{Energetic formulation of active biological networks}\label{sec:energy}

We build on the properties outlined in the previous section and the model developed in \cite{Paparini:2025:PGM}, which establishes a mathematical framework for the coupling between mesogen disclinations and the polymer network. This model proved effective in capturing the emergence of post-instability shapes in close agreement with experimental results, thus providing a robust foundation for describing morphing driven by the interplay between nematic and elastic interactions. Nevertheless, effects such as growth and stress relaxation arising from material reorganization were entirely absent.

Macroscopic shape deformations in ABNs arise from the tissue response to four distinct mechanisms:
\begin{itemize}
\item[(a)] Mechanical stress that develops in the vicinity of topological defects.
\item[(b)] Activity-induced ordering: activity drives local variations in $s$, promoting alignment of the material away from topological defects and thereby increasing $s$.
\item[(c)] Stress-driven growth.
\item[(d)] Structural relaxation of the tissue aimed at relieving internal stresses.
\end{itemize}
Therefore, the total free energy $\free$ will contain two contributions,
\begin{equation}
\free[\bm\varphi]=\free_{\mathrm{n}}[\bm \varphi]+\free_{\mathrm{e}}[\bm \varphi].
\end{equation}
The nematic contribution $\free_{\mathrm{n}}$ captures mechanism (a) and is essential for a comprehensive treatment of defects. The elastic contribution $\free_{\mathrm{e}}$ describes the response of the biological fiber network to deformation, accounting for mechanisms (b), (c), and (d).

\subsection{Nematic energy}\label{sec:BLCE_energy_N}

The first contribution $\free_{\mathrm{n}}[\bm\varphi]$ reflects the nematic nature of the ABN and accounts for spatial variations in the director field $\n$. In the one-constant approximation, it takes the form
\begin{equation}
\label{eq:free_nematic}
\free_{\mathrm{n}}[\bm\varphi]=\int_\body \frac{\kappa}{2}s(\bm\varphi^{-1}(\bm x))^2\left|\nabla_{\bm x}\n(\bm\varphi^{-1}(\bm x))\right|^2\dd\bm x,
\end{equation}
where $\kappa>0$ is an elastic constant characteristic of the material. It quantifies the energetic cost of distortions from the natural state, which, for nematic systems, corresponds to any uniform director field. The functional $\free_{\mathrm{n}}$ is frame-indifferent and even in $\n$, reflecting the head-tail symmetry typical of nematic liquid crystals (LCs). Although the integral is defined on the deformed configuration and depends only on the spatial gradient of the current director $\n$ and the order parameter $s$, we recall that both fields are expressed in terms of the material coordinates $\bm X\in\body_0$. Specifically, $s$ is prescribed on $\body_0$, while $\n$ is determined by the macroscopic deformation through equation \eqref{eq:n_deformed}.
The formulation \eqref{eq:free_nematic} follows Ericksen's continuum theory for nematic LCs \cite{ericksen:liquid}, which provides a unified treatment of defects in arbitrary spatial dimensions. When $s$ vanishes, the molecular orientation becomes fully disordered, indicating a local transition to the isotropic phase without a temperature change; in this framework, defects naturally correspond to localized isotropic regions. While the classical theory, which assumes constant $s$, adequately describes point defects in three-dimensional (3D) geometries, it fails for line and plane defects, where it yields infinite energy. Allowing the degree of orientation $s$ to vary spatially regularizes these singularities in $\n$, enabling the LC to locally transition to the isotropic phase wherever the classical theory predicts divergent energy. Finally, while Ericksen's theory includes an additional energy term proportional to $|\nabla_{\bm X}s(\bm X)|^2$ to account for spatial variations in $s$, this term would merely act as an additive constant in our formulation (since $s$ is prescribed on $\body_0$) and is therefore neglected.

\subsection{Elastic energy with growth and relaxation effects}\label{sec:BLCE_energy_EL}

The second contribution $\free_{\mathrm{e}}$ extends the well-known phenomenological \emph{neoclassical strain-energy function} for liquid crystal elastomers (LCEs) \cite{bladon:deformation,Mihai:2022,warner1988:theory,Warner:2007:WT,warner1991:elasticity}, to account not only for the anisotropic monomer distribution before and after the change in the order parameter from $s\equiv 0$ to $s(\bm X)$, but also for how non-uniform growth (c) and relaxation (d) drive the system out of equilibrium, leading to shape changes. To capture intermediate responses, we must describe both the evolution of the elastically relaxed shape and the actual deformation. A key feature of the neoclassical strain-energy function for LCEs is the multiplicative decomposition of the macroscopic deformation gradient $\bF$ from the reference isotropic state $\body_0$ to the current configuration $\body$ into an elastic distortion followed by a natural (stress-free) shape change. Consequently, the strain-energy density depends only on the elastic part of the deformation and matches that of an isotropic polymer network. We adopt this framework to model the mechanical behavior of ABNs. Within this setting, identifying the elastic distortion tensor is essential, as it determines the stress response of the material. To this end, we draw on theories of thermoelasticity, anelasticity, and growth \cite{Goodbrake:2021:GGY,goriely2017:growth,Lubarda:2004,Sadik:2017:SY}, since the elastic distortion is independent of activation mechanisms associated with changes from $s\equiv0$ to $s(\bm X)\neq0$, and is followed by a stress-free (natural) shape change.

\subsubsection{Multiplicative deformation decomposition}

\begin{figure}[t]
\centering
\includegraphics[width=0.9\linewidth]{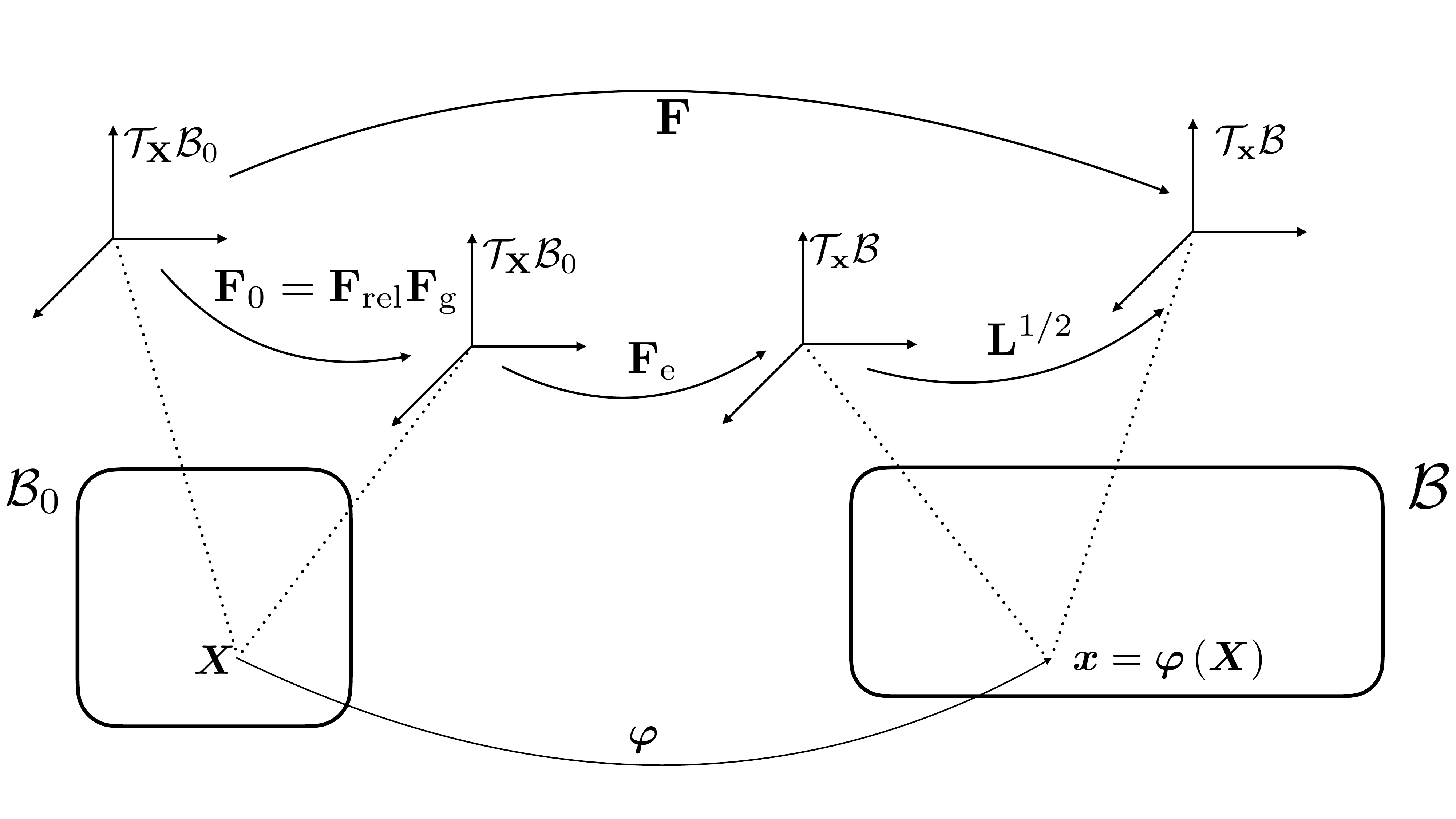}
\caption{Schematics of the multiplicative decomposition of $\bF$ in \eqref{eq:multiplicative_decomposition}.}
\label{fig:LCE_deformation}
\end{figure}

The decomposition of the macroscopic deformation gradient takes the form (see Figure \ref{fig:LCE_deformation})
\begin{equation}
\label{eq:multiplicative_decomposition}
\bF=\bL^{1/2}\Fel\bF_0.
\end{equation}
In the above equation, $\bL$ denotes the nematically induced step-length tensor 
\begin{equation}
\label{eq:step_tensor}
\bL=a^{-1/3}\left[\left(a-1\right)\n\otimes\n+\bm{\mathrm{I}}\right],
\end{equation}
where $a(\bm X)>1$ is a prescribed function of $s(\bm X)$, while $\bm{\mathrm{I}}$ is the identity tensor. The tensor $\bm{\mathrm{L}}^{1/2}$ represents a stress-free geometric transformation induced by mesoscopic reorientation of LC domains and describes a change of frame from the isotropic to the nematic phase. The tensor field $\bF_0$ maps isotropic material line elements into elastically relaxed isotropic line elements. To account for both growth and relaxation of residual stresses generated by local growth and deformation, we further decompose
\begin{equation}
\bF_0=\Frel\bG
\end{equation}
where $\bG$ is the local stress-free volumetric growth tensor described by \eqref{eq:Fg_isotropic}, and $\Frel$ accounts for rearrangements and the resulting relaxation of the polymer network. Notice that $\bF_0$ need not be the gradient of a deformation: it provides a local description of the strain preferred by the material to relax elastic stresses. The remaining elastic tensor $\Fel$ is the local elastic deformation gradient, such that $\det\Fel=1$. The elastic stresses then depend on this tensor, which is given by
\begin{equation}
\Fel=J_{\mathrm{g}}^{-1/3}\bL^{-1/2}\bF\Frel^{-1}.
\end{equation}

Therefore, for our ideal uniaxial ABN, we consider a strain-energy density of the form
\begin{equation}
W_{\mathrm{ABN}}(\bF,\n,s)=W(\Fel),
\end{equation}
where $W(\Fel)$ depends only on the local elastic deformation tensor $\Fel$. This revisits the neoclassical model in \cite{bladon:deformation,warner1988:theory,warner1991:elasticity}, now extended to incorporate both growth and tissue relaxation. The elastic contribution to the free energy is then
\begin{equation}
\label{eq:free_elastic}
\free_{\mathrm{e}}[\bm \varphi]=\int_{\body_0}W_{\mathrm{ABN}}(\bF,\n,s)\dd \bm X=\frac{\mu}{2}\int_{\body_0}\left[\tr\left(\Fel^{\mathrm{T}}\Fel\right)-3\right]\dd \bm X,
\end{equation}
where $\mu$ is the shear modulus at infinitesimal strain. For Hydra tissue, Young's modulus was measured as $E=440$ Pa \cite{Perros:2024:etal}. Then $\mu=E/3=147$ Pa, assuming the tissue is incompressible. The subtraction of $3$ in \eqref{eq:free_elastic} ensures that the stored elastic energy vanishes in the elastic ground state, namely when $\Fel = \mathbf{I}$.

\subsection{Total free energy and coupled mechanisms}

According to \eqref{eq:free_nematic} and \eqref{eq:free_elastic}, we can express the total free energy associated to the deformations of a ABN as
\begin{equation}
\label{eq:free_energy}
\free[\bvarphi]=\int_{\body_0}\left[\frac{\kappa}{2}s(\bm X)^2\left|\nabla\n\bF^{-1}\right|^2 J +\frac{\mu}{2}\left(\tr\left(\Fel^\transp\Fel\right)-3\right)\right]\dd\bm X,
\end{equation}
where the nematic contribution in \eqref{eq:free_energy},
originally expressed as an integral over the deformed configuration, is here rewritten in terms of the reference
configuration. The interplay between elasticity, relaxation, growth, and spontaneous anisotropic distortions in \eqref{eq:free_energy} endows ABNs with mechanical and actuation properties reminiscent of growing biological tissues, thus distinguishing them from conventional elastic materials. 

We find it useful to rescale all lengths according to a characteristic length scale set by $R_{0}$, typically associated with the diameter of $\body_0$. Calling $\overline{\mathcal{B}}_0$ the rescaled initial domain, and letting
\begin{equation}
\label{eq:k_s_def}
\overline{\kappa}=\dfrac{\kappa}{\mu R_{0}^2},
\end{equation}
we arrive at the following dimensionless form of $\free[\bm\varphi]$:
\begin{equation}
\label{eq:free_rescaled}
\overline{\mathcal{F}}[\bm\varphi]=\frac{2\free[\bm\varphi]}{\mu R_{0}^3}=\int_{\mathcal{\overline{B}}_0}\left[\overline{\kappa} s(\overline{\bm X})^2\left|\nabla\n\bF^{-1}\right|^2 J +\tr\left(\Fel^\transp\Fel\right)\right]\dd\overline{\bm X}.
\end{equation}
The constant $\overline{\kappa}$ defined in \eqref{eq:k_s_def} is a measure of the balance between the nematic elasticity, encoded in the elastic constant $\kappa$, and the material stiffness, described by $\mu$.

\section{Application to Hydra regeneration: Defects as morphogenetic drivers}\label{sec:hydra}

We now explore a paradigmatic example of developing organisms that fits within the mathematical framework of ABNs, namely Hydrae. These aquatic animals have a cylindrical body with several tentacles extending from the sides of the feeding end. Hydra develops from a spheroidal shape by activating nematic interactions between its cells, with centers of maximal proliferation located around topological defects, driving elongation of the central body and extension of tentacles. The biological drive to establish a global tangent nematic order on a surface with spherical topology inevitably leads to the emergence of topological defects. According to the global Poincar\'{e}-Hopf theorem, their total topological charge must equal the Euler–Poincar\'{e} characteristic of the surface, which is $+2$ for a sphere. Recent studies on Hydra show a strong correlation between the position of these defects in the actin fibers and the emergence of specific morphological features. In standard uniaxial regeneration, two $+1$ defects appear at opposite poles of the body axis, fixing the sites of the nascent mouth and foot. Moreover, local $+1$ defects occur at the tips of emerging tentacles, while pairs of $-1/2$ defects lie at their bases, acting as topological balancers. 

Interestingly, isotropic spherical shells can regenerate not only into standard uniaxial forms; additional defects in the nematic field can also trigger biaxial regeneration, yielding organisms with two distinct heads.
To address this morphogenetic complexity, we consider the effect of an active impulse on an initially isotropic spherical shell of inner radius $R_{\mathrm{int}}$ and outer radius $R_{\mathrm{out}}$, with holes corresponding to the assumed mouth locations. In accord to \eqref{eq:free_rescaled}, we rescale all lengths by $R_{\mathrm{int}}$ to define the geometry of the dimensionless material domain $\bodynorm$. In this setting, the inner radius becomes $1$, while the dimensionless outer radius is given by $\bar{R} = R_{\mathrm{out}}/R_{\mathrm{int}}>1$.
Based on observations in \cite{maroudas:topological}, we assume that the active impulse, whose mechanism remains unknown, induces order in the ABN, leading to a nematic state characterized by $s(\bm X)\not\equiv 0$ and an orientation field $\n(\bm x)$ with localized topological defects, corresponding to regions where the degree of orientation vanishes. We denote by $\n_0(\bm X)$ a prescribed director field on $\bodynorm$ reproducing the experimentally observed defects that act as organizing centers of the developing body plan; $\n(\bm X)$ then satisfies \eqref{eq:n_deformed}. This setup can be interpreted as cells adopting a degree of orientation set by $s$, with defects forming where $s$ vanishes.

\begin{figure}[htbp]
	\centering
	\begin{subfigure}[c]{\linewidth}
		\centering
		\includegraphics[width=0.7\linewidth]{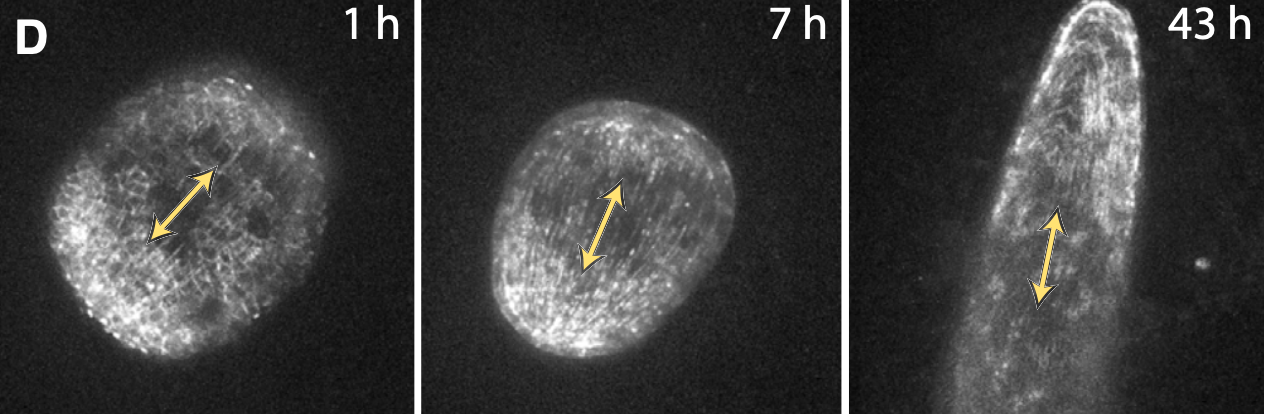}
		\caption{Configuration of topological defects in a uniaxial Hydra that has one head and one foot. The head and the foot each have a $+1$ defect 		at their tip. The arrows indicate the alignment of actin fibers.} 
		\label{fig:hydraonehead}
	\end{subfigure}
	\begin{subfigure}[c]{\linewidth}
		\centering
		\includegraphics[width=0.7\linewidth]{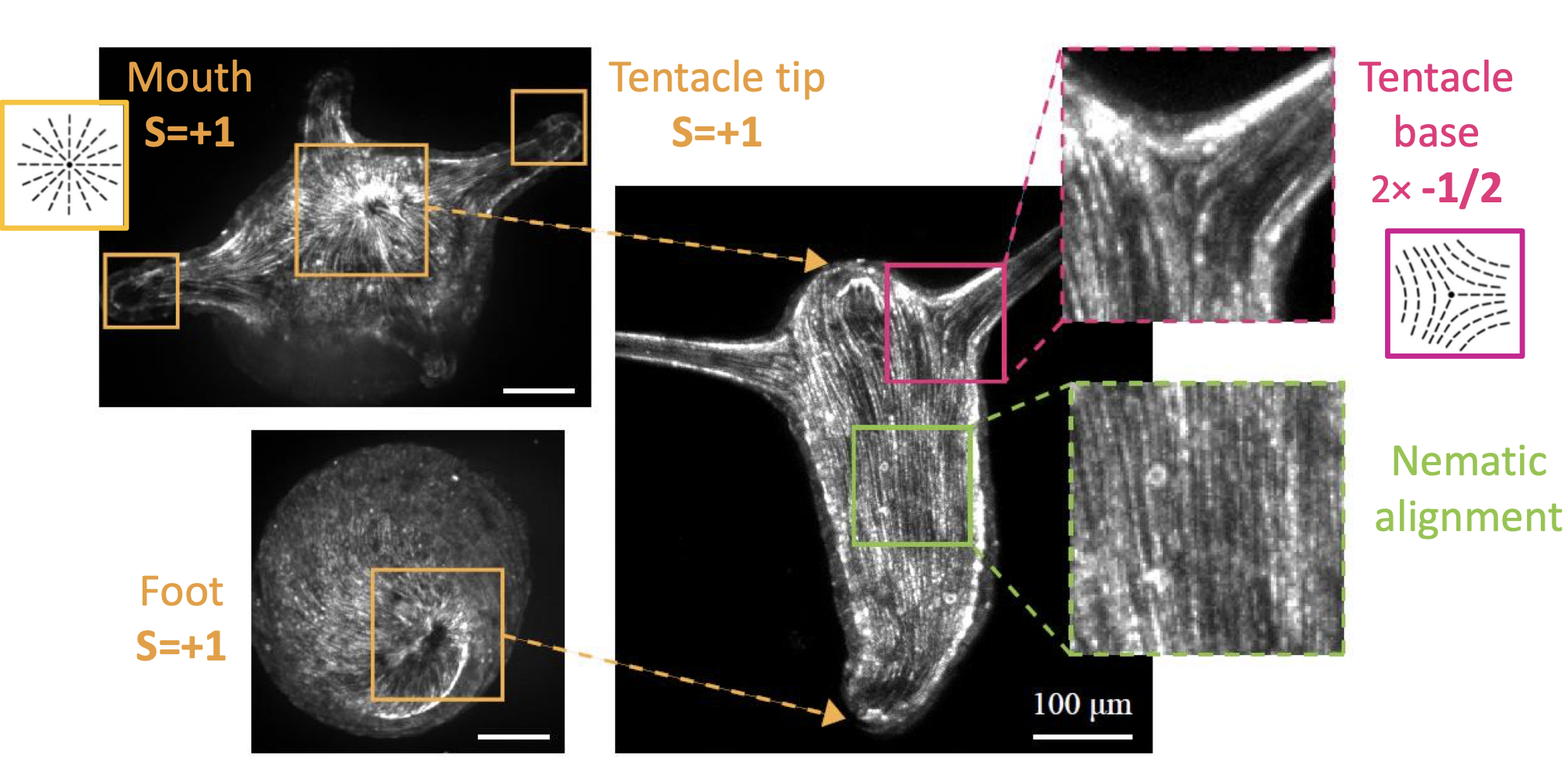}
		\caption{Configuration of topological defects in a uniaxial Hydra that has one head, one foot and two tentacles. The head and the foot each have 		a $+1$ defect at their tip, whereas the tentacles have a $+1$ at the tip, and two $-1/2$ at the base.} 
		\label{fig:hydraexperimental}
	\end{subfigure}
	\begin{subfigure}[c]{\linewidth}
		\centering
		\includegraphics[width=0.7\linewidth]{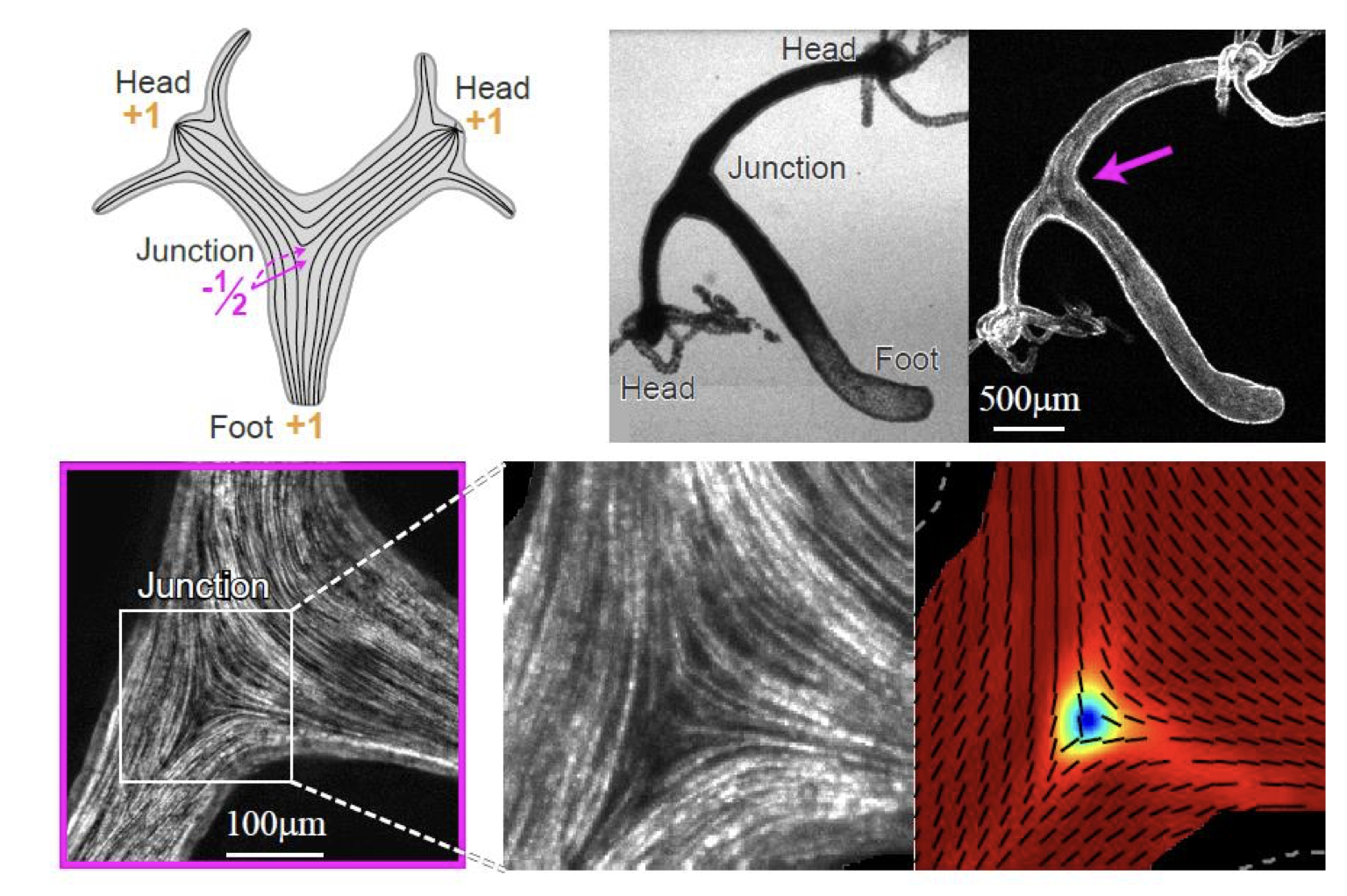}
		\caption{Configuration of topological defects in a multi-axis Hydra that has two heads and a foot. The two heads and the foot each have a $+1$ 		defect at their tip, whereas the junction between the two axes has two $-1/2$ defects.}
		\label{fig:hydratwoheads}
	\end{subfigure}
	\caption{Figures (a)-(c) reprinted or adapted from \cite{livshits:structural} and \cite{maroudas:topological}.}
	\label{fig:regeneratinghydra}
\end{figure}

\subsection{Quasi-static evolution of nematic order and growth in Hydra}

As described in our general ABN framework, tissue regeneration involves the concurrent injection of nematic order and volumetric growth, leading to structural remodeling. Because regeneration proceeds slowly, inertial and viscous effects are neglected. This quasi-static assumption allows the system to be treated as a sequence of mechanical equilibrium states.

More precisely, as active impulses progressively order the shell, we impose a fixed increase in the scalar order parameter $s$ away from defects at each step, together with a reduction in defect core radius. Growth is implemented sequentially by adding new cells at each step, with their distribution tightly correlated with regions of high mechanical stress, concentrated where the director field is most distorted (i.e., near topological defects). This is consistent with experimental evidence showing that stress at defect sites strongly influences Hydra regeneration. As nematic order increases and mass accumulates, the material deforms away from its stress-free spherical reference state, while relaxation continuously updates the elastically relaxed configuration.

Each intermediate step is denoted by $k$, with corresponding quantities marked by a superscript $(k)$. The model thus yields a sequence of equilibrium shapes describing the transition from an initial isotropic spheroid to a mature Hydra. Morphogenetic features such as protrusions and valleys emerge naturally as mechanisms to relieve the mechanical stress generated by growth and topological defects.

\subsection{Finite element formulation and numerical implementation}\label{sec:FEM}

The rescaled reference domain $\bodynorm$ is discretized using a non-uniform tetrahedral mesh $\mathcal{M}_h$. To accurately resolve the high concentration of elastic energy near topological defects, the mesh is locally refined in their vicinity, while remaining comparatively coarse elsewhere to reduce computational cost.
To solve the coupled nonlinear problem in \eqref{eq:free_rescaled}, we employ a mixed finite element formulation implemented in the FEniCSx framework. 

The discrete physical state is represented by the tuple $\bm{w} = (\bm{u}_h, \n_h, \boldsymbol{\lambda}_h, \Lambda_h)$, corresponding to the displacement field, the nematic director, and two Lagrange multipliers. These variables are defined over an extended Taylor–Hood mixed function space, $\mathcal{W}^h = \mathcal{V}_{\bm{u}}^h \times \mathcal{V}_{{\n}}^h \times \mathcal{V}_{\boldsymbol{\lambda}}^h \times \mathcal{V}_{\mu}^h$, with all fields discretized using continuous Lagrange elements. Specifically, second-order elements are used for the displacement $\bm{u}_h$, while first-order elements are employed for $\n_h$, $\boldsymbol{\lambda}_h$, and $\Lambda_h$.
Within this setting, the discrete deformation gradient is given by
\begin{equation} \mathbf{F}_h = \mathbf{I} + \nabla\bm{u}_h.
\end{equation}
The Lagrange multipliers $\boldsymbol{\lambda}_h$ and $\Lambda_h$ enforce the physical constraints of the system. The vector multiplier $\boldsymbol{\lambda}_h$ imposes the kinematic relation
\begin{equation}
\label{eq:n_deformedh}
\n_h = \frac{\mathbf{F}_h \n_0}{|\mathbf{F}_h \n_0|},
\end{equation}
consistent with \eqref{eq:n_deformed}, while the scalar multiplier $\Lambda_h$ acts as a hydrostatic pressure enforcing the volumetric constraint $\det \mathbf{F}_h = J_{\mathrm{g}}$, in agreement with \eqref{eq:J_detG}.
Incorporating these constraints into the scaled functional \eqref{eq:free_rescaled} yields the discrete Lagrangian functional 
\begin{align} \mathcal{L}_h(\bm{u}_h, {\n}_h, \boldsymbol{\lambda}_h, \Lambda_h) &= \int_{\mathcal{M}_h} \left[\Psi_{\mathrm{e}}(\nabla\bm{u}_h,{\n}_h)+\Psi_{\mathrm{n}}(\nabla\bm{u}_h,\nabla{\n}_h)\right] \, d\bm X_{h}\nonumber\\ &+ \int_{\mathcal{M}_h} \left[\boldsymbol{\lambda}_h \cdot \left({\n}_h - \frac{\mathbf{F}_h {\n}_0}{|\mathbf{F}_h {\n}_0|}\right) + \Lambda_h \left( \det(\mathbf{F}_h) - J_{\mathrm{g}} \right)\right] d\bm X_{h} \end{align}
where 
\begin{subequations} \begin{gather} 
\Psi_{\mathrm{e}}=\frac{1}{2}\left[\tr\left({\Fel}_h^\transp{\Fel}_h\right)-3\right],\\
\Psi_{\mathrm{n}}=\frac{\overline{\kappa}}{2}s(\bm X_h)^2\left|\nabla\n_h\bF_h^{-1}\right|^2 J_h
\end{gather} \end{subequations}
denote the discrete elastic and nematic energy densities defined in \eqref{eq:free_elastic} and \eqref{eq:free_nematic}, respectively. 

The dimensionless constant $\overline{\kappa}$, defined in \eqref{eq:k_s_def}, is estimated from experimental data. Although the Frank constant for cells exhibiting nematic order in tissues has not been directly measured \cite{pearce:passive}, several studies have inferred its value from indirect observations. Its magnitude depends on the cell type; for epithelial cells, the product $ks^2$ typically lies in the range $10^3$–$10^4\, \mathrm{pN}$ \cite{lee:crawling,perez:activewetting}. In our simulations, the target value of the order parameter away from defects is $s = 0.4$, which yields an estimate $\kappa \approx 0.625\times 10^{-8}$–$10^{-7}\, \mathrm{N}$ for Hydra tissue.

Finally, using the measured shear modulus $\mu = 1.47\times 10^{2}\, \mathrm{N}/\mathrm{m}^2$ from \cite{Perros:2024:etal} and a spherical geometry with radius $R_0 \approx 160\, \mu\mathrm{m}$—taken as the characteristic length $\radius$ in \eqref{eq:k_s_def}—we obtain $\overline{\kappa} \approx 1.6\times 10^{-3}$–$10^{-2}$.

\subsubsection{Initialization: Stress-free reference state (Step 0)}

The simulation is initialized from a stress-free, undeformed configuration. The tissue is modeled as an isotropic spherical shell, possibly featuring circular apertures representing the organism’s oral poles. In this reference state, neither nematic order nor residual stress is present, and the configuration is specified by
\begin{subequations}
\begin{align}
\mbox{STEP 0:}\quad &\qquad \Frel^{(0)} = \mathbf{I}, \quad &\mbox{(no residual stress)} \\
&\qquad J_{\mathrm{g}}^{(0)} = 1, \quad &\mbox{(absence of growth)} \\
&\qquad s^{(0)}(\bm{X}_h) \equiv 0, \quad \forall\bm{X}_h\in\mathcal{M}_h. \quad &\mbox{(isotropic state)}
\end{align}
\end{subequations}
For the iterative solution of the discrete problem, we initialize the displacement and Lagrange multiplier fields as $\bm{u}_h = \mathbf{0}$, $\boldsymbol{\lambda}_h = \mathbf{0}$, $\Lambda_h = 0$. The director field is initialized as $\n_h = \n_0$, providing the reference orientation. Consistent with the undeformed configuration, no initial volumetric growth is prescribed.

\subsubsection{Boundary conditions and elimination of rigid-body modes}

The total energy \eqref{eq:free_rescaled} is invariant under rigid translations, i.e., it remains unchanged under the addition of a constant vector to $\bm{u}$. As a result, the discrete system exhibits three translational zero-energy modes, which must be removed to avoid singularity of the global stiffness matrix.

To eliminate these modes without over-constraining the morphogenetic dynamics, we impose minimal Dirichlet boundary conditions on the displacement field $\bm{u}_h$. No additional boundary conditions are required for $\n_h$, since its dependence on the deformation gradient and the reference field $\n_0$ is already enforced weakly through the Lagrange multiplier $\boldsymbol{\lambda}_h$ via \eqref{eq:n_deformedh}.

Concretely, we constrain all components of the displacement to vanish within a small neighborhood $\mathcal{N}_h^{\text{SP}}$ of nodes located near the South Pole. Let $\bm{X}_{\text{pole}} = (0, 0, -R_{\text{out}})$ denote the pole position in the reference configuration. The imposed constraints are
\begin{equation}
\label{eq:bc_south_pole}
(\bm{u}_h)_X = (\bm{u}_h)_Y = (\bm{u}_h)_Z = 0
\quad \mbox{on}\quad
\mathcal{N}_h^{\text{SP}} \coloneqq \left\{ \bm{X}_h \in \mathcal{M}_h \;\text{such that}\; |\bm{X}_h - \bm{X}_{\text{pole}}| < \gamma \right\},
\end{equation}
where $\gamma$ is a numerical tolerance proportional to the local mesh size.

\subsubsection{Incremental nonlinear solution strategy}

The equilibrium configuration of the regenerating Hydra is obtained by seeking stationary points of the discrete Lagrangian $\mathcal{L}_h$ with respect to the coupled state variables $\bm{w}_h = (\bm{u}_h, \n_h, \boldsymbol{\lambda}_h, \Lambda_h)$. This leads to a system of nonlinear algebraic equations, which we solve using a Newton method with line search, as implemented in PETSc. To improve robustness in the presence of large deformations and strong nonlinearities associated with the constrained nematic energy, we employ a backtracking line search strategy (\texttt{bt}). Convergence is assessed using both absolute and relative residual norms of the nonlinear system, with tolerances typically set to $10^{-5}$.

\subsubsection{Constitutive modeling}

The mechanical response of the tissue is governed by the following set of constitutive elements, summarized also at the end of this section:
\begin{itemize}
\item \textsc{Nematic Activation:} As specified in \text{STEP 0}, the system initially exhibits a vanishing nematic order parameter, $s^{(0)} \equiv 0$, corresponding to a perfectly isotropic state. For subsequent stages of the morphogenetic process, we prescribe a quasi-static activation law $s^{(k)} = s^{(k)}(\bm{X}_h)$. This spatial distribution describes the gradual emergence of nematic order across the shell, while remaining strictly zero in the vicinity of the oral poles. The specific law for $s^{(k)}$ is detailed according to the specific case under consideration (see Section~\ref{sec:order_parameter}). We further assume that the function $a = a(s)$ in the step-length tensor \eqref{eq:step_tensor} follows the classical constitutive law for thermotropic nematic elastomers. Interpreting the onset of nematic order in Hydra fibers as an effective cooling process, the parameter at step $k$ is given by
\begin{equation} 
\label{eq:aN} a^{(k)}=\frac{1+2s^{(k)}}{1-s^{(k)}}. 
\end{equation}

\item \textsc{Volumetric growth:} Morphogenesis is further driven by stress-dependent volumetric growth. Consistent with the initial condition $J_{\mathrm{g}}^{(0)} = 1$, the system starts from a non-growing configuration. For $N > 0$, the growth factor evolves according to
\begin{equation} 
\label{eq:J_alpha_constitutive}
J_{\mathrm{g}}^{(k)} = J_{\mathrm{g}}^{(k-1)}\exp\left[\alpha\frac{\left(\Psi_e^{(k-1)}\right)^2}{\Psi_0^2+\left(\Psi_e^{(k-1)}\right)^2}\right], 
\end{equation}
where $\alpha$ is a small parameter and $\Psi_e^{(k-1)}$ denotes the elastic energy density at the previous step. In this formulation, elastic energy serves as the driving signal for growth, so that the spatial distribution of newly formed tissue is dynamically governed by the preceding stress field. This choice is consistent with experimental observations in Hydra regeneration, where stress localization—particularly near topological defects—is believed to direct growth and tissue remodeling.
The constitutive dependence on $\Psi_e^{(k-1)}$ in \eqref{eq:J_alpha_constitutive} ensures that growth is negligible at low stress levels; consequently, it becomes progressively localized in regions of high stress, notably at the mouth, foot, and tentacles. The parameter $\Psi_0$ is prescribed to promote gradual growth.

\item \textsc{Relaxation Tensor:} At \text{STEP 0}, the isotropic spherical configuration is stress-free, and the relaxation tensor is initialized as $\mathbf{F}_{\mathrm{rel}}^{(0)} = \mathbf{I}$. As the simulation evolves, nematic ordering and volumetric growth induce deformations that are incompatible with the current relaxed configuration, thereby generating residual stresses \cite{huang2025:modelling}. This is reflected in the elastic deformation tensor,
\begin{equation} 
\Fel^{(k)}=(J^{(k)})^{-1/3}\left(\bL^{(k)}\right)^{-1/2}\bF^{(k)}\left(\Frel^{(k-1)}\right)^{-1}\neq\Id .
\end{equation}
To account for material reorganization, we update the mechanical ``memory'' incrementally. At each step $k$, we solve the coupled variational problem to obtain the equilibrium fields $\bm{u}_h^{(k)}$ and $\n_h^{(k)}$, which define the current geometry and internal structure. Assuming a separation of timescales, where mechanical relaxation is fast compared to growth and nematic activation, the system reaches a local energy minimum before further evolution. We then assume that the material relaxes internally, adopting the current configuration as a new stress-free state for step $k+1$. This is enforced by requiring
\begin{equation}
\Fel^{(k)}|_{\Frel^{(k-1)}=\Frel^{(k)}}\equiv\Id
\end{equation}
which yields the update rule
\begin{equation}
\label{eq:Fpl_stepN}
\Frel^{(k)} = (J^{(k)})^{-1/3} \left(\mathbf{L}^{(k)}\right)^{-1/2} \mathbf{F}^{(k)}.
\end{equation}
All quantities on the right-hand side are known from the solution at step $k$.
\end{itemize}

\subsubsection{Incremental solution procedure (Step $k$)}
The simulation proceeds through an incremental loop of $\overline{N}$ steps. At each step $k$, the solver performs the following sequence of operations:
\begin{subequations}
\begin{align}
\mbox{STEP }k: \quad & \mbox{INPUT} \nonumber\\[2pt]
& \Frel^{(k-1)}, \nonumber\\
& \Psi_{\mathrm{e}}^{(k-1)} \longrightarrow J_{\mathrm{g}}^{(k)} = e^{\alpha N \Psi_{\mathrm{e}}^{(k-1)}} 
\quad \mbox{(stress-driven exponential growth)}, \nonumber\\
& s^{(k)} = s^{(k)}(\bm{X}_h), 
\quad \forall\, \bm{X}_h \in \mathcal{M}_h,
\quad s^{(k)} \equiv 0\ \mbox{near the poles}. \nonumber\\[6pt]
& \mbox{OUTPUT} \nonumber\\[2pt]
& \left\{
\begin{aligned}
\mathbf{F}^{(k)}(\bm{X}) \\
\mathbf{L}^{(k)}
\end{aligned}
\right.
\;\longrightarrow\;
\Frel^{(k)} 
\;\text{(see \eqref{eq:Fpl_stepN})}, 
\quad \Psi_{\mathrm{e}}^{(k)}.
\end{align}
\end{subequations}
To enhance the robustness and convergence of the Newton–line search algorithm, we employ a numerical continuation strategy. For each step $k > 0$, the initial guess for the state variables $\bm{w}_h^{(k)} = (\bm{u}_h^{(k)}, \n_h^{(k)}, \boldsymbol{\lambda}_h^{(k)}, \Lambda_h^{(k)})$ is taken as the solution from the previous step, $\bm{w}_h^{(k-1)}$. This choice is justified by the quasi-static nature of the morphogenetic process: because the updates in volumetric growth $J_{\mathrm{g}}$ and nematic order $s$ are applied incrementally, the equilibrium configuration at step $k$ is expected to remain close to that at step $k-1$. 

\subsection{Initialization of topological defect configurations}

To reproduce the experimental scenarios of interest, we prescribe an initial director field $\n_{0,h}(\bm{X}_h)$ on the computational mesh $\mathcal{M}_h$ that captures the topology of the observed defect configurations, which play a key role in driving the emergence of specific morphological features. Specifically, the discrete field $\n_{0,h}$ is obtained by interpolating an analytically defined vector field $\n_0$ onto the reference mesh. This field is tangent to the outer spherical boundary and is extended uniformly across the shell thickness. Consistently with the continuum framework, the prescribed order parameter $s^{(k)}(\bm{X}_h)$ is required to vanish at the locations of the topological defects on $\mathcal{M}_h$.

\subsubsection{Analytical construction of the initial director field} \label{sec:director_field}

To describe orientational order on the surface of a sphere of radius $R$, we define the director field $\n_0$ in three-dimensional (3D) Cartesian space. On closed surfaces such as the sphere, topological defects are unavoidable: by the Poincar\'{e}–Hopf theorem, the sum of the topological charges $\chi$ of any tangent vector field must equal the Euler characteristic, i.e., $\sum_j \chi_j = +2$.

To prescribe defect positions and charges analytically, we employ a \emph{stereographic projection} from the South Pole ($Z = -R$). As detailed in Appendix~\ref{sec:stereographic}, this maps a point $\mathbf{p} = (X, Y, Z)$ on the sphere to planar coordinates $(X_s, Y_s)$ via
\begin{equation}
\label{eq:projector_def}
    X_s = \frac{X}{R + Z}, \quad Y_s = \frac{Y}{R + Z}.
\end{equation}

Rather than using the spherical basis $(\mathbf{e}_\vartheta, \mathbf{e}_\varphi)$, which introduces coordinate singularities at both poles, we pull back the constant Cartesian basis of the projection plane onto the sphere. This construction yields a smooth orthonormal tangent basis $(\mathbf{e}_1, \mathbf{e}_2)$ at each point $\mathbf{p}$, expressed in the Cartesian frame $(\mathbf{e}_x, \mathbf{e}_y, \mathbf{e}_z)$ as
\begin{equation}
\label{eq:E_12_def}
    \e_1 = \left( 
    \begin{array}{c}
    1 - \dfrac{X^2}{R(R+Z)}\\ 
    \\
    -\dfrac{XY}{R(R+Z)}\\ 
    \\
    -\dfrac{X}{R}
    \end{array}
    \right), \qquad 
    \e_2 = \left(
    \begin{array}{c}
    -\dfrac{XY}{R(R+Z)}\\
    \\
    1 - \dfrac{Y^2}{R(R+Z)}\\
    \\
    -\dfrac{Y}{R}
    \end{array}
    \right).
\end{equation}
This basis is smooth everywhere except at the South Pole, where the stereographic projection is singular; the singularity manifests as a background topological charge of $+2$. In particular, the basis is regular at the North Pole, ensuring that no spurious defects arise there (e.g., in the two-head configuration).

We now construct a director field $\n_0$ containing $N$ prescribed defects away from the South Pole, together with an additional defect that emerges at the South Pole. Mapping the defect locations onto the stereographic plane as $(X_{s,i}, Y_{s,i})$, we define an angle field $w$ by superposing their contributions:
\begin{equation}
\label{eq:omega_def}
    {w}(X_s,Y_s) = \sum_{i=1}^{N} q_i \arctan\left(Y_s - Y_{s,i}, X_s - X_{s,i}\right) + {w}_0,
\end{equation}
where $q_i$ are the nominal defect charges and $w_0$ sets the global orientation.

The director field is then obtained as a unit vector in the local tangent plane,
\begin{equation}
\label{eq:n_0_def}
    \n_0 = \cos{w} \e_1 + \sin{w} \e_2.
\end{equation}

Because the chosen basis contributes an intrinsic charge of $+2$ at the South Pole and zero at the North Pole, the physical defect charges $C_i$ on the sphere need not coincide with the nominal planar charges $q_i$. Specifically:
\begin{itemize}
    \item \emph{Away from the South Pole (including the North Pole):} the physical and nominal charges coincide, $C_i = q_i$. In particular, the North Pole corresponds to $(X_s, Y_s) = (0,0)$ and is treated as a regular defect location if included in equation~\eqref{eq:omega_def}.
    \item \emph{At the South Pole:} since this point maps to infinity in the stereographic plane, its effective charge arises from both the intrinsic $+2$ contribution of the basis and the far-field behavior of $w$, giving 
    \begin{equation}
      C_{\text{south}} = 2 - \sum_{i=1}^{N} q_i.
    \end{equation}
    This defect need not be explicitly introduced; it emerges automatically from the construction.
\end{itemize}
Consequently, by appropriately choosing the charges $q_i$ and their positions $(X_{s,i}, Y_{s,i})$ in the stereographic plane, one can realize arbitrary defect configurations on the sphere with precise control over their locations and total topological charge.

\subsubsection{Spatial distribution of the scalar order parameter}\label{sec:order_parameter}

The scalar order parameter $s^{(k)}(\bm{X}_h)$ quantifies the local degree of nematic alignment at iteration step $k$. Away from defects, it assumes a uniform bulk value,
\begin{equation}
\label{eq:bar_s}
s_0^{(k)}=\overline s\frac{k}{\overline N},
\end{equation}
where $\overline{s}$ is the target order parameter and $\overline{N}$ the total number of steps. As $k$ increases, $s_0^{(k)}$ grows linearly, modeling the progressive establishment of nematic order within the tissue.

At defect cores, however, orientational order necessarily breaks down and the material locally melts, implying $s \to 0$. To capture this behavior, we introduce a smooth spatial damping around each defect. Let $\bm{X}_i = (X_i, Y_i, Z_i)$ denote the position of the $k$-th defect on the sphere. The scalar order parameter at a mesh point $\bm{X}_h$ is then defined as the bulk value \eqref{eq:bar_s} modulated by Gaussian-like attenuation factors associated with all defects:
\begin{equation}
\label{eq:sN_def}
    s^{(k)}(\bm{X}_h) = s_0^{(k)} \prod_{i} \left( 1 - e^{-\delta \|\bm{X}_h - \bm{X}_i\|^2} \right)
\end{equation}
where the constant parameter $\delta$ controls the defect core size, with larger values corresponding to more localized (narrower) cores.

The product runs over all physical defects with nonzero topological charge $C_i$. In the head–tail configuration (Section~\ref{sec:HeadTail}), this includes the explicit planar defect at the North Pole together with the intrinsic defect at the South Pole. In the \emph{two-head} configuration (Section~\ref{sec:TwoHeads}), it comprises the four imposed planar defects and the emergent defect at the South Pole. Finally, for the \textit{Hydra with tentacles} configuration (Section~\ref{sec:Tentacles}), the product extends over the six planar defects along with the intrinsic defect at the South Pole.

\subsection{Tensorial representation of the nematic director field}

A defining feature of nematic systems is their \emph{head-tail symmetry}, whereby the configurations $\n$ and $-\n$ are physically indistinguishable. As a consequence, topological defects may carry half-integer charges: for instance, encircling a $-1/2$ defect results in a rotation of the director by $\pi$. Any continuous representation of the director field $\n$ must therefore include a discontinuity line (branch cut) across which the field flips sign, $\n \to -\n$. Although this discontinuity is physically irrelevant—since the energy is invariant under this transformation—it poses a numerical challenge in finite element frameworks such as FEniCSx. When standard continuous Lagrange elements are used to interpolate $\n$ across such a sign change, the solver enforces continuity by smoothly interpolating between $\n$ and $-\n$. This forces the field through zero, leading to an artificial suppression of its magnitude ($|\n| = 0$) along the branch cut and degrading the quality of the solution, particularly in visualization.

To circumvent this issue, we reformulate the problem in terms of a symmetric rank-2 tensor field defined by the dyadic product
\begin{equation}
\label{eq:M_def}
\mathbf{M} = \n \otimes \n.
\end{equation}
By construction, this representation is invariant under sign reversal, since $(-\n) \otimes (-\n) = \n \otimes \n$. As a result, the tensor field $\mathbf{M}$ remains continuous across branch cuts associated with half-integer defects, eliminating spurious discontinuities. This smooth representation prevents the finite element interpolation from artificially driving the field to zero.

Accordingly, in FEniCSx, we define the function space over continuous tensor fields, discretized using first-order (degree-1) Lagrange elements.

\subsubsection{Construction of the initial tensor field $\mathbf{M}_0$}

If the orientational component of the mixed formulation were expressed directly in terms of the director field $\n$, a natural initial condition would simply be $\n = \n_0$ as defined in equation~\eqref{eq:n_0_def}. In the tensor-based formulation, this translates to prescribing the initial state as the dyadic product
\begin{equation}
\label{eq:M_0omega}
\mathbf{M}_0 = \n_0 \otimes \n_0.
\end{equation}
However, since the director angle $w$ is defined only modulo $\pi$, a direct expansion of this expression in terms of $w$ may introduce artificial branch cuts. To consistently account for this $\pi$-periodicity, we instead rewrite the components using double-angle trigonometric identities. This yields a smooth, branch-cut-free representation on the sphere, expressed entirely in terms of $2w$:
\begin{equation}
        \mathbf{M}_0=\frac{1 + \cos(2{w})}{2} (\e_1 \otimes \e_1) + \frac{\sin(2{w})}{2} (\e_1 \otimes \e_2 + \e_2 \otimes \e_1) + \frac{1 - \cos(2{w})}{2} (\e_2 \otimes \e_2).
 \end{equation}
With the director now represented by $\mathbf{M}$, the inverse shape tensor $\mathbf{L}^{-1}$ is obtained by the substitution $\n \otimes \n \to \mathbf{M}$, giving
\begin{equation}
    \mathbf{L}^{-1} = (a^{-2/3} - a^{1/3})\mathbf{M} + a^{1/3}\mathbf{I}.
\end{equation}
To express the Frank elastic energy in terms of $\mathbf{M}$, we use the identity
\begin{equation}
\label{eq:identity_grads}
    |\nabla_{\bm{x}} \n|^2 = \frac{1}{2} |\nabla_{\bm{x}} \mathbf{M}|^2,
\end{equation}
where $\bm{x}$ denotes coordinates in the deformed configuration (see Appendix~\ref{sec:identity_grads} for a proof). Pulling this expression back to the reference configuration $\body_0$, the nematic free energy \eqref{eq:free_nematic} for the active body takes the form
\begin{equation}
\label{eq:free_nematic_active}
    \free_{\mathrm{n}}[\bvarphi] = \int_{\body_0} \frac{\kappa}{4} s(\bm{X})^2 \left| (\nabla\mathbf{M})\mathbf{F}^{-1} \right|^2 J \dd\bm{X}.
\end{equation}

\subsubsection{Enforcement of the nematic kinematic constraint}

In nematic polymer networks, the director field is intrinsically coupled to the macroscopic deformation of the polymer matrix. In particular, the reference director $\n_0$ is transported by the deformation gradient $\mathbf{F}$ into its current configuration $\n$, according to the kinematic relation given in equation~\eqref{eq:n_deformed}.

Substituting this relation into the definition $\mathbf{M} = \n \otimes \n$ yields a deformation-induced tensor, which we describe by
\begin{equation}
    \mathbf{M}_{\mathrm{kin}} = \frac{1}{|\mathbf{F}\n_0|^2} (\mathbf{F}\n_0) \otimes (\mathbf{F}\n_0).
\end{equation}
Using standard properties of tensor products, the numerator can be expressed entirely in terms of the reference tensor $\mathbf{M}_0$ (defined in equation~\eqref{eq:M_0omega}) as $\mathbf{F}\mathbf{M}_0\mathbf{F}^T$. The normalization factor in the denominator is simply the trace of this push-forward tensor. Accordingly, $\mathbf{M}_{\mathrm{kin}}$ can be written in the compact form
\begin{equation}
\label{eq:M_kin}
    \mathbf{M}_{\mathrm{kin}} = \frac{\mathbf{F} \mathbf{M}_0 \mathbf{F}^T}{\mathrm{tr}(\mathbf{F} \mathbf{M}_0 \mathbf{F}^T)}.
\end{equation}
In the mixed formulation, $\mathbf{M}$ is treated as an independent continuous field. The original vectorial kinematic constraint \eqref{eq:n_deformed} is therefore replaced by its tensorial counterpart, $\mathbf{M} = \mathbf{M}_{\mathrm{kin}}$. To enforce this condition weakly within the energy functional, we introduce a tensor-valued Lagrange multiplier $\boldsymbol{\lambda}$, discretized with first-order (degree-1) Lagrange elements, leading to the contribution
\begin{equation}
    \Psi_\lambda = \int_{\body_0} \boldsymbol{\lambda} \cdot \left( \mathbf{M} - \mathbf{M}_{\mathrm{kin}} \right) \dd\bm{X}.
\end{equation}

\section{Specific configurations}\label{sec:config}

By appropriately selecting the planar charges $q_i$ and their positions $(X_{s,i}, Y_{s,i})$, we can realize a range of distinct morphological states. We focus here on three configurations of interest: the head–foot configuration, the two-head configuration, and the tentacular configuration.

\subsection{Uniaxial (head-foot) configuration}\label{sec:HeadTail}

Here, the computational domain represents a uniaxial regenerating Hydra tissue fragment. Two $+1$ topological defects emerge at opposite poles, corresponding to the sites of the future mouth and foot during regeneration \cite{maroudas:topological,maroudas2025:mechanical}. Guided by the inherited tissue polarity, the apical defect determines the precise location of tissue rupture and the subsequent formation of the hypostome (oral pore), whereas the basal pole remains sealed, giving rise to the foot. To capture the biomechanical state of the tissue during uniaxial regeneration, we model the domain as a thick spherical shell with inner radius $R_{\mathrm{in}} = 1.0$ and outer radius $R_{\mathrm{out}} = 1.1$. A circular aperture of radius $R_{\mathrm{small}} = 0.1$ is introduced at the apical pole, representing the nascent oral pore located precisely at the +1 defect. As nematic order progressively develops, distortions in the director field generate a localized concentration of mechanical stress near the defect. A central result of our simulations is that this stress accumulation naturally drives the gradual enlargement of the apical aperture.

\subsubsection{Initial defect topology for the head-foot configuration}

The initial head–tail configuration is characterized by two $+1$ topological defects located at the North and South poles. Following the mapping procedure outlined above, the discrete initial director field $\n_{0,h}$ on the reference mesh is constructed by interpolating the analytical field $\n_0$ defined in equation~\eqref{eq:n_0_def}. This construction is implemented by introducing a single nominal defect ($N = 1$) with topological charge $q_1 = +1$ at the North Pole. Under stereographic projection, this corresponds to positioning the defect at the origin of the plane, $(X_{s,1}, Y_{s,1}) = (0,0)$. The geometry of the pulled-back basis then automatically enforces the correct effective charge at the South Pole, yielding $C_{\text{south}} = 2 - q_1 = +1$.
Setting the global phase to ${w}_0 = 0$ ensures that the resulting orientational field is everywhere tangent to the meridians along each radial section.

\subsubsection{Nematic ordering law and spatial activation}

The general expression for the scalar order parameter at step $k$ is given in equation~\eqref{eq:sN_def}. At the initial stage ($k = 0$), $s^{(0)}(\bm X_h) \equiv 0$, corresponding to a fully disordered, isotropic tissue.
For the head–tail configuration, we model the experimentally observed scenario in which an excised Hydra tissue fragment folds into a spheroid while nematic order remains spatially localized. Specifically, order does not emerge uniformly across the domain, but is initially confined to a central latitudinal band, with the polar caps remaining fully disordered ($s = 0$).

To capture this behavior, at each step $k$, the order parameter $s^{(k)}(\bm X_h)$ applied to the computational mesh is obtained by modulating the baseline value from equation~\eqref{eq:sN_def} with a spatial mask. This mask is constructed to ensure that $s^{(k)}$ coincides with equation~\eqref{eq:sN_def} within the active nematic band, vanishes identically at the polar caps, and varies smoothly between these regions.

As $k$ increases, two coupled processes take place. First, the magnitude of nematic order grows locally, reflecting the progressive establishment of orientational order as prescribed by equation~\eqref{eq:sN_def}. Second, the ordered region expands toward the poles. The extent of the isotropic caps is controlled by
\begin{equation}
\varepsilon^{(k)}=\frac{\pi}{2}\mathrm{exp}\left(-\frac{k^2}{\overline{N}}\right)
\end{equation}
such that the disordered regions correspond to polar angles $\vartheta(\bm{X}_h) \in [0, \varepsilon^{(k)}) \cup (\pi - \varepsilon^{(k)}, \pi]$, where $\bm{X}_h=(X_h,Y_h,Z_h)$. 

As the simulation progresses, $\varepsilon^{(k)}$ decreases monotonically in $k$, approaching $0$ at sufficiently advanced stages. Physically, this evolution represents the gradual retraction of the isotropic caps and the concomitant expansion of the ordered phase.

With these boundaries defined, the spatial mask can be written explicitly. To ensure a physically consistent nematic–isotropic transition, the mask enforces a smooth decay of the order parameter near the edges of the active band. The resulting field on the discrete mesh is
\begin{equation} 
\label{eq:sN_field_arctan}
s^{(k)} \, \to \, \frac{s^{(k)}}{2} \left[ \tanh\big(k(\cos\vartheta_h + \cos\varepsilon^{(k)})\big) - \tanh\big(k(\cos\vartheta_h - \cos\varepsilon^{(k)})\big) \right],
\end{equation}
where $\cos\vartheta_h = Z_h/\sqrt{X_h^2+Y_h^2+Z_h^2}$. The constant parameter $k > 0$ is chosen sufficiently large to produce a sharp yet smooth transition between the ordered nematic band and the disordered polar caps.

\begin{figure}[htbp]
    \centering
    \captionsetup[subfigure]{justification=centering} 
 \includegraphics[width=\textwidth]{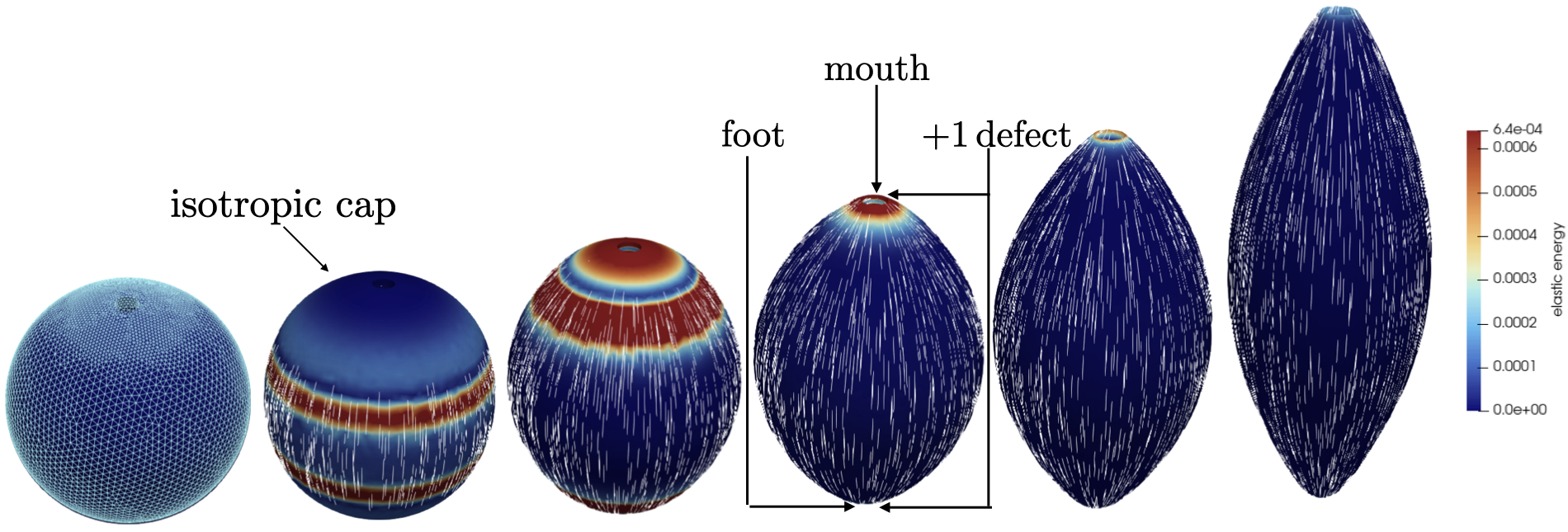}
    \vspace{0.3cm}    
    \includegraphics[width=\textwidth]{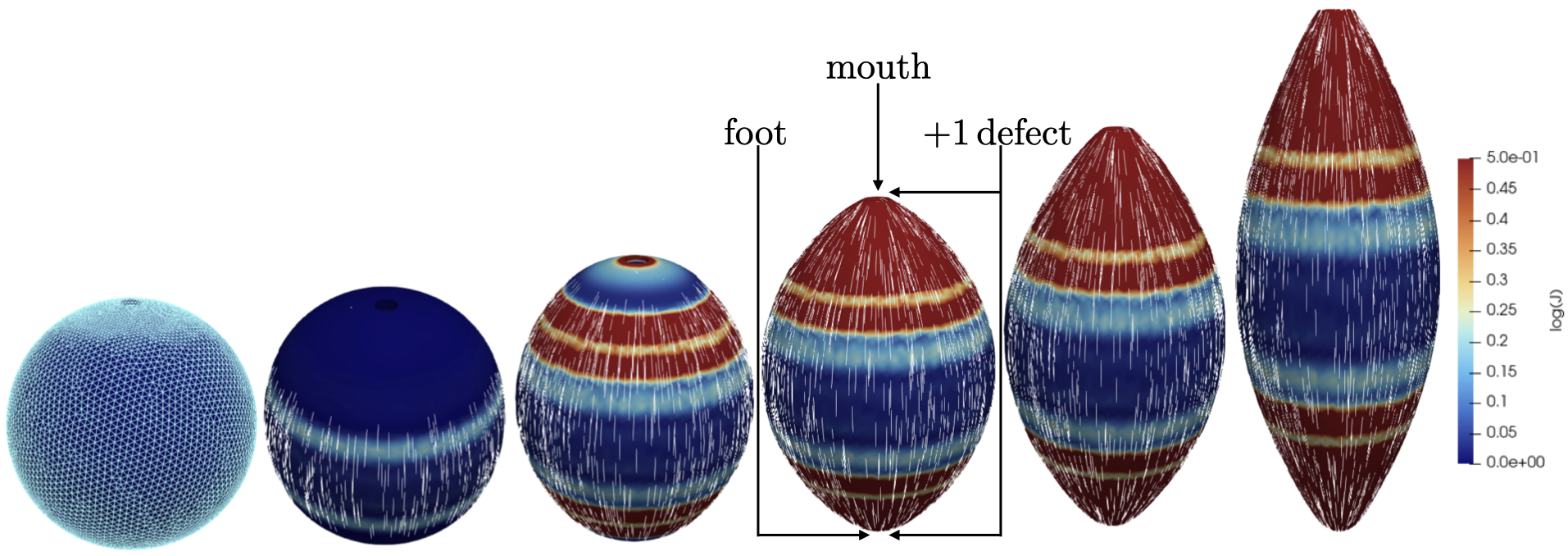}
    \caption{$\overline N=25$, $\overline s=0.4$, $\kappa=0.01$. Successive equilibrium configurations illustrating the quasi-static morphological deformation of the Hydra tissue fragment (from left to right) for steps $k \in \{0, 3, 5, 8, 15, 25\}$. The corresponding parameters take the values $s_0^{(k)} \in \{0, 0.04, 0.1, 0.14, 0.25, 0.4\}$ and $\varepsilon^{(k)} \in \{\pi/2, 1.34, 0.578, 0.121, 0, 0\}$, respectively. The sequences show the stepwise expansion of the active nematic band, with local nematic alignment visualized using short rod elements. In parallel, the isotropic polar caps progressively shrink, disappearing at the prospective mouth while persisting as a small residual region at the prospective foot. This evolution is accompanied by a gradual elongation of the spheroid along its principal body axis. The color map in the top panel shows the spatial distribution of local mechanical stress. Two high-stress halos are observed to propagate symmetrically from the equator toward the apical and basal poles, closely following the outward expansion of the active nematic band. The color map of $\log(J_{\mathrm{g}}^{(k)})$ in the bottom panel, instead, shows material accumulation near the newly formed nematic regions as the isotropic caps shrink toward the poles, with subsequent localization around the mouth and foot. Indeed, growth is stress-driven: new material is preferentially incorporated in regions of elevated mechanical stress.}
    \label{fig:extending_active_band}
\end{figure}

\begin{figure}[htbp]
        \centering
    \captionsetup[subfigure]{justification=centering} 
        \begin{subfigure}[b]{0.32\textwidth}
        \centering
        \includegraphics[width=\linewidth]{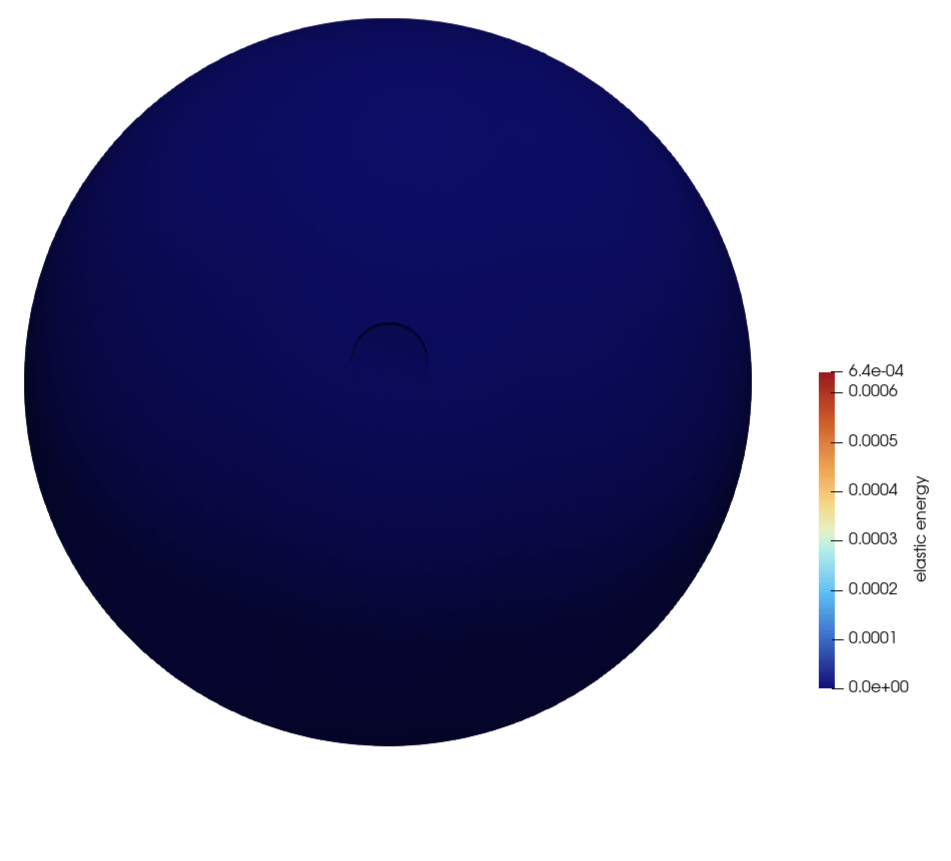}
        \caption{$k=0$, $s_0^{(0)}\equiv0$,\\
         $\varepsilon^{(0)}=\pi/2$.}
    \end{subfigure}
    \hfill
    \begin{subfigure}[b]{0.32\textwidth}
        \centering
        \includegraphics[width=\linewidth]{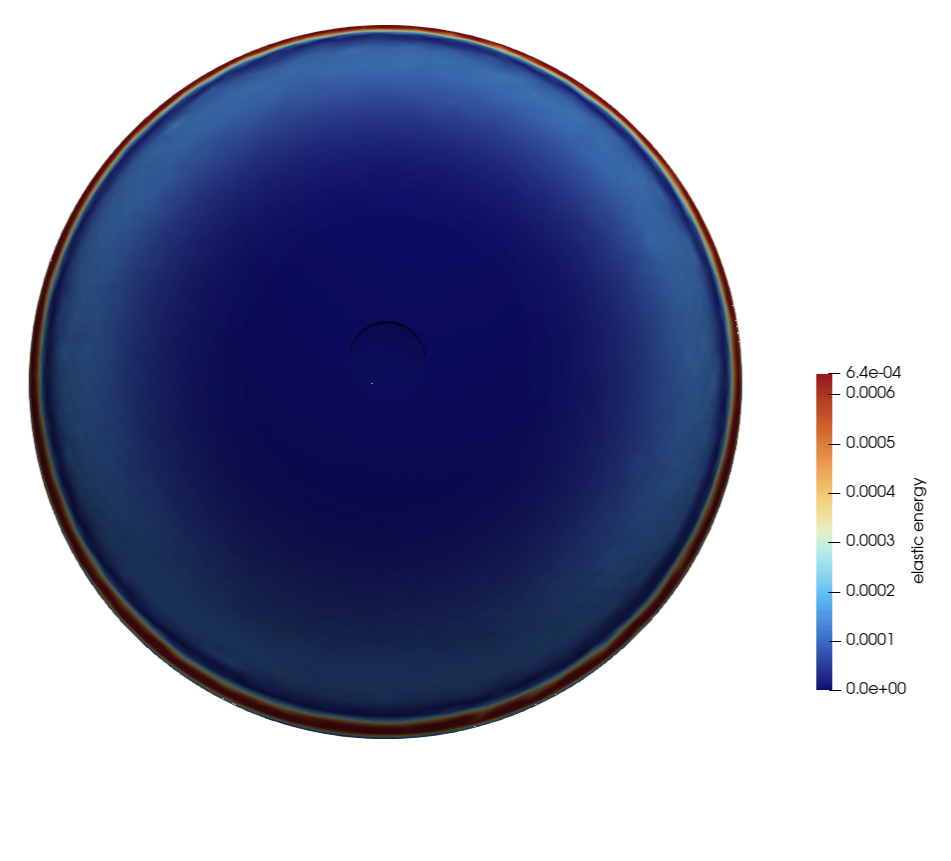}
        \caption{$k=3$, $s_0^{(2)}=0.04$, \\
        $\varepsilon^{(2)}=1.34$}
    \end{subfigure}
    \hfill
    \begin{subfigure}[b]{0.32\textwidth}
        \centering
        \includegraphics[width=\linewidth]{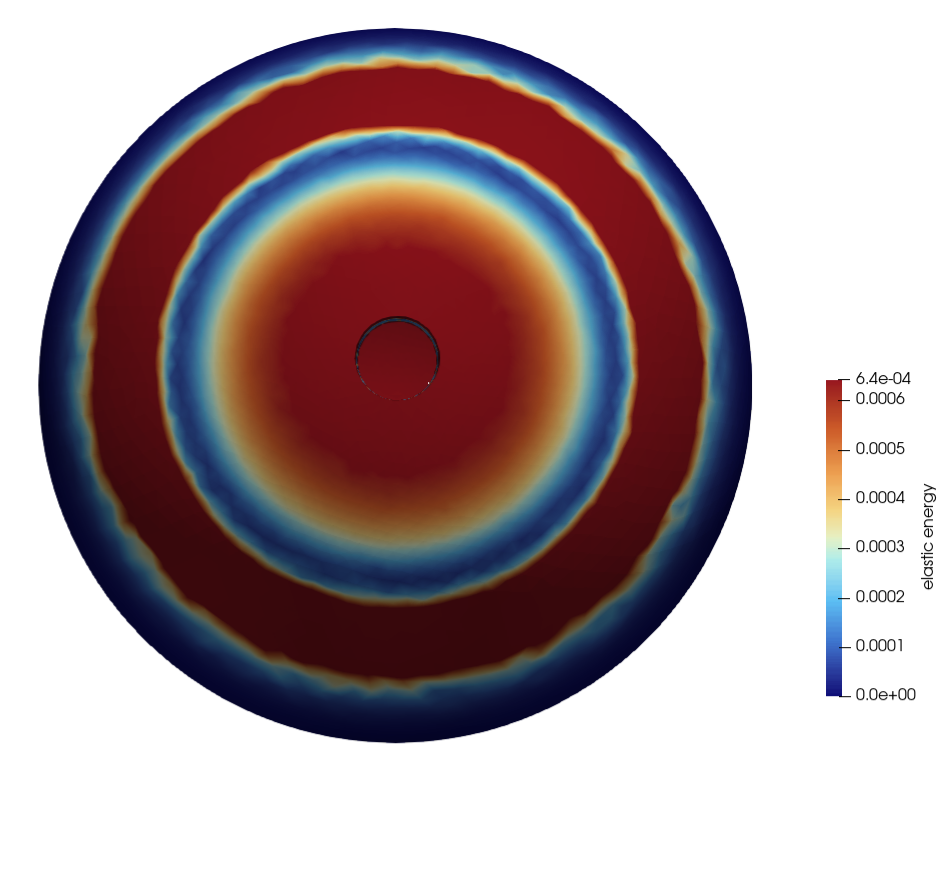}
        \caption{$k=5$, $s_0^{(5)}=0.1$, \\
        $\varepsilon^{(5)}=0.578$}
    \end{subfigure}
    \vspace{1.5cm} 
    \begin{subfigure}[b]{0.32\textwidth}
        \centering
        \includegraphics[width=\linewidth]{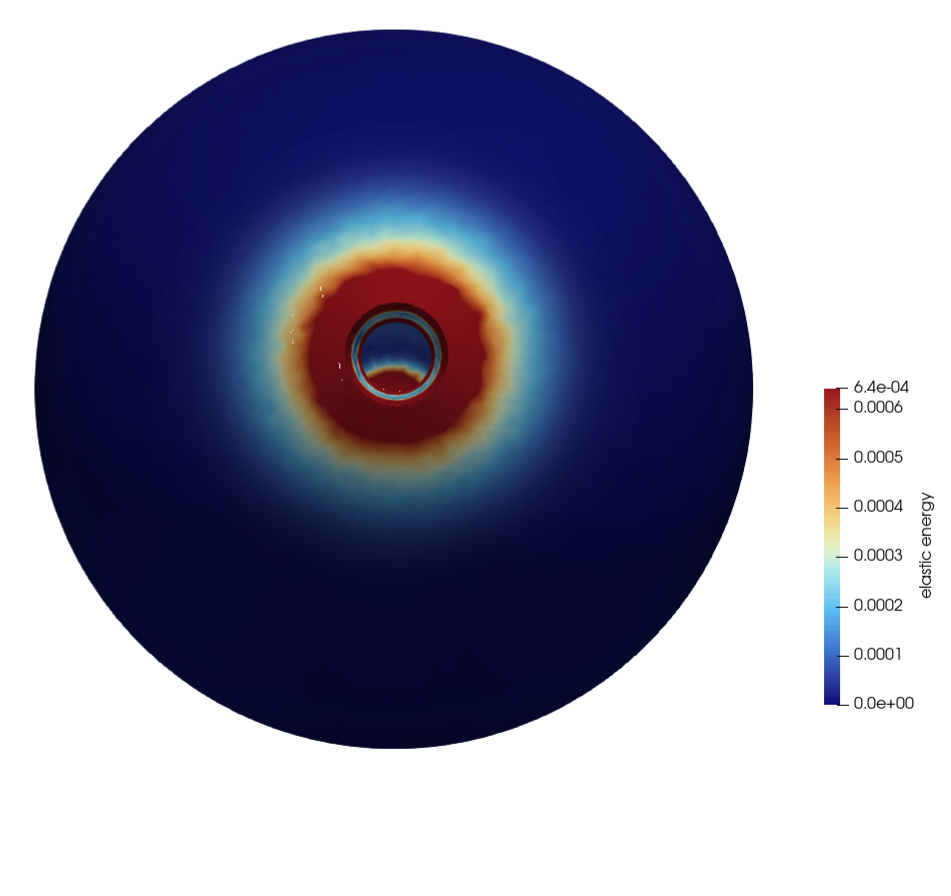}
        \caption{$k=8$, $s_0^{(8)}=0.14$,\\
        $\varepsilon^{(8)}=0.121$}
    \end{subfigure}
    \hfill
    \begin{subfigure}[b]{0.32\textwidth}
        \centering
        \includegraphics[width=\linewidth]{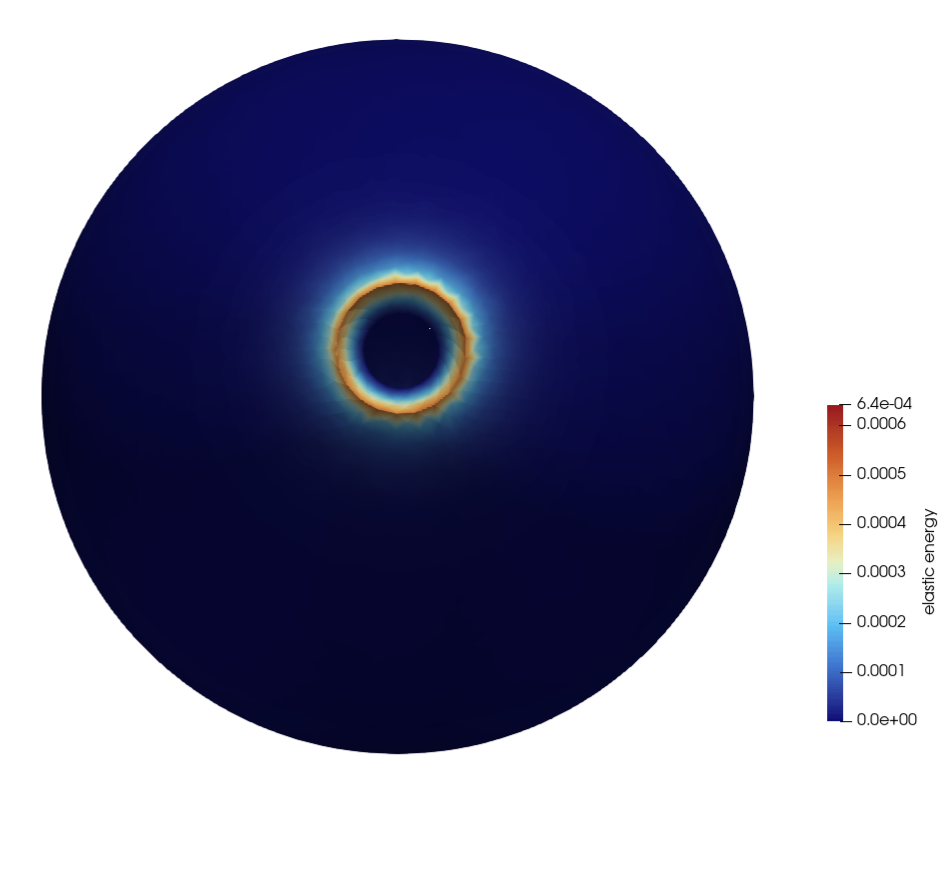}
        \caption{$k=15$, $s_0^{(15)}=0.25$,\\
        $\varepsilon^{(15)}=0$}
    \end{subfigure}
    \hfill
    \begin{subfigure}[b]{0.32\textwidth}
        \centering
        \includegraphics[width=\linewidth]{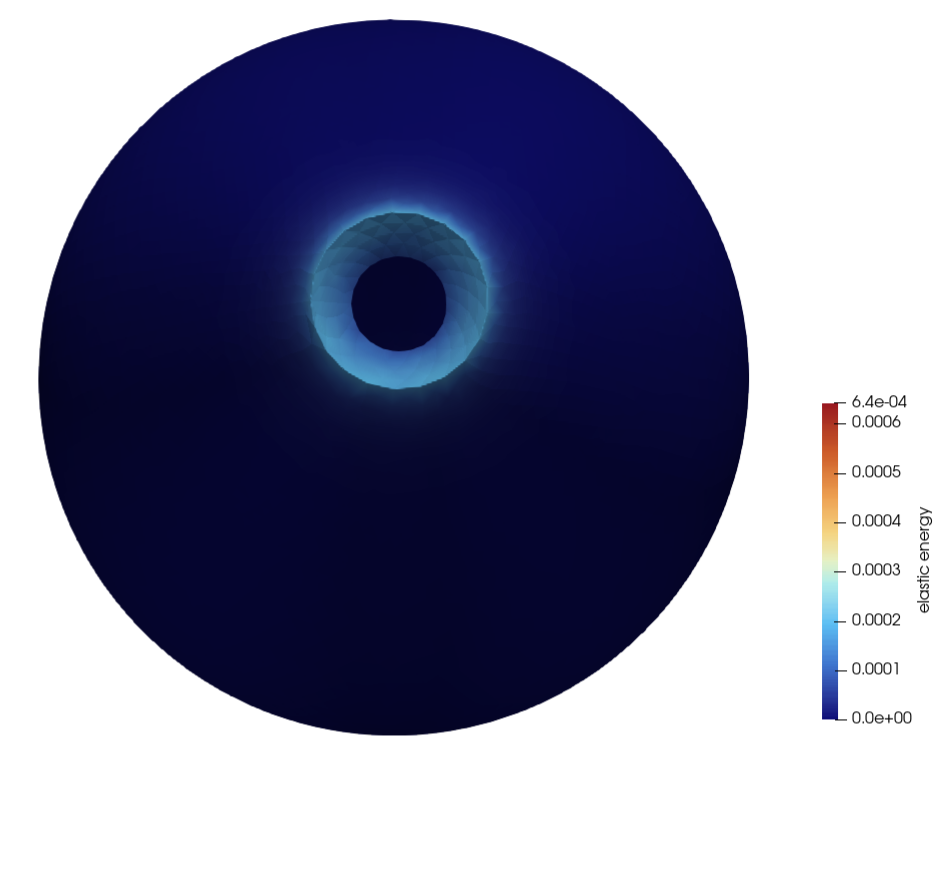}
        \caption{$k=25$, $s_0^{(25)}=0.4$, \\
        $\varepsilon^{(25)}=0$.}
        \end{subfigure}
    \caption{$\overline N=25$, $\overline s=0.4$, $\kappa=0.01$. \textbf{Apical view (prospective mouth).} At each simulation step $k$, the colormap shows the spatial distribution of local mechanical stress. A pronounced high-stress halo is observed to propagate toward the apical pole, closely following the outward expansion of the active nematic band. In our stress-driven growth framework, new material is preferentially incorporated in regions of elevated mechanical stress. Ultimately, at the apical pole, the stress generated by nematic distortions is progressively relaxed, leading to the gradual enlargement of the oral aperture.}
    \label{fig:extending_stress}
\end{figure}

\begin{figure}[htbp]
       \centering
    \captionsetup[subfigure]{justification=centering} 
        \begin{subfigure}[b]{0.32\textwidth}
        \centering
        \includegraphics[width=\linewidth]{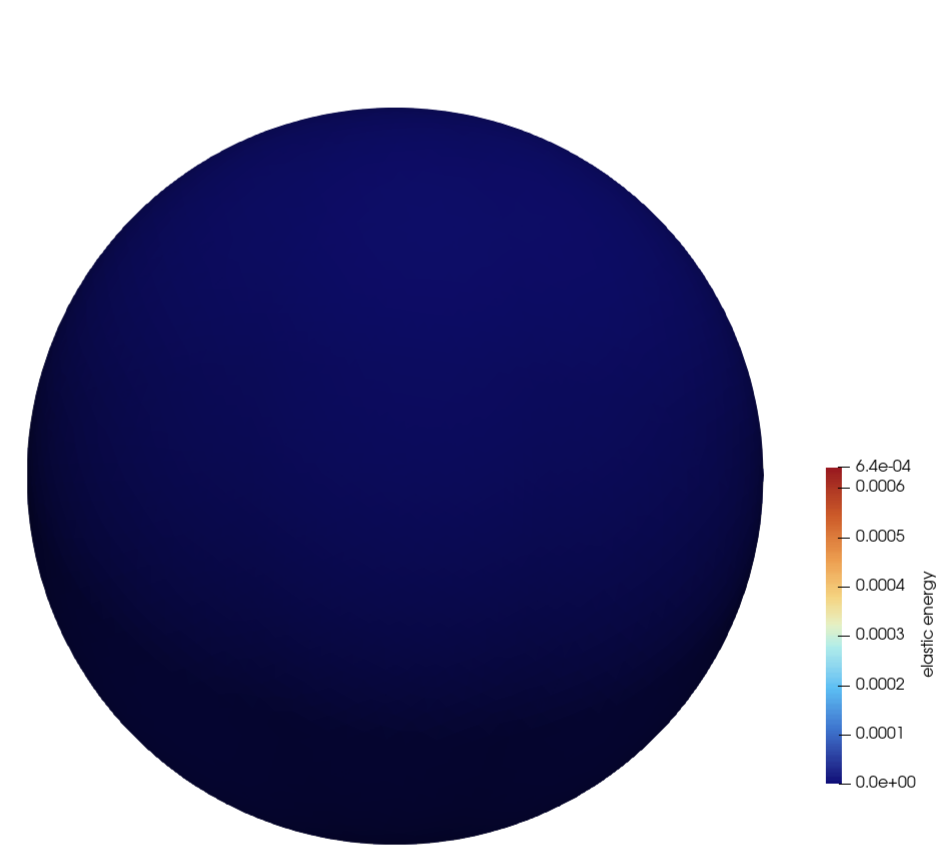}
        \caption{$k=0$, $s^{(0)}\equiv0$,\\
         $\varepsilon^{(0)}=\pi/2$.}
    \end{subfigure}
    \hfill
    \begin{subfigure}[b]{0.32\textwidth}
        \centering
        \includegraphics[width=\linewidth]{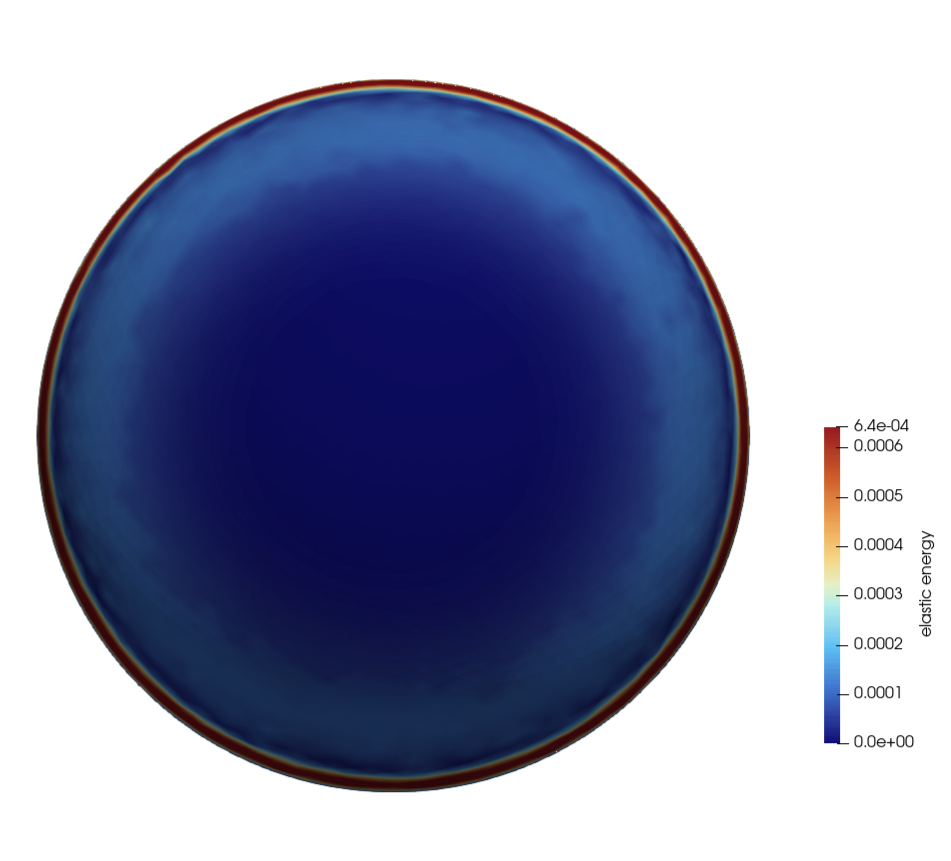}
        \caption{$k=2$, $s^{(2)}=0.04$, \\
        $\varepsilon^{(2)}=1.34$}
    \end{subfigure}
    \hfill
    \begin{subfigure}[b]{0.32\textwidth}
        \centering
        \includegraphics[width=\linewidth]{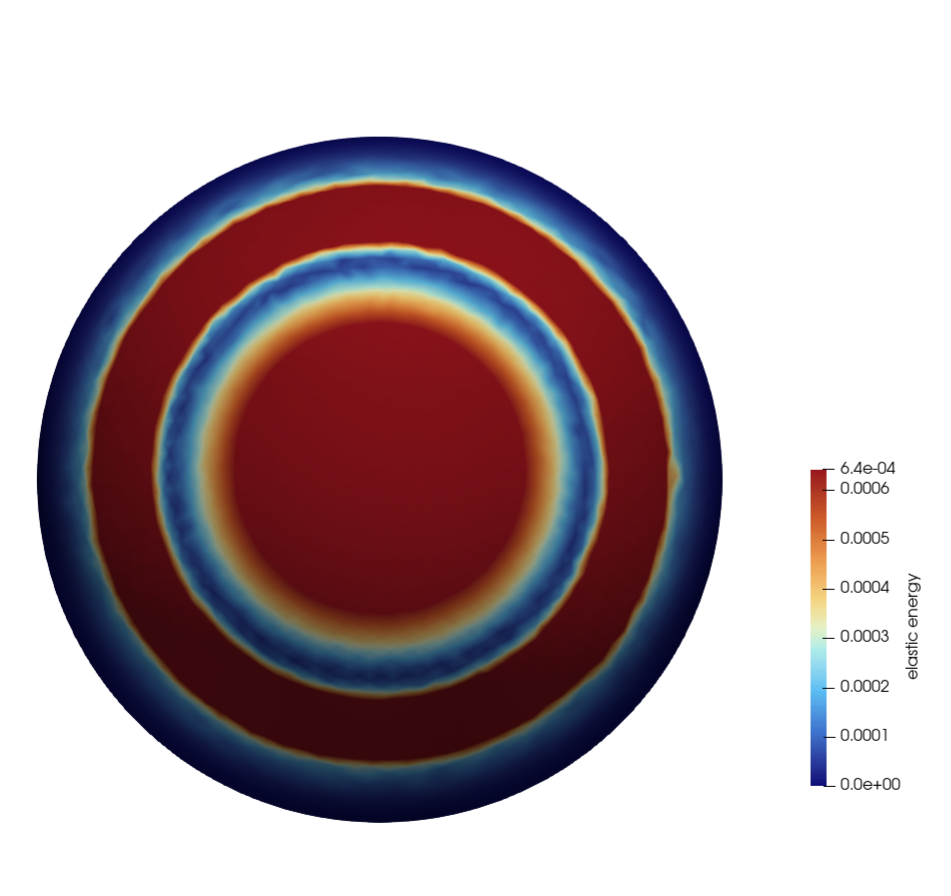}
        \caption{$k=5$, $s^{(5)}=0.1$, \\
        $\varepsilon^{(5)}=0.578$}
    \end{subfigure}
    \vspace{1.5cm} 
    \begin{subfigure}[b]{0.32\textwidth}
        \centering
        \includegraphics[width=\linewidth]{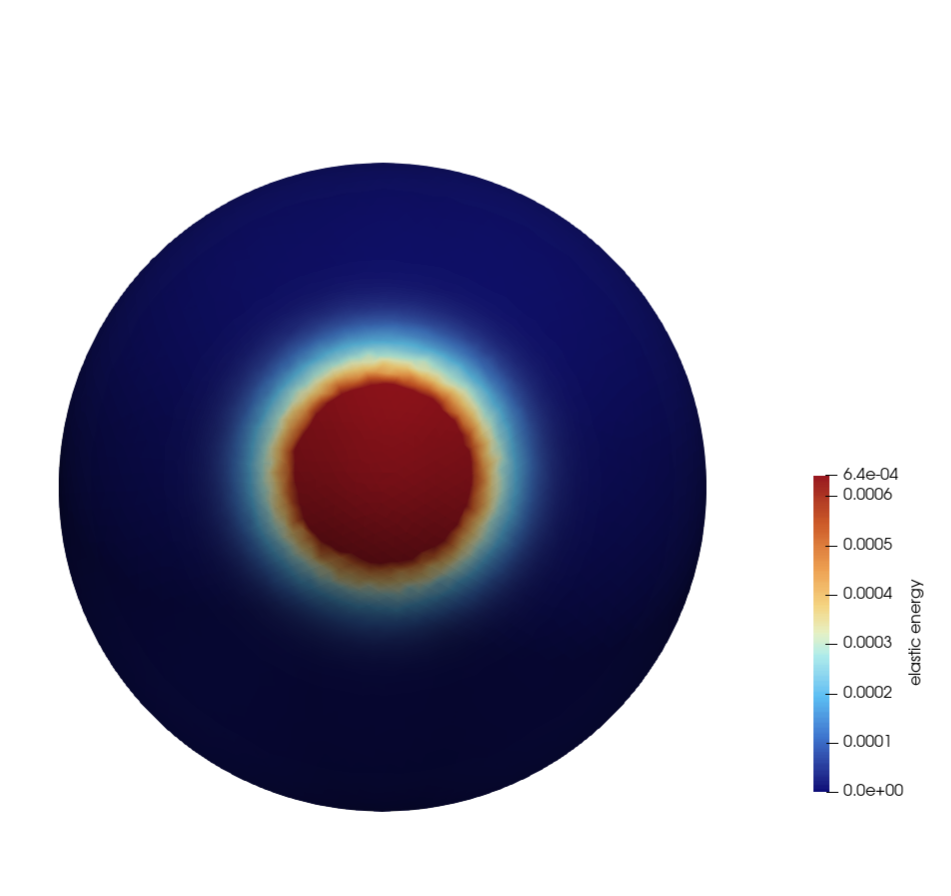}
        \caption{$k=8$, $s^{(8)}=0.14$,\\
        $\varepsilon^{(8)}=0.121$}
    \end{subfigure}
    \hfill
    \begin{subfigure}[b]{0.32\textwidth}
        \centering
        \includegraphics[width=\linewidth]{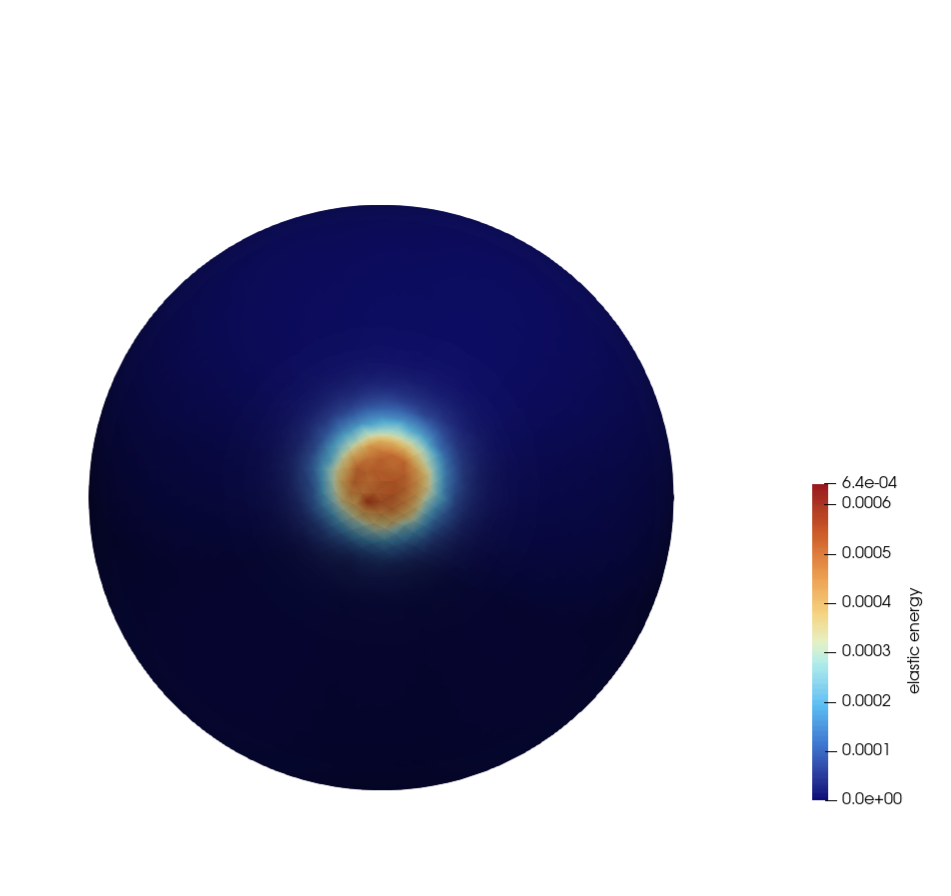}
        \caption{$k=15$, $s^{(15)}=0.25$,\\
        $\varepsilon^{(15)}=0$}
    \end{subfigure}
    \hfill
    \begin{subfigure}[b]{0.32\textwidth}
        \centering
        \includegraphics[width=\linewidth]{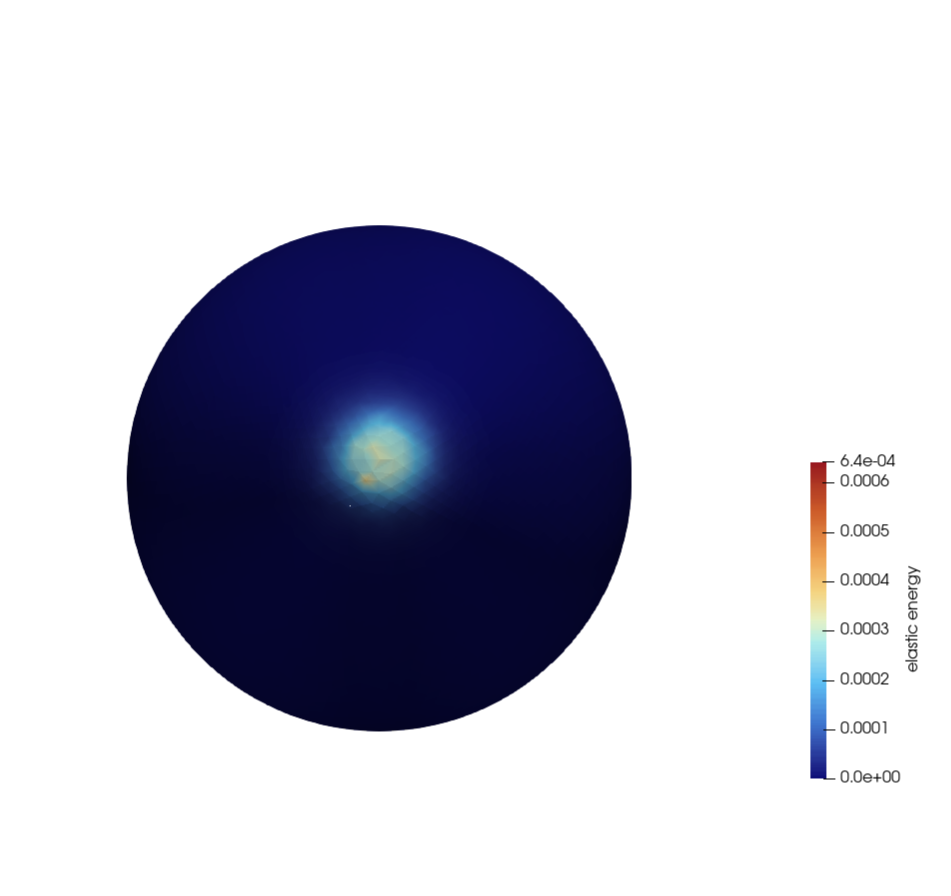}
        \caption{$k=25$, $s^{(25)}=0.4$, \\
        $\varepsilon^{(25)}=0$.}
         \end{subfigure}
     \caption{$\overline N=25$, $\overline s=0.4$, $\kappa=0.01$. \textbf{Basal view (prospective foot).} At each simulation step $k$, the colormap shows the distribution of local mechanical stress. A pronounced high-stress halo propagates toward the basal pole, closely following the outward expansion of the active nematic band. Within the stress-driven growth framework, new cells are preferentially incorporated in regions of elevated mechanical stress. However, because the basal pole remains structurally closed, it cannot undergo the geometric relaxation observed at the mouth. As a result, once the expanding nematic band reaches this region, stress persists and accumulates, leaving the tissue at the foot in a sustained high-stress state.}
    \label{fig:extending_stress_foot}
\end{figure}

\subsubsection{Simulation results: Morphological evolution of the head-foot configuration}\label{sec:FEM_headfoot}

As shown in Figures~\ref{fig:extending_active_band} and \ref{fig:extending_stress}, the simulations capture the quasi-static morphological evolution of an initially isotropic spheroid representing an excised Hydra tissue fragment. This evolution is driven jointly by tissue growth and the progressive emergence of nematic order within the fiber network. The resulting local alignment of nematic directors is visualized via short rod elements embedded within the active band.

We set $\overline{N}=25$, representing a natural process of approximately 40 hours. It is therefore appropriate to treat the model as quasi-static, allowing for full tissue mechanical relaxation between successive steps. For the nematic order parameter, we set a limiting value of $\overline{s}=0.4$ to ensure a well-established nematic phase away from the defects. Although specific upper limiting values for $s$ are not clearly identified in the literature, we follow the estimation in Section \ref{sec:FEM}, assuming that for $\overline{s}=0.4$, the choice $\kappa=0.01$ is physically consistent.

As the simulation advances (Figure~\ref{fig:extending_active_band}), the active nematic region expands progressively toward the poles, accompanied by a corresponding contraction of the disordered isotropic caps. Within this growing band, nematic order becomes increasingly pronounced, with the scalar order parameter $s$ rising monotonically at each step, reflecting enhanced fiber alignment. We impose that, as $k$ increases, the isotropic caps shrink until they vanish at both the apical pole (mouth) and the basal pole (foot). Beyond this point, further evolution proceeds through continued increase in the scalar order parameter $s$ away from defects, according to \eqref{eq:sN_field_arctan}. 

To characterize the mechanical state of the tissue, the color map in the first panel of Figure~\ref{fig:extending_active_band} and in Figures~\ref{fig:extending_stress} and \ref{fig:extending_stress_foot} shows the distribution of the energy density $\Psi_{\mathrm{e}}^{(k)}$ at each step, providing a direct measure of the local stress distribution. A pronounced high-stress halo is observed to propagate toward the poles, closely following the advancing front of the active nematic band. This behavior reflects the fact that distortions in the nematic director field generate mechanical stresses, which are then redistributed and partially relaxed through subsequent morphological deformation of the tissue. 

Additionally, the color map in the second panel of Figure~\ref{fig:extending_active_band} represents the logarithm of the stress-dependent volumetric factor, $\log(J_{\mathrm{g}}^{(k)})$, highlighting material accumulation primarily near the newly formed regions of nematic order as the isotropic caps progressively shrink toward the poles. Subsequently, the accumulation becomes localized in the vicinity of the mouth and the foot. Importantly, a comparison of the color maps between the two panels in Figure~\ref{fig:extending_active_band} reveals a behavior consistent with experimental observations: tissue growth is explicitly coupled to the energy density field. As a result, growth is stress-driven: new material is preferentially incorporated in regions of elevated mechanical stress.

The combined effects of nematic alignment and stress-mediated growth drive a progressive elongation of the tissue along its main body axis, terminating at the two topological defects located at opposite poles. The model further reproduces a mechanical asymmetry between these sites, associated with rupture at the mouth. At the apical pole, the localized stress generated by nematic distortion is progressively relaxed, leading to the enlargement of the oral aperture (Figure~\ref{fig:extending_stress}). By contrast, the basal pole remains structurally closed and cannot undergo comparable geometric relaxation. Consequently, once the expanding band reaches this region, stress accumulates and persists at the foot (Figure~\ref{fig:extending_stress_foot}).

\subsection{Bicephalic (two-head) configuration}\label{sec:TwoHeads}

In contrast to the uniaxial case, we model the quasi-static mechanics of a \emph{bicephalic} (two-headed) regenerating Hydra by introducing two $+1$ topological defects in the upper hemisphere. In our simulations, these apical defects have already prescribed the sites of tissue rupture. Accordingly, the initial configuration explicitly includes the open oral pores associated with the two emerging mouths. To represent this biomechanical state, we consider a thick spherical shell with inner radius $R_{\mathrm{in}} = 1.0$ and outer radius $R_{\mathrm{out}} = 1.1$, from which two circular apertures of radius $R_{\mathrm{hole}} = 0.15$ are removed.

\begin{figure}[htbp]
    \centering
    \begin{subfigure}[c]{\linewidth}
        \centering
        \includegraphics[width=0.55\linewidth]{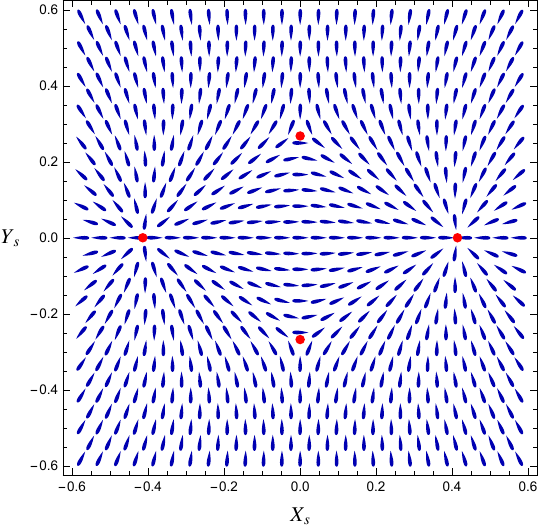}
        \caption{Projected nematic director field $\n_0$ on the stereographic plane. Red markers denote prescribed locations of topological defects.} 
        \label{fig:stereo_planeTwoHeads}
    \end{subfigure}
    \bigskip
    
    \begin{subfigure}[c]{\linewidth}
        \centering
        \includegraphics[width=0.55\linewidth]{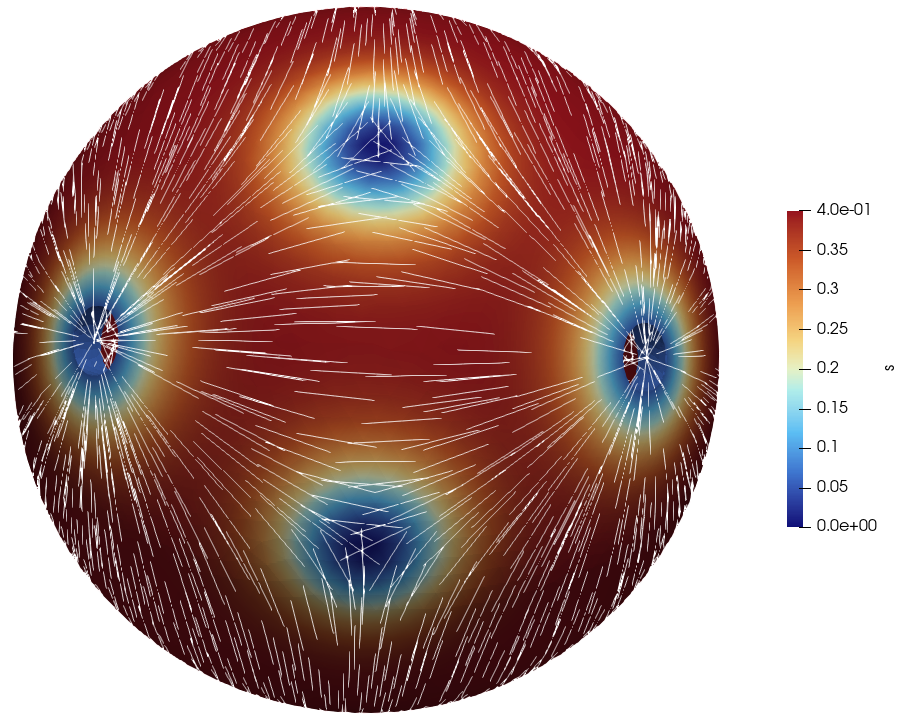}
        \caption{Top-down view of the spherical configuration. Color map shows the magnitude of the scalar order parameter $s$, with blue regions indicating isotropic melting at defect cores ($s = 0$). The two structural openings correspond to the mouths. For illustration, $\delta = 15.0$ and $s_0^{(k)} = 0.4$ in equation~\eqref{eq:sN_def}.} 
        \label{fig:3d_sphereTwoHeads}
    \end{subfigure}
    \caption{Hydra with two heads configuration: (a) stereographic projection showing the planar defect locations, (b) three-dimensional spherical configuration. Superimposed rods indicate the prescribed initial director field $\n_0$. The field contains two $+1$ defects at the mouths, one $+1$ defect at the foot (South Pole), and two $-1/2$ topological balancers located closer to the North Pole. Here, $\vartheta_{\text{mouth}} = \pi/4$ and $\vartheta_{\text{half}} = \pi/6$.}
    \label{fig:two_heads_configuration}
\end{figure}

\subsubsection{Initial defect topology for the two-head configuration}

We design the nematic director field to topologically mimic an organism with two distinct mouths and a single foot. 
Since the intended configuration contains three $+1$ defects (two apical mouths and one basal foot), an additional compensating negative charge is required to balance the net $+3$ contribution. To construct a realizable configuration, we therefore prescribe four explicit defects in the stereographic plane:
\begin{itemize}
\item \emph{Two mouths ($+1$ defects):} located at polar angle $\vartheta_{\text{mouth}}$, with azimuthal positions $\varphi = 0$ and $\varphi = \pi$. These points coincide with the physical apertures cut into the spherical shell.
\item \emph{Topological balancers ($-1/2$ defects):} positioned at a smaller polar angle $\vartheta_{\text{half}}$ closer to the North Pole, with azimuthal angles $\varphi = \pi/2$ and $\varphi = 3\pi/2$.
\end{itemize}
In this construction, the total charge in the stereographic representation is $\sum q_i = 2(+1) + 2(-1/2) = +1$. Upon mapping back to the sphere, an additional intrinsic contribution arises at the South Pole, yielding a physical defect of strength $C_{\text{south}} = 2 - 1 = +1$, which naturally generates the basal foot defect through geometric necessity. The resulting configuration therefore carries a total topological charge of $+2$, in agreement with the Poincar\'{e}–Hopf theorem.

The corresponding director field $\n_0$ on the stereographic plane is shown in Figure~\ref{fig:stereo_planeTwoHeads}. Its full realization on the spherical surface is depicted in Figure~\ref{fig:3d_sphereTwoHeads}.

\subsubsection{Evolution of defect core size and nematic order}

To capture the quasi-static remodeling during the transition to the two-head configuration, the defect core-size parameter $\delta$ in equation~\eqref{eq:sN_def} is not held constant, but instead updated at each simulation step as
\begin{equation}
\label{eq:deltak_twoheads}
\delta^{(k)} = 0.5 + \frac{k}{\overline N} (150 - 0.5).
\end{equation}
Physically, this formulation reflects a progressive contraction of the isotropic (melted) defect cores. At the initial stage ($k = 0$), the small value of $\delta^{(0)}$ produces an extended region of reduced nematic order, representing a diffuse zone of active structural reorganization in which the nematic phase has not yet fully enveloped the defects. As the system evolves ($k\to\overline N$), $\delta^{(k)}$ increases, progressively localizing the nematic order and driving the structure to consolidate more tightly around the defect centers.

\subsubsection{Simulation results: Morphological evolution of the two-head configuration}

\begin{figure}[htbp]
   \centering
    \captionsetup[subfigure]{justification=centering} 
 \includegraphics[width=\textwidth]{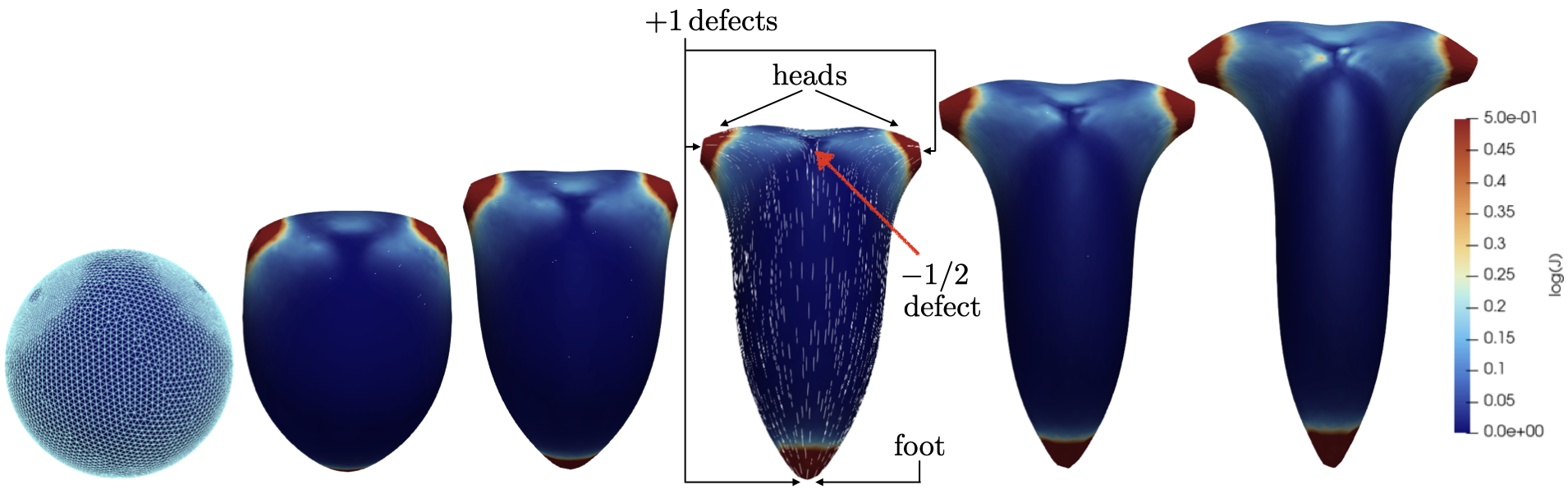}
      \caption{$\overline N=50$, $\overline s=0.6$, $\kappa=0.01$. Successive equilibrium configurations illustrating the quasi-static morphological deformation of the Hydra bicephalic configuration (from left to right) for steps $k \in \{0, 10, 20, 30, 40, 50\}$. The corresponding parameters take the values $s_0^{(k)} \in \{0, 0.12, 0.24, 0.36, 0.48, 0.6\}$ and $\delta^{(k)} \in \{5, 30.4, 60.3, 90.2, 120, 150\}$, respectively. The color map displays the logarithm of the stress-dependent volumetric factor, $\log(J_{\mathrm{g}}^{(k)})$, showing that growth is localized at the heads and the basal foot. The $-1/2$ topological defects at the junctions exhibit negligible growth.}
\label{fig:frontview_twoheads}
\end{figure}

\begin{figure}[htbp]
    \centering
    \captionsetup[subfigure]{justification=centering} 
        \begin{subfigure}[b]{0.32\textwidth}
        \centering
        \includegraphics[width=\linewidth]{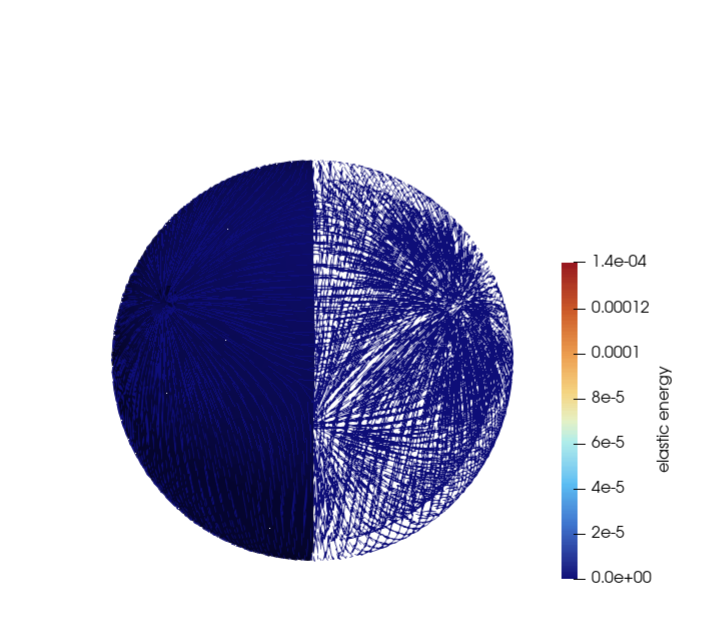}
        \caption{$k=0$, $s_0^{(0)}\equiv0$,\\
        $\delta^{(0)}=0.5$.}
    \end{subfigure}
    \hfill
    \begin{subfigure}[b]{0.32\textwidth}
        \centering
        \includegraphics[width=\linewidth]{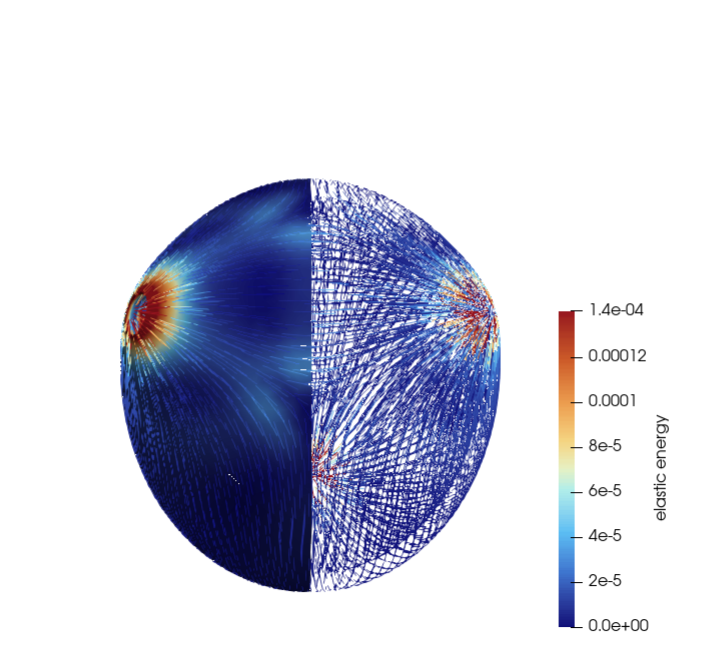}
        \caption{$k=10$, $s_0^{(10)}=0.12$,\\
        $\delta^{(10)}=30.4$.}
    \end{subfigure}
    \hfill
    \begin{subfigure}[b]{0.32\textwidth}
        \centering
        \includegraphics[width=\linewidth]{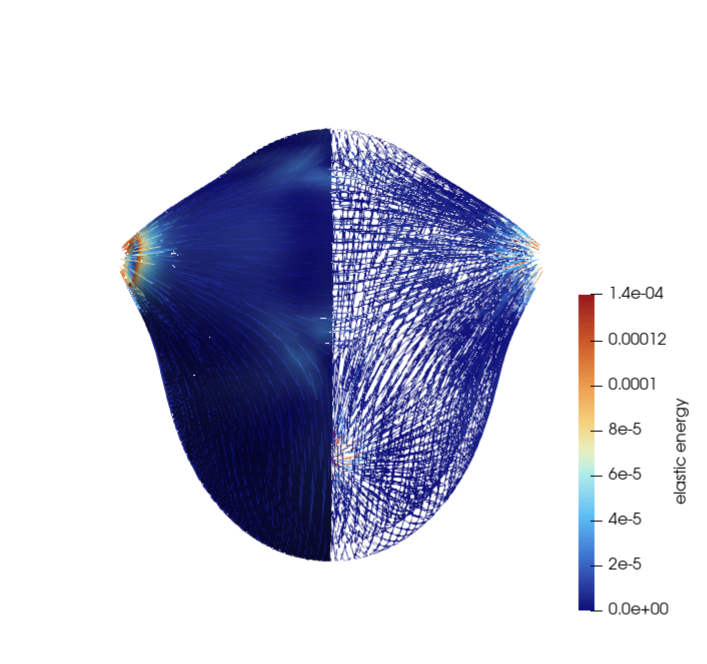}
        \caption{$k=20$, $s_0^{(20)}=0.24$,\\
        $\delta^{(20)}=60.3$.}
    \end{subfigure}
    \vspace{1.5cm} 
    \begin{subfigure}[b]{0.32\textwidth}
        \centering
        \includegraphics[width=\linewidth]{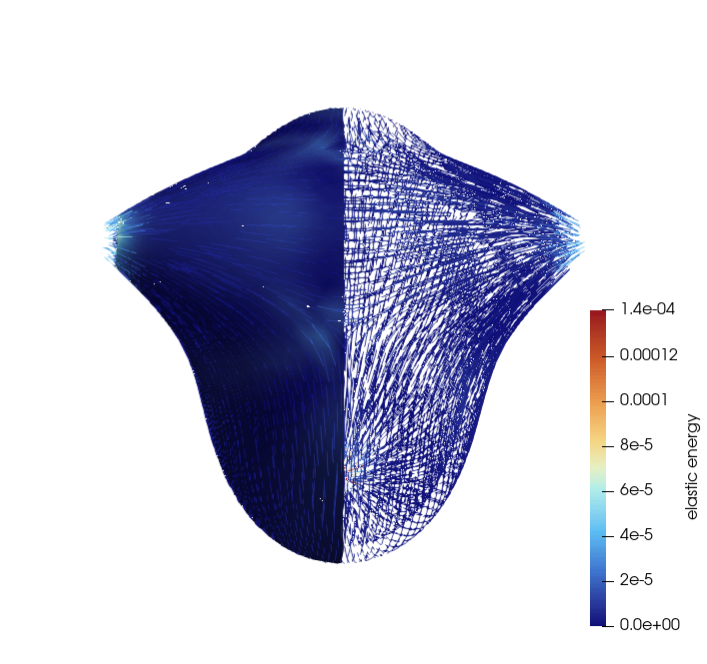}
        \caption{$k=30$, $s_0^{(30)}=0.36$,\\
        $\delta^{(30)}=90.2$.}
    \end{subfigure}
    \hfill
    \begin{subfigure}[b]{0.32\textwidth}
        \centering
        \includegraphics[width=\linewidth]{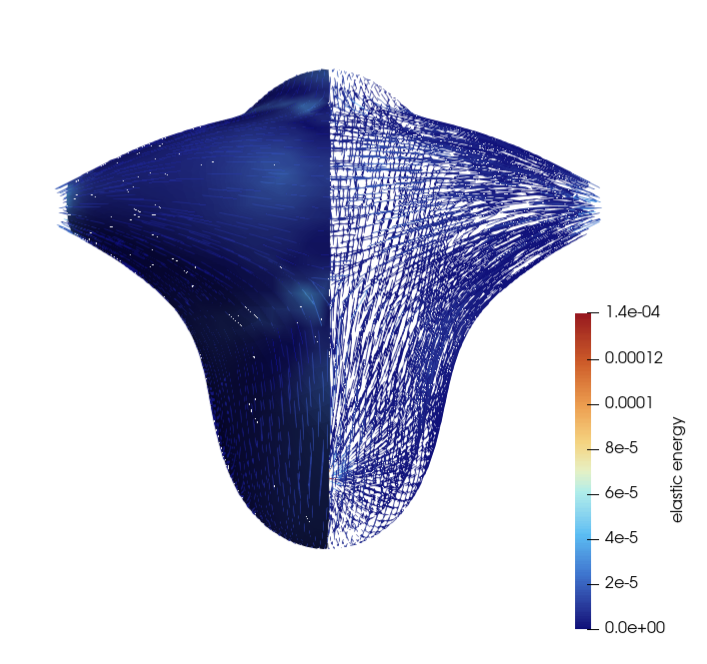}
        \caption{$k=40$, $s_0^{(40)}=0.48$,\\
        $\delta^{(40)}=120$.}
    \end{subfigure}
    \hfill
    \begin{subfigure}[b]{0.32\textwidth}
        \centering
        \includegraphics[width=\linewidth]{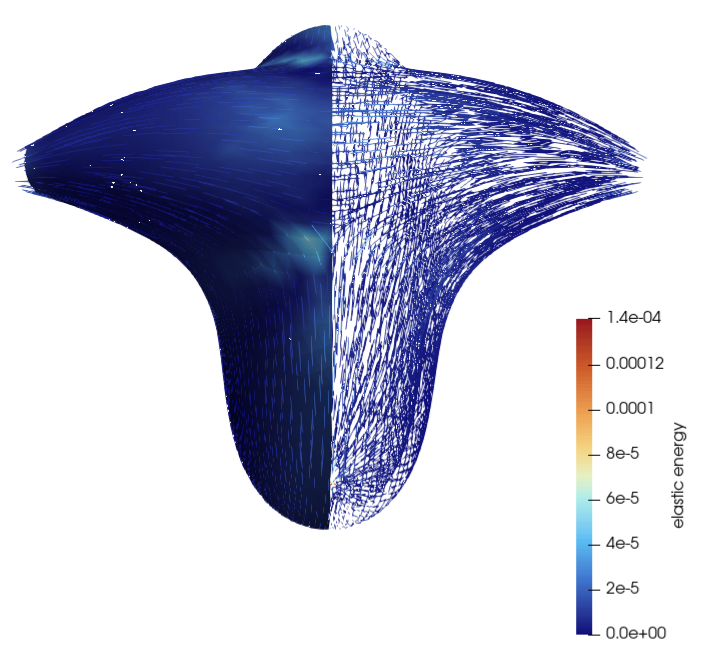}
        \caption{$k=50$, $s_0^{(50)}=0.6$,\\
        $\delta^{(50)}=150$.}
    \end{subfigure}
         \caption{$\overline N=50$, $\overline s=0.6$. \textbf{Apical view.} At each simulation step $k$, the color map shows the spatial distribution of local mechanical stress $\Psi_{\mathrm{e}}^{(k)}$. High-stress regions locate around the heads and the foot. At the apical poles, the stress generated by nematic distortions is progressively relaxed, leading to the gradual enlargement of the oral apertures, while the foot remains the most stressed region. The simulation results recover the expected $Y$-shaped morphology.}
         \label{fig:apicalview_twoheads}
\end{figure}

In Figures~\ref{fig:frontview_twoheads} and \ref{fig:apicalview_twoheads}, the simulations capture the quasi-static morphological evolution of an initially isotropic spheroid featuring two holes in its upper hemisphere. This specific geometry, along with the initial configuration of defects, is set to model an excised tissue fragment developing into a bicephalic Hydra. 

\begin{figure}[htbp]
    \centering
    \includegraphics[width=0.45\textwidth]{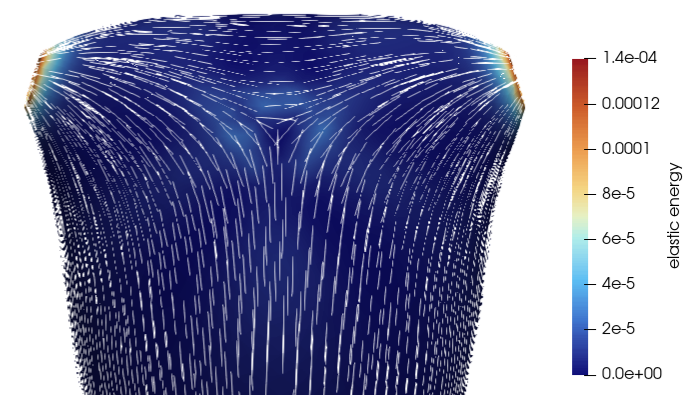}
    \caption{$\overline N=50$, $\overline s=0.6$, $k=20$. Magnified view of the actin fiber organization at the branching junction between the two heads and the main trunk.}
\label{fig:zooming_junctions}
\end{figure}

We investigate how tissue growth and the progressive establishment of nematic order within the fiber network affect its morphology. In this case, there are no shrinking isotropic caps toward the poles; instead, we simulate a progressive reduction of the disordered cores surrounding the defects dictated by $\delta^{(k)}$ in \eqref{eq:deltak_twoheads}. We set $\overline{N}=50$ to represent a natural process of approximately 50 hours. We then treat the model as quasi-static, allowing for full tissue mechanical relaxation between successive steps. For the nematic order parameter, we set a limiting value of $\overline{s}=0.6$ to ensure a well-established nematic phase away from the defects. Following the estimation in Section~\ref{sec:FEM}, we choose $\kappa=0.01$, for physical consistency.

In Figure~\ref{fig:frontview_twoheads}, the color map shows the stress-dependent volumetric factor $\log(J_{\alpha}^{(k)})$, suggesting that growth is concentrated in the vicinity of the two heads and the basal foot. 
As seen from the color maps in Figure~\ref{fig:apicalview_twoheads}, these regions correspond to where the elastic stress $\Psi_{\mathrm{e}}^{(k)}$ is primarily localized at each step, driving the growth in agreement with experimental data. Note that negligible growth occurs at the $-1/2$ defects. Instead, these defects act solely as topological balancers to maintain the total charge of the system, without direct morphological consequences. In particular, Figure~\ref{fig:apicalview_twoheads} illustrates the director field tangent to the inner and outer surfaces of the deformable shell. The color map indicates the elastic energy density $\Psi_{\mathrm{e}}^{(k)}$, providing a direct measure of the local stress distribution. During the initial steps, high-stress regions are observed, localized around the heads and the foot. As the simulation progresses and the tissue deforms, driven by both growth and the mechanical stress generated by the nematic director distortion, a partial relaxation of the stress occurs in the vicinity of the heads, facilitating the enlargement of the oral apertures. Conversely, the foot remains the most stressed region, a feature also observed in the head-tail configuration discussed in Section~\ref{sec:FEM_headfoot}.

The simulation results recover well the $Y$-shaped morphology shown in Figure~\ref{fig:hydratwoheads} where the growth of the heads proceeds distally. Specifically, the two heads push outward from the primary body axis along the trajectories defined by the cellular fiber alignment. The highlighted junctions are associated with the two $-1/2$ topological defects which serve as branching points between the heads and the main trunk. The model captures the configuration characteristic of these defects as evidenced by the enlarged image in Figure~\ref{fig:zooming_junctions} detailing the actin fiber organization near the junction. Growth is entirely stress-driven rather than externally prescribed; new material is preferentially incorporated into regions of elevated mechanical stress. 

\subsection{Tentacular configuration}\label{sec:Tentacles}

Returning to the uniaxial Hydra model, the computational domain represents a tissue fragment with a single aperture at the North Pole, corresponding to the apical mouth. While the overall geometry matches the standard head–foot configuration, a thick spherical shell with a single apical opening, the spatial discretization is adapted accordingly. In particular, we employ a non-uniform mesh that is locally refined in regions where tentacle outgrowth is expected, thereby improving the resolution of highly localized strain fields.

\begin{figure}[htbp]
    \centering
    \begin{subfigure}[c]{\linewidth}
        \centering
        \includegraphics[width=0.65\linewidth]{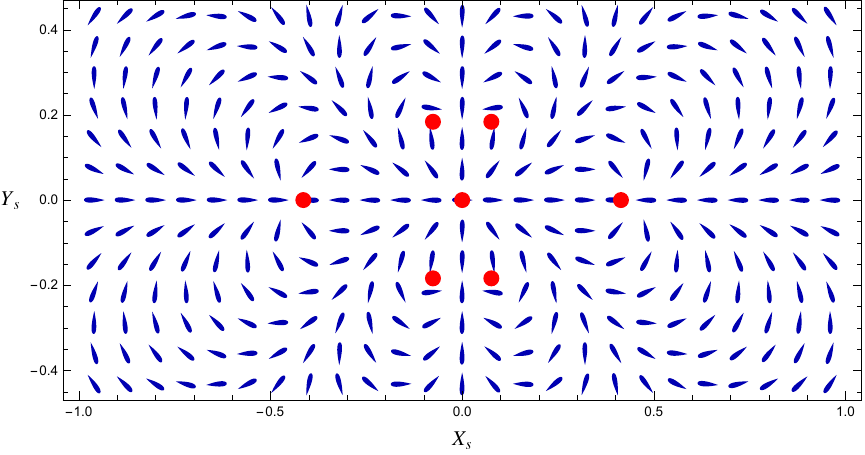}
        \caption{Projected nematic director field $\n_0$ on the stereographic plane. Red markers denote prescribed locations of topological defects.} 
        \label{fig:stereo_planeTentacles}
    \end{subfigure}
    \bigskip
    
    \begin{subfigure}[c]{\linewidth}
        \centering
        \includegraphics[width=0.55\linewidth]{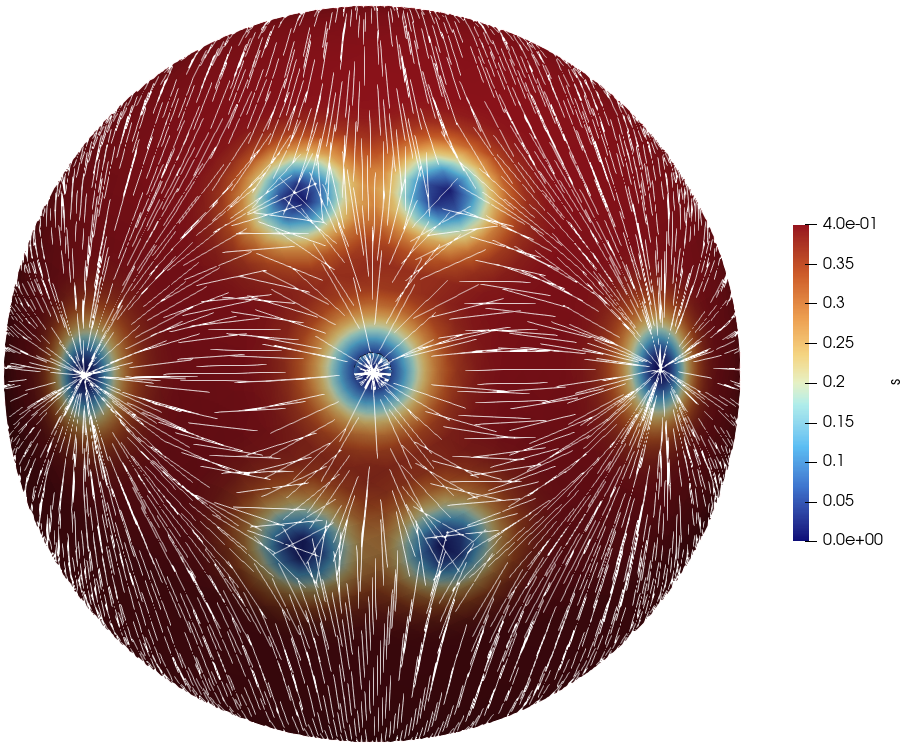}
        \caption{Side view of the spherical configuration. Color map shows the magnitude of the scalar order parameter $s$, with blue regions indicating isotropic melting at the defect cores ($s = 0$). The structural opening corresponds to the mouth. For illustration, $\delta = 50.0$ and $s_0^{(k)} = 0.4$ in equation~\eqref{eq:sN_def}.} 
        \label{fig:3d_sphereTentacles}
    \end{subfigure}
    \caption{Hydra with tentacles configuration: (a) stereographic projection showing planar defect locations, (b) spherical configuration. Superimposed rods represent the prescribed initial director field $\n_0$. The field contains a $+1$ defect at the mouth, one $+1$ defect at the foot (South Pole), two $+1$ defects corresponding to the future tentacle tips, and four $-1/2$ defects defining their bases. Here, $\vartheta{\text{eq}} = 2\pi/8$, $\vartheta_{\text{def}} = \pi/8$, and $\Delta\varphi = 3\pi/8$.}
    \label{fig:tentacles_configuration}
\end{figure}

\subsubsection{Initial defect topology for tentacle formation}

To drive the formation of these structures, we prescribe a specific defect topology. In addition to the $+1$ defect at the North Pole representing the mouth, we introduce six further topological defects. These govern the emergence of two distinct tentacles: two $+1$ defects define the future tips, while four $-1/2$ defects determine their bases. Geometrically, the two tip defects are placed at a polar angle $\vartheta_{\text{eq}}$ and opposite azimuthal positions $\varphi = 0$ and $\varphi = \pi$. Each tip defect is flanked by a pair of $-1/2$ base defects located at an azimuthal offset $\Delta\varphi$ and at a nearby polar angle $\vartheta_{\text{def}}$, chosen in the vicinity of $\vartheta_{\text{eq}}$. The explicit locations $(\vartheta_i, \varphi_i)$ and corresponding charges $q_i$ are given as follows:
\begin{itemize}
\item \emph{North Pole (mouth):} a single $+1$ defect located at $\vartheta = 0$ (origin of the stereographic plane).
\item \emph{Tentacle tip ($+1$) defects:} located at $(\vartheta_{\text{eq}}, 0)$ and $(\vartheta_{\text{eq}}, \pi)$.
\item \emph{Tentacle base ($-1/2$) defects:} clustered around the first tip at $(\vartheta_{\text{def}}, \Delta\varphi)$ and $(\vartheta_{\text{def}}, 2\pi - \Delta\varphi)$, and around the second tip at $(\vartheta_{\text{def}}, \pi - \Delta\varphi)$ and $(\vartheta_{\text{def}}, \pi + \Delta\varphi)$.
\end{itemize}
These coordinates are mapped to the stereographic plane $(X_s, Y_s)$ via $X_{s,i} = \tan(\vartheta_i/2)\cos\varphi_i$ and $Y_{s,i} = \tan(\vartheta_i/2)\sin\varphi_i$. Including the North Pole defect, the total nominal charge in the projected plane is $\sum q_i = 1(+1){\text{north}} + 2(+1){\text{tips}} + 4(-1/2){\text{bases}} = +1$. Upon mapping back to the sphere, an intrinsic contribution at the South Pole yields $C_{\text{south}} = 2 - 1 = +1$, which naturally acts as the basal foot of the Hydra. The full configuration therefore carries a total topological charge of $+2$.

The reference director field $\n_0$ on the stereographic plane is shown in Figure~\ref{fig:stereo_planeTentacles}, while its corresponding three-dimensional realization on the spherical surface is shown in Figure~\ref{fig:3d_sphereTentacles}. Prescribing a distinct polar angle $\vartheta_{\text{def}} \neq \vartheta_{\text{eq}}$ for the $-1/2$ base defects controls the emergent three-dimensional morphology of the tentacles, which typically develop a curved rather than an orthogonally protruding shape.

\subsubsection{Simulation results: Morphological evolution of tentacle formation}

Similarly to the bicephalic case, to capture the quasi-static remodeling during the transition to the tentacle configuration, the defect core-size parameter $\delta$ in equation~\eqref{eq:sN_def} is updated at each simulation step via~\eqref{eq:deltak_twoheads} instead of being held constant, reflecting the progressive contraction of the isotropic (melted) defect cores. 

\begin{figure}[htbp]
    \centering
    \captionsetup[subfigure]{justification=centering} 
\includegraphics[width=\textwidth]{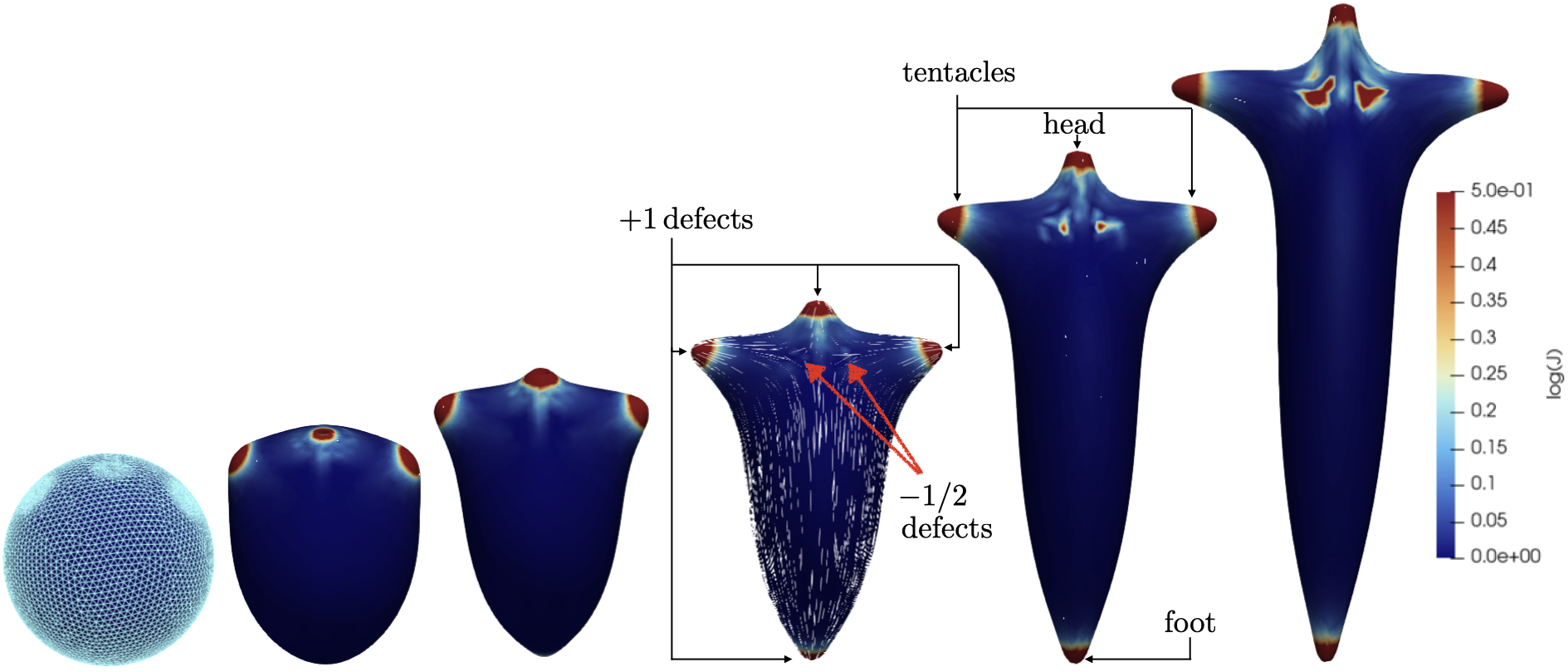}
    \caption{$\overline N=80$, $\overline s=0.6$, $\kappa=0.01$. Successive equilibrium configurations illustrating the quasi-static morphological deformation of the Hydra tissue fragment (from left to right) for steps $k \in \{0, 16, 32, 48, 64, 80\}$. The corresponding parameters take the values $s_0^{(k)} \in \{0, 0.12, 0.24, 0.36, 0.48, 0.6\}$ and $\delta^{(k)} \in \{5, 34, 63, 92, 121, 150\}$, respectively. The sequence shows successive equilibrium configurations illustrating the quasi-static morphological deformation of the tentacular configuration. The color map represents the stress-dependent volumetric growth factor $\log(J_{\mathrm{g}}^{(k)})$ during regeneration. Early in the process ($k \lesssim 48$), tissue growth is confined to the apical head, basal foot, and emerging tentacles. In the later stages, a distinct localization of the growth factor emerges also at the $-1/2$ defect cores, following the elongation of the tentacles and the neck region. Growth is stress-driven: new material is preferentially incorporated in regions of elevated mechanical stress.}
    \label{fig:frontview_tentacles}
\end{figure}

\begin{figure}[htbp]
    \centering
    \captionsetup[subfigure]{justification=centering} 
        \begin{subfigure}[b]{0.32\textwidth}
        \centering
        \includegraphics[width=\linewidth]{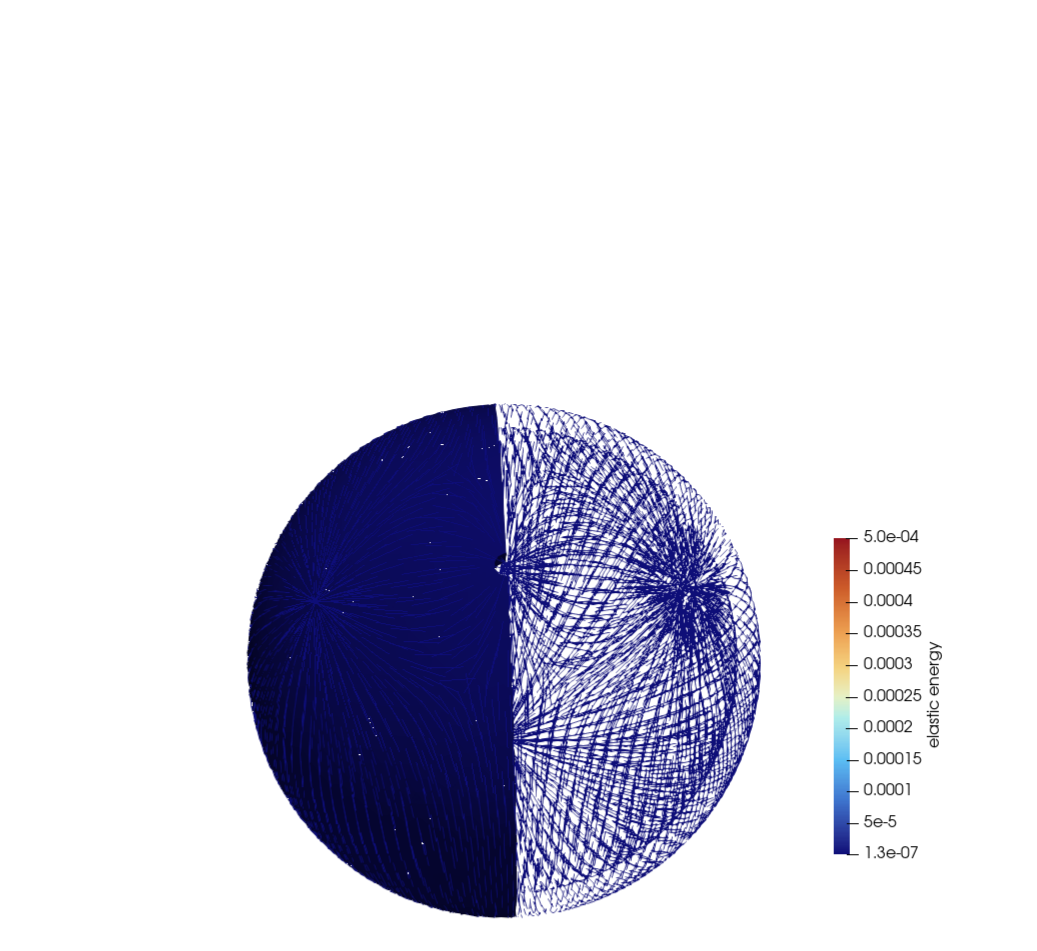}
        \caption{$k=0$, $s_0^{(0)}\equiv0$,\\
        $\delta^{(0)}=5$.}
    \end{subfigure}
    \hfill
    \begin{subfigure}[b]{0.32\textwidth}
        \centering
        \includegraphics[width=\linewidth]{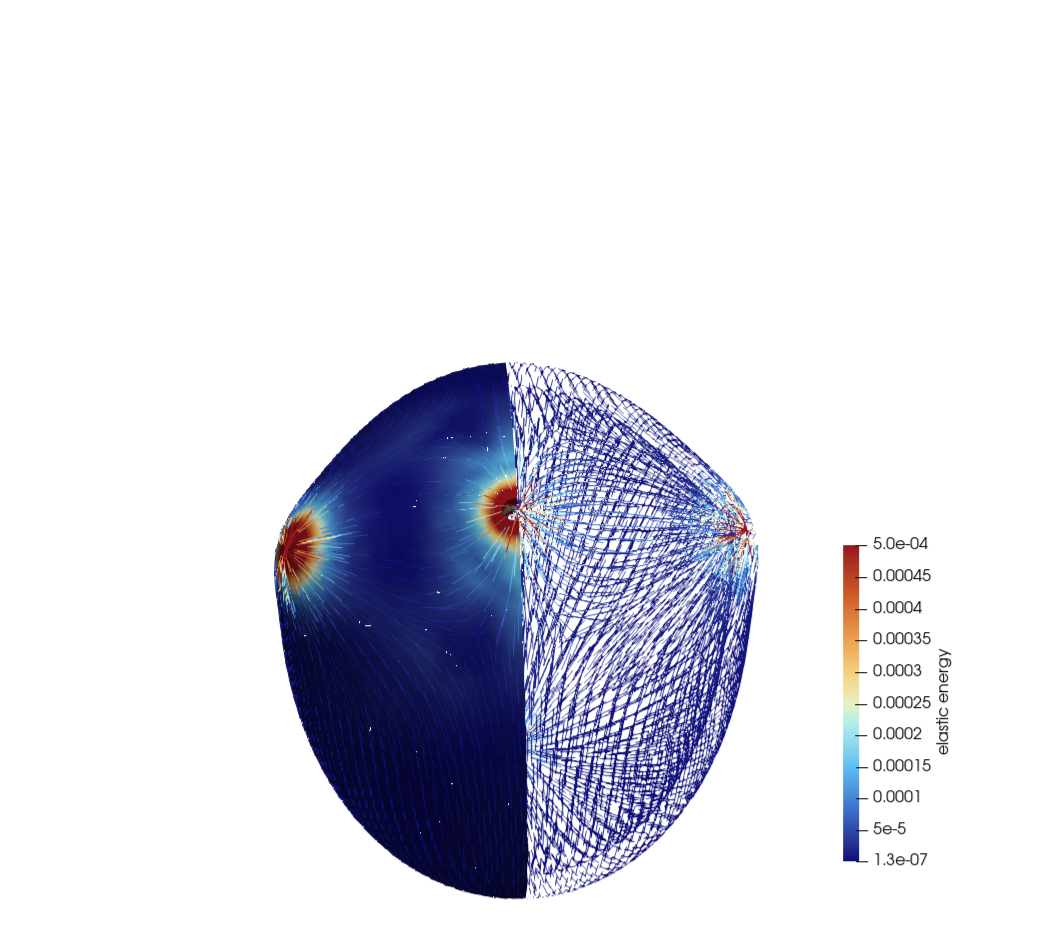}
        \caption{$k=16$, $s^{(16)}=0.12$,\\
        $\delta^{(16)}=34$.}
    \end{subfigure}
    \hfill
    \begin{subfigure}[b]{0.32\textwidth}
        \centering
        \includegraphics[width=\linewidth]{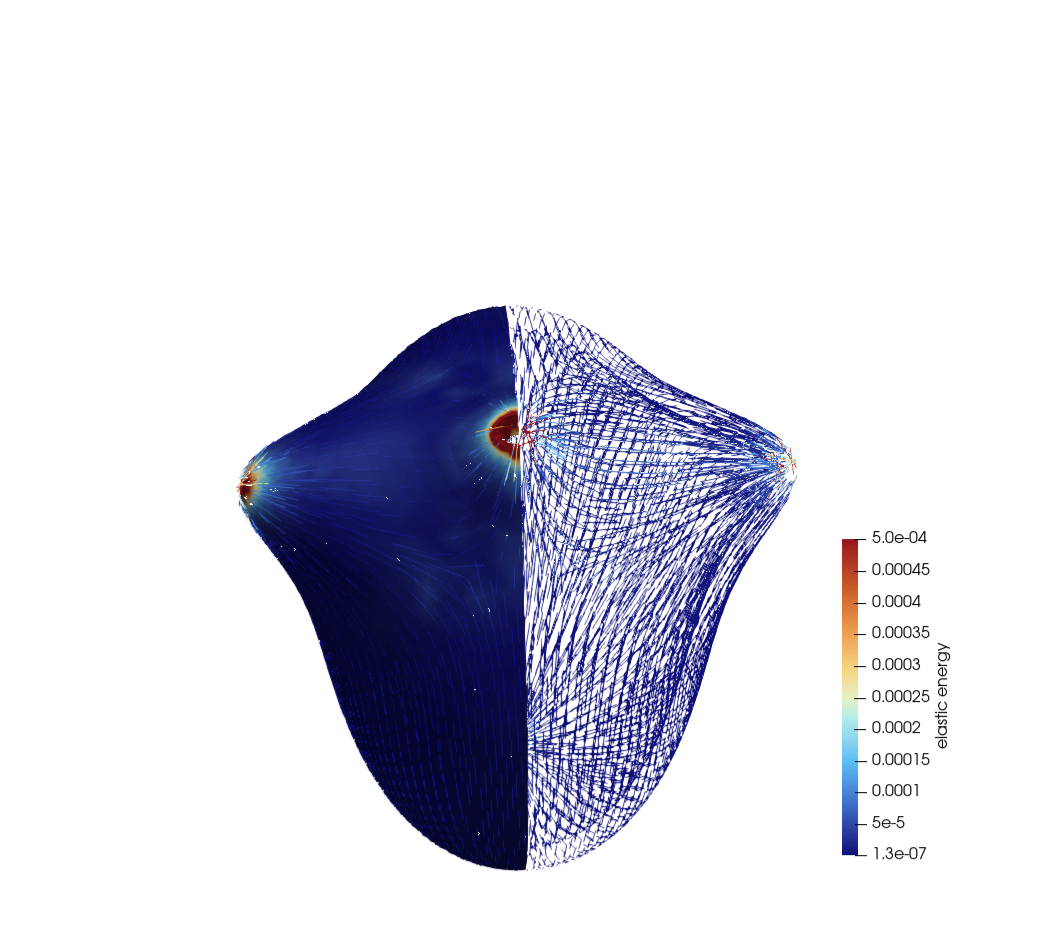}
        \caption{$k=20$, $s^{(32)}=0.24$,\\
        $\delta^{(32)}=63$.}
    \end{subfigure}
    \vspace{1.5cm} 
    \begin{subfigure}[b]{0.32\textwidth}
        \centering
        \includegraphics[width=\linewidth]{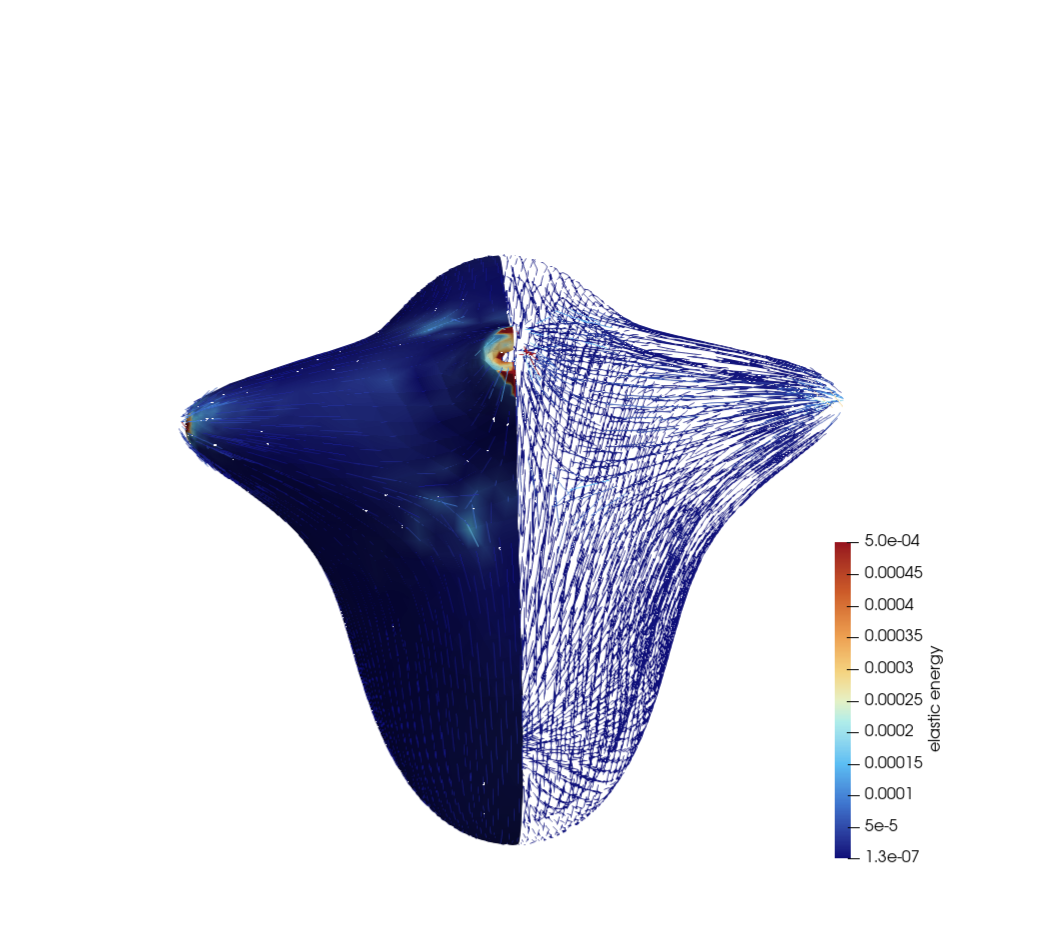}
        \caption{$k=48$, $s^{(48)}=0.36$,\\
        $\delta^{(48)}=92$.}
    \end{subfigure}
    \hfill
    \begin{subfigure}[b]{0.32\textwidth}
        \centering
        \includegraphics[width=\linewidth]{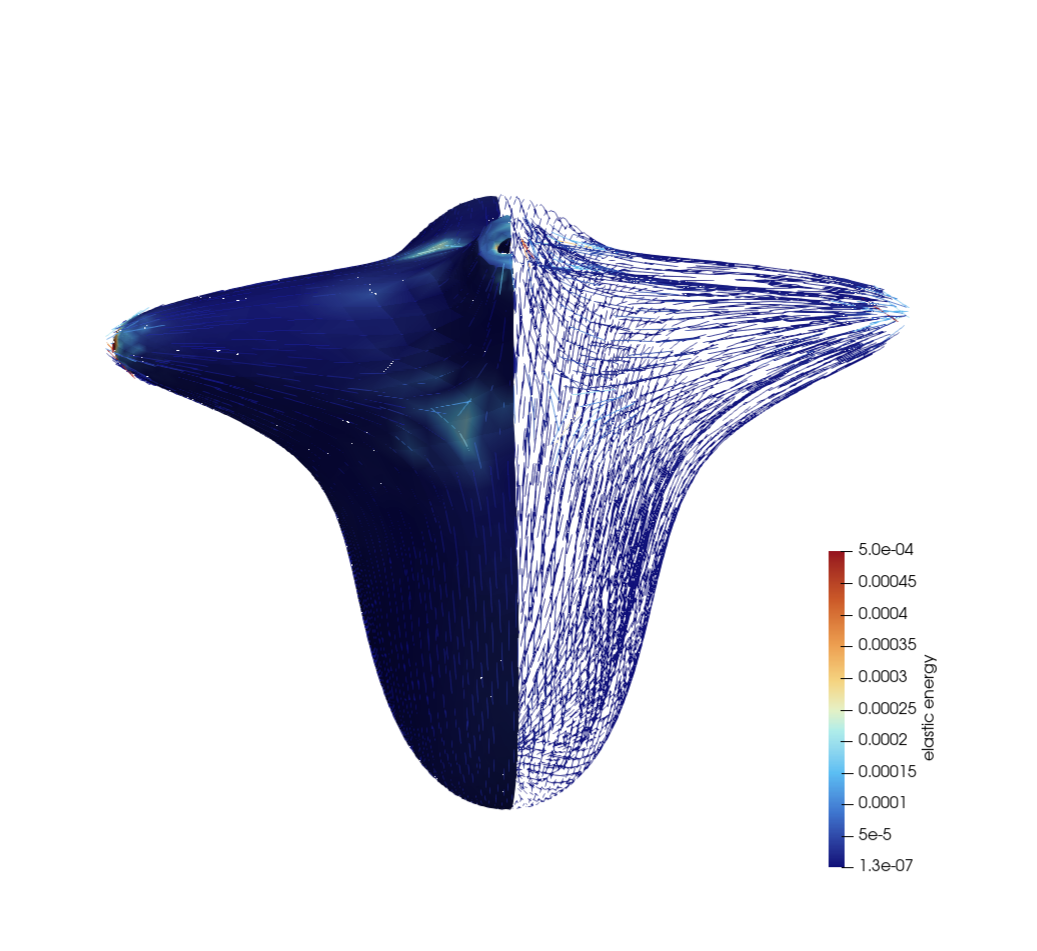}
        \caption{$k=64$, $s^{(64)}=0.48$,\\
        $\delta^{(64)}=121$.}
    \end{subfigure}
    \hfill
    \begin{subfigure}[b]{0.32\textwidth}
        \centering
        \includegraphics[width=\linewidth]{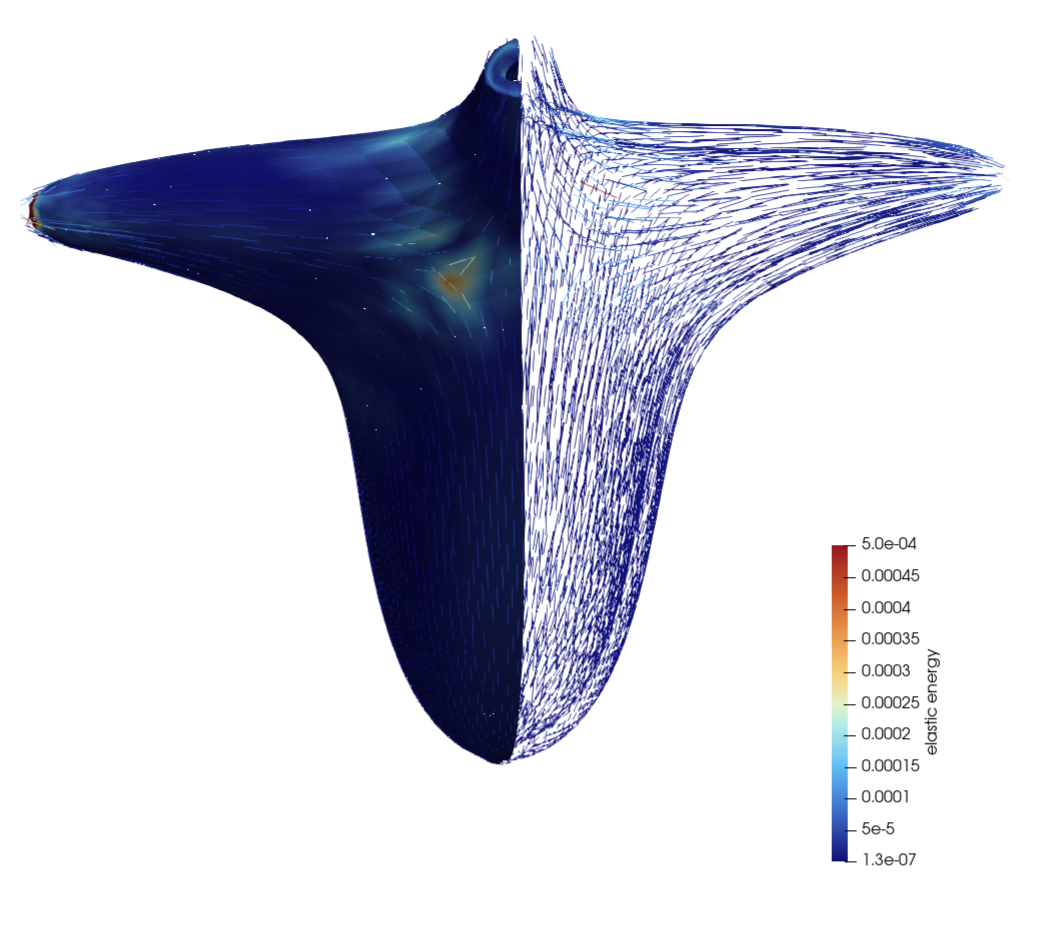}
        \caption{$k=80$, $s^{(80)}=0.6$,\\
        $\delta^{(80)}=150$.}
    \end{subfigure}
         \caption{$\overline N=80$, $\overline s=0.6$. \textbf{Apical view.} The right half of the regenerating tissue displays the director field tangent to the inner and outer surfaces of the shell. The color map represents the elastic energy density $\Psi_{\mathrm{e}}^{(k)}$, providing a direct measure of the local stress distribution. During the initial steps, high-stress regions are observed, localized around the head and the tentacles. As the simulation progresses and the tissue deforms, driven by both growth and the mechanical stress generated by the nematic director distortion, a partial relaxation of the stress occurs in the vicinity of the head, facilitating the enlargement of the oral aperture. Moreover, while a relaxation occurs along the tentacle shafts, the tentacle tips remain stressed.}
         \label{fig:apicalview_tentacles}
\end{figure}

\begin{figure}[htbp]
    \centering
    \captionsetup[subfigure]{justification=centering} 
        \begin{subfigure}[b]{0.32\textwidth}
        \centering
        \includegraphics[width=\linewidth]{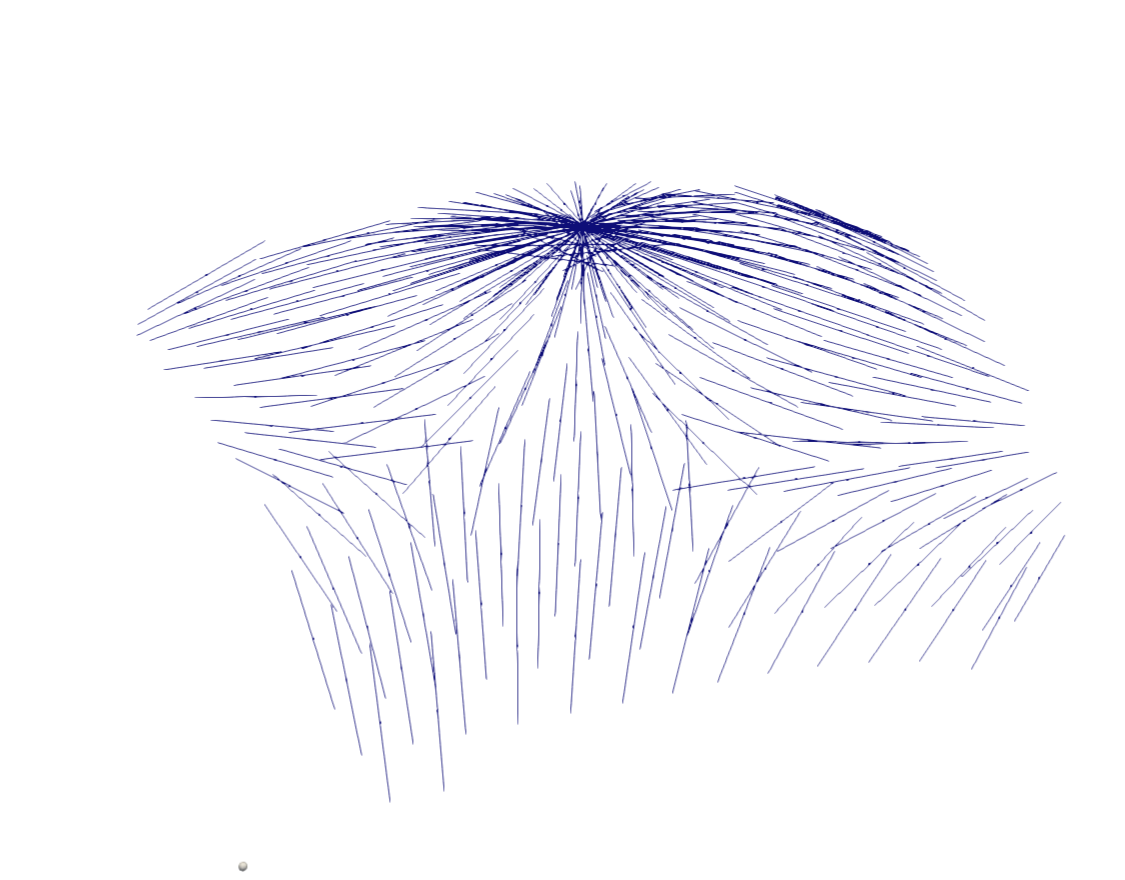}
        \caption{$k=0$, $s_0^{(0)}\equiv0$,\\
        $\delta^{(0)}=5$.}
    \end{subfigure}
    \hfill
    \begin{subfigure}[b]{0.32\textwidth}
        \centering
        \includegraphics[width=\linewidth]{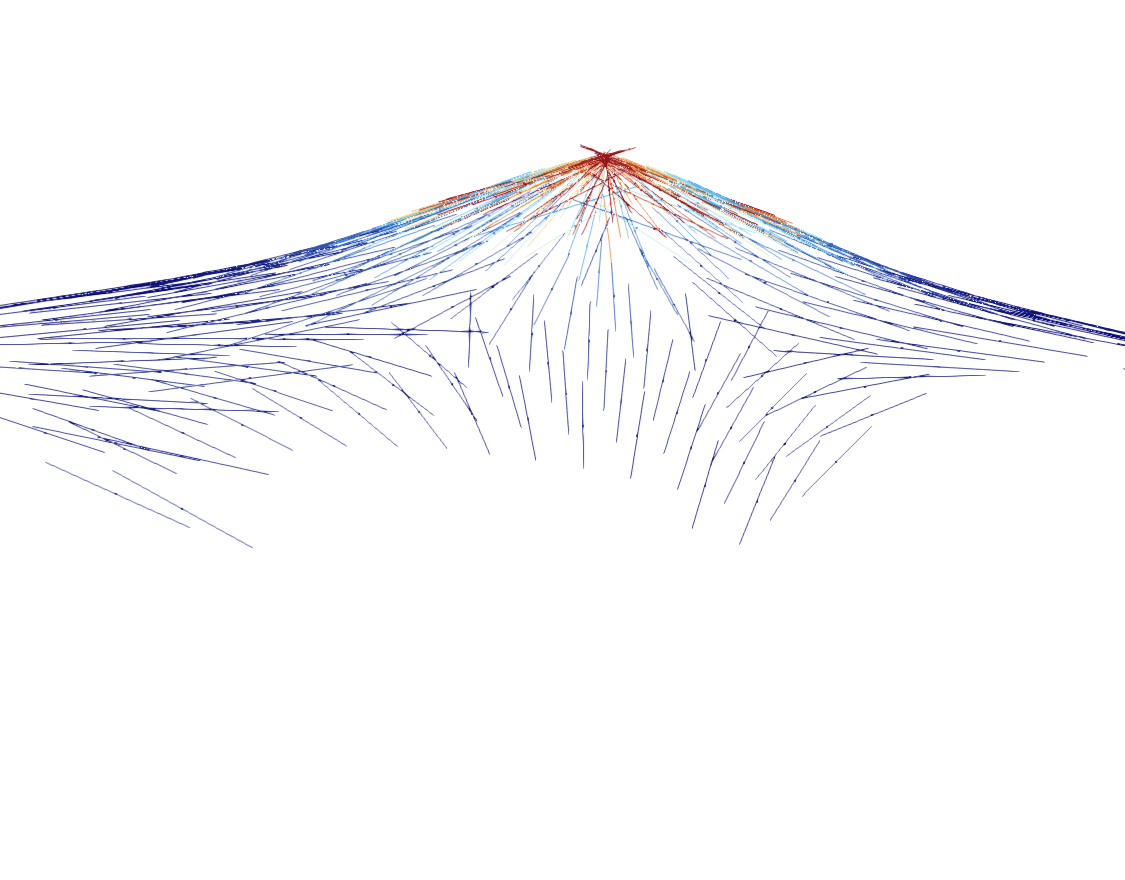}
        \caption{$k=16$, $s^{(16)}=0.12$,\\
        $\delta^{(16)}=34$.}
    \end{subfigure}
    \hfill
    \begin{subfigure}[b]{0.32\textwidth}
        \centering
        \includegraphics[width=\linewidth]{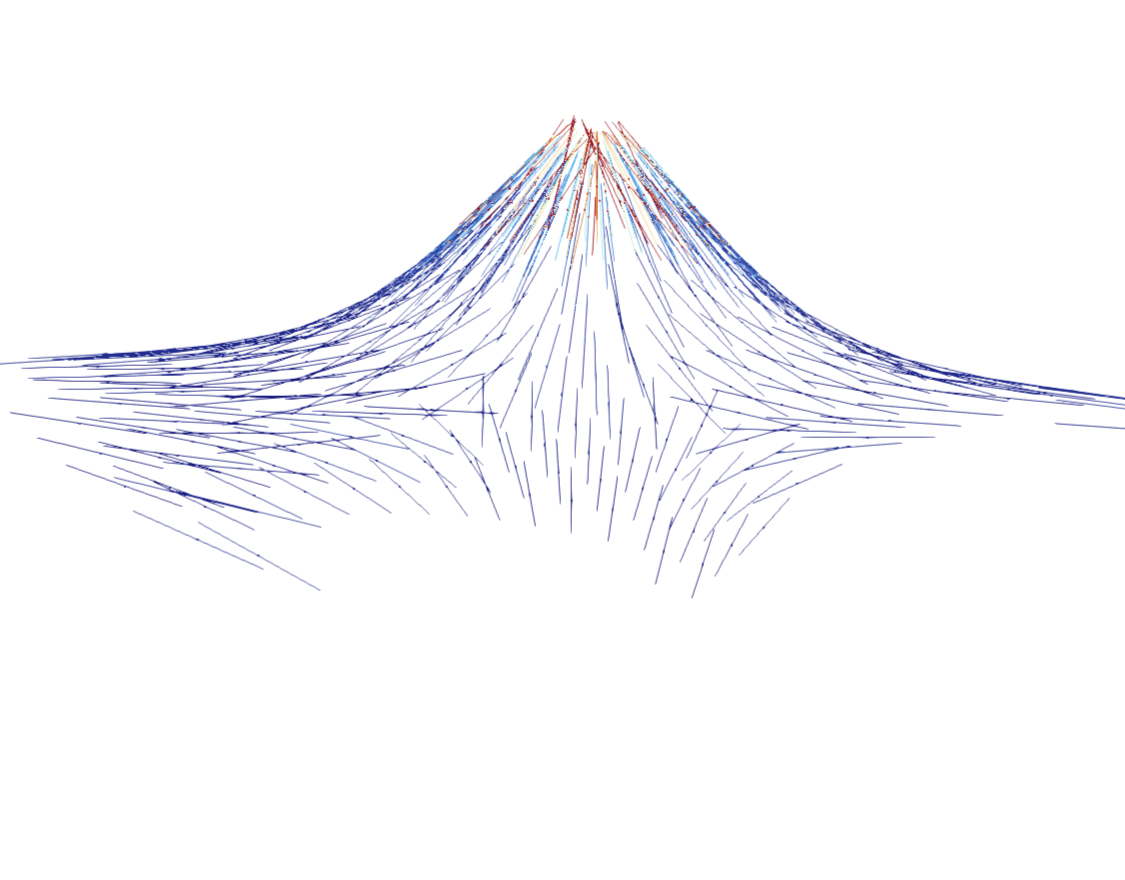}
        \caption{$k=20$, $s^{(32)}=0.24$,\\
        $\delta^{(32)}=63$.}
    \end{subfigure}
    \vspace{1.5cm} 
    \begin{subfigure}[b]{0.32\textwidth}
        \centering
        \includegraphics[width=\linewidth]{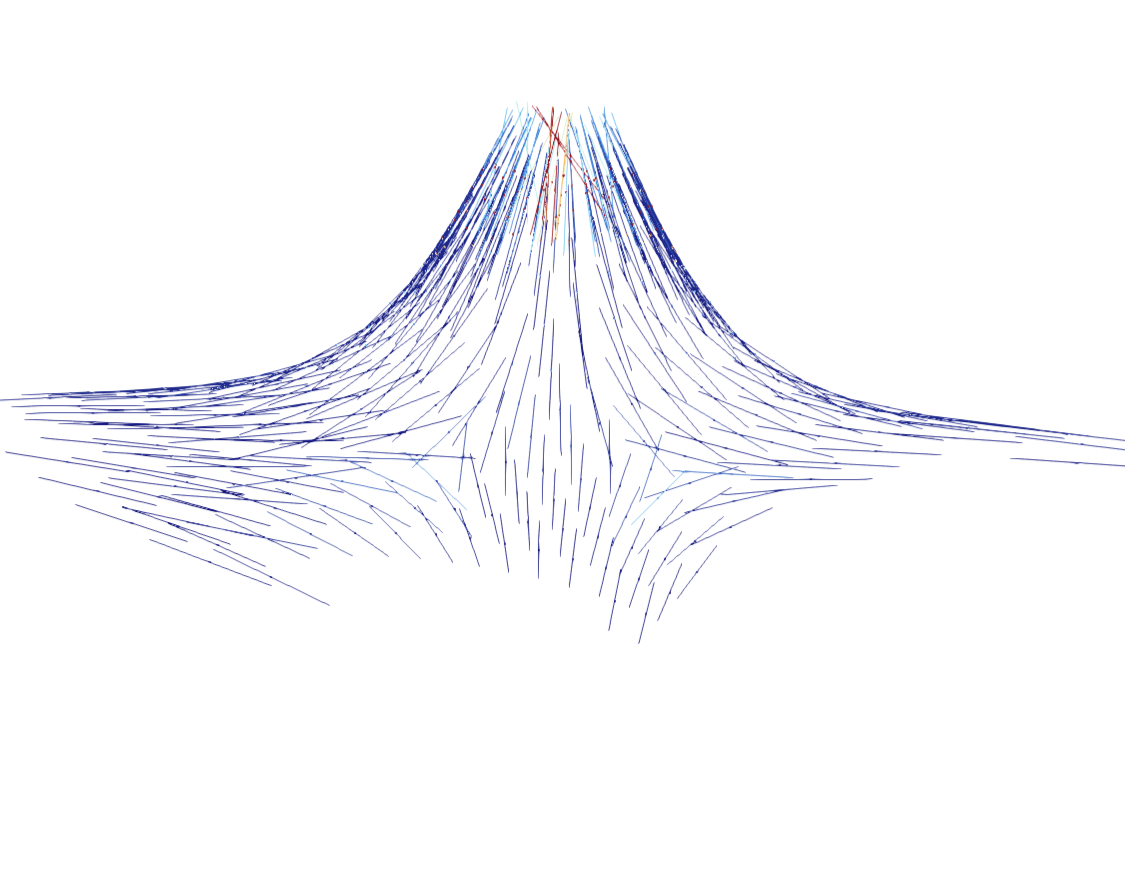}
        \caption{$k=48$, $s^{(48)}=0.36$,\\
        $\delta^{(48)}=92$.}
    \end{subfigure}
    \hfill
    \begin{subfigure}[b]{0.32\textwidth}
        \centering
        \includegraphics[width=\linewidth]{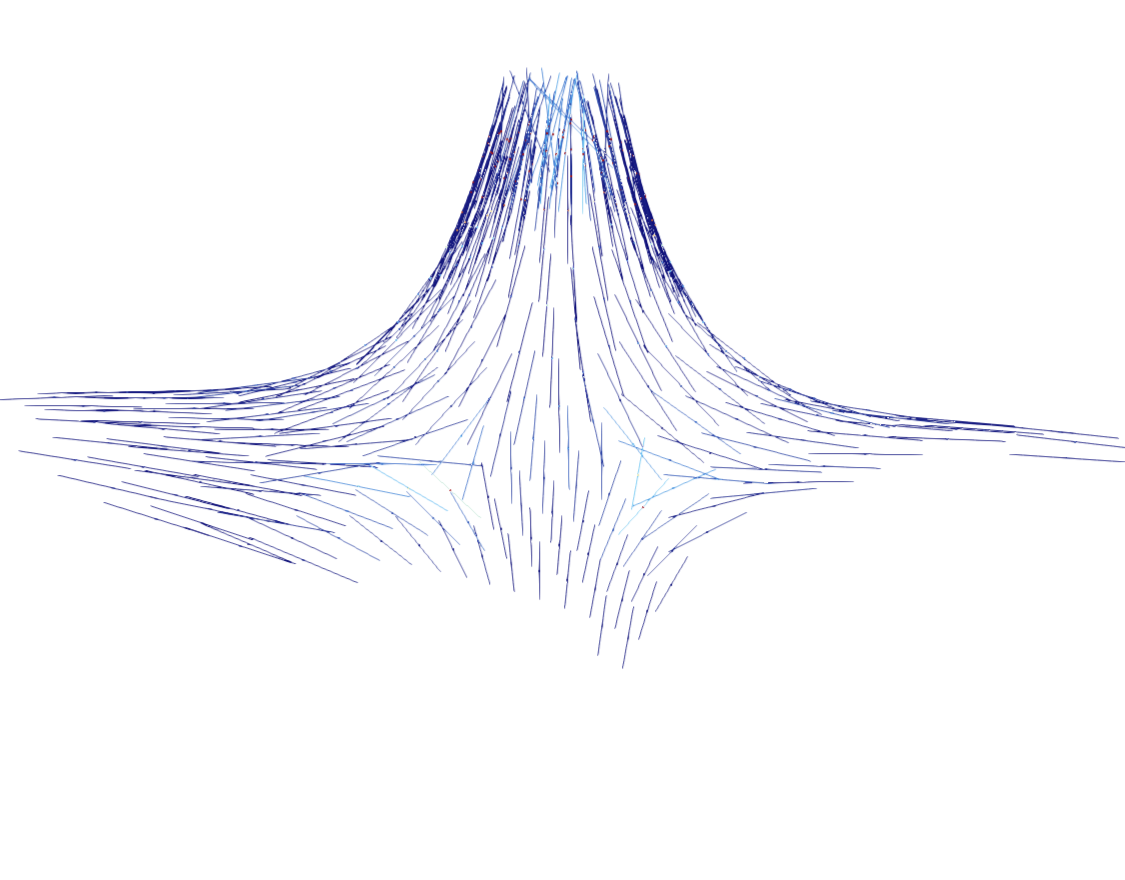}
        \caption{$k=64$, $s^{(64)}=0.48$,\\
        $\delta^{(64)}=121$.}
    \end{subfigure}
    \hfill
    \begin{subfigure}[b]{0.32\textwidth}
        \centering
        \includegraphics[width=\linewidth]{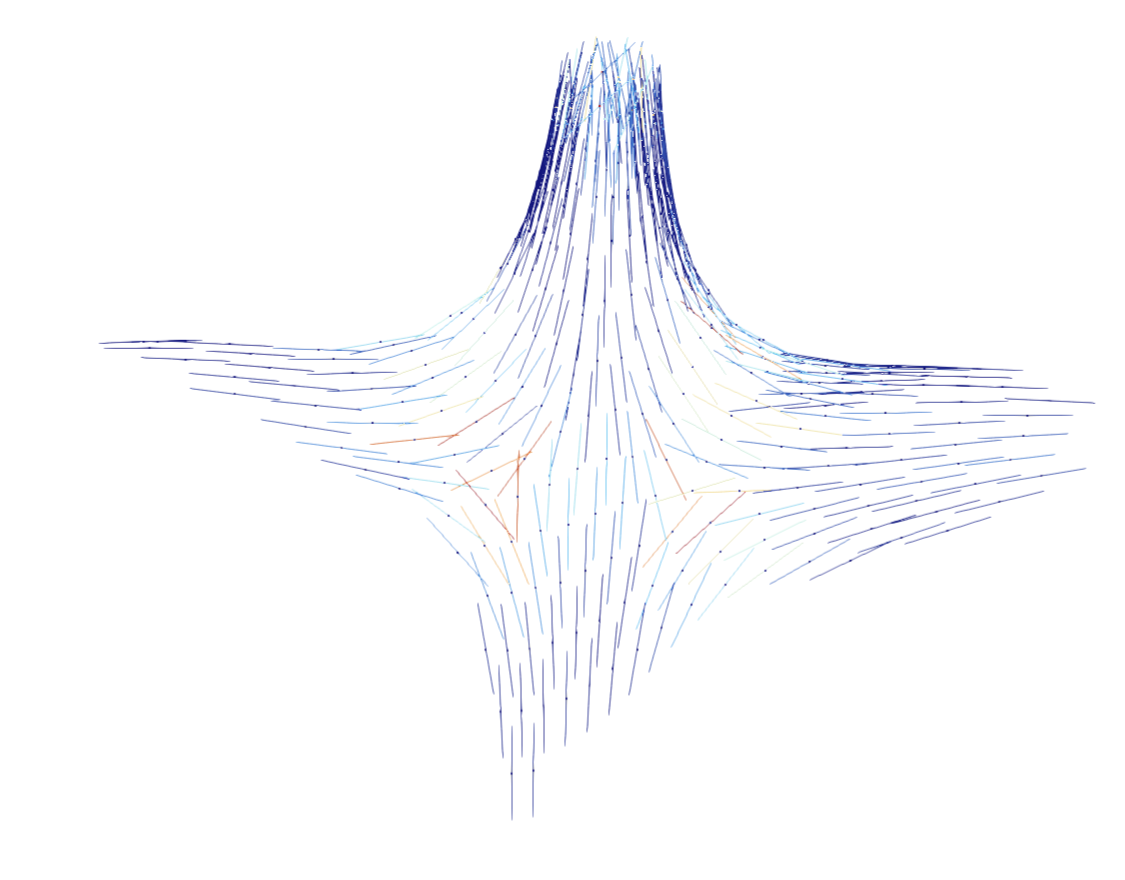}
        \caption{$k=80$, $s^{(80)}=0.6$,\\
        $\delta^{(80)}=150$.}
    \end{subfigure}
         \caption{$\overline N=80$, $\overline s=0.6$. The sequence shows a detailed view of the director field around the frontal $-1/2$ defects, color-coded by the local elastic stress distribution $\Psi_{\mathrm{e}}^{(k)}$. As regeneration proceeds, the characteristic three-fold nematic configuration becomes progressively stretched along the direction of the head and the adjacent tentacle. The localization of elastic energy around the $-1/2$ defects at advanced simulation steps shows that, for the tentacular configuration, these defects do not act solely as balancers of the system's total topological charge. Instead, the resulting stress accumulation directly triggers localized material growth at the defect cores.}
\label{fig:director1o2}
\end{figure}

Figures~\ref{fig:frontview_tentacles} and \ref{fig:apicalview_tentacles} show how the simulations track the quasi-static morphological evolution of an initially isotropic spheroid featuring a single hole in its upper hemisphere. This specific geometry, combined with the initial configuration of the defects, is designed to model an excised tissue fragment developing into a mature uniaxial Hydra with two tentacles. 

We run the simulations until a late stage in the regeneration of the tissue where the tentacles are elongated and the nematic order is well established. We again examine how the combination of tissue growth and the gradual establishment of nematic order within the fiber network shapes the overall morphology. 

In Figure~\ref{fig:frontview_tentacles}, the color map represents the stress-dependent volumetric factor $\log(J_{\alpha}^{(k)})$, indicating that growth is initially concentrated in the vicinity of the head, the basal foot, and the tentacles. As for the previous configurations, at initial stages ($k \lesssim 48$), the $-1/2$ defects do not involve any associated material growth. As simulation progresses, however, the elongation of the tentacles and the neck region below the head eventually drives growth at the $-1/2$ defects as well. This effect appears to be linked to the regions in which the elastic stress $\Psi_{\mathrm{e}}^{(k)}$ localizes. As the tentacles elongate, the three-fold nematic field characteristic of the $-1/2$ defects becomes stretched along the direction of the head and the associated tentacle, concentrating stress around the $-1/2$ defects core and consequently triggering the stress-dependent growth factor. 

Figure~\ref{fig:apicalview_tentacles} then illustrates how this elastic stress distribution directly drives tissue growth observed in Figure~\ref{fig:frontview_tentacles}. 
The evolution of the corresponding director field configuration $\n$ around the $-1/2$ defects and progressive stretching are shown in Figure~\ref{fig:director1o2}.

\section{Conclusions}

Over the last two decades, a number of two-dimensional realizations of biological systems have demonstrated that long-range ordering of cells and cytoskeletal structures can give rise to a nematic phase, in which elongated constituents align parallel to one another, generating a partial orientational order analogous to that observed in nematic liquid crystals (LCs). In these systems, topological defects in the nematic order can act as organizing centers, enabling organisms to develop protrusions or locally deplete material in order to relieve mechanical stress.

In a recent work \cite{Paparini:2025:PGM}, we addressed the open problem of theoretically modeling the coupling between mesogenic disclinations and polymeric networks by developing a mathematical framework for describing the out-of-plane shape transformations of initially flat liquid-crystalline network sheets containing a central topological defect. The model successfully captured the emergence of post-instability morphologies in good agreement with experimental observations, providing a solid foundation for describing shape morphing driven by the interplay between nematic ordering and elasticity.

Here, we build on this framework by incorporating stress-driven growth and adaptive relaxation resulting from material reorganization, with the aim of capturing morphogenetic processes in living tissues through a sequence of quasi-static states. These tissues are modeled as \emph{active biological networks} (ABNs). Rather than explicitly describing the underlying activation mechanisms---namely, how active impulses generated by contractile actin fibers drive the emergence of order from disorder---we focus on their macroscopic manifestation: changes in the degree of orientational order, quantified by the scalar order parameter $s$. The tissue is assumed to be isotropic prior to activation, whereas upon activation $s$ becomes a prescribed field defined on the reference configuration. The induction of orientational order consequently generates a non-uniform nematic alignment characterized by topological defects, where $s$ vanishes. The locations of these defects in the undeformed configuration are prescribed according to the morphological features expected in the mature organism. The resulting macroscopic shape changes emerge from the tissue response to four interacting mechanisms: mechanical stresses developing in the vicinity of topological defects, activity-induced ordering, stress-driven growth, and structural relaxation aimed at relieving internal stresses.

A paradigmatic example is provided by Hydra, simple aquatic organisms characterized by a cylindrical body and tentacles extending from the oral end. Hydra develop from an initially spheroidal morphology through the activation of nematic interactions among their cells. Regions of enhanced cell proliferation are localized around topological defects, driving the progressive elongation of the body axis and the extension of tentacles.

We employ finite element simulations, which we implement using the library FEniCSx, to track the quasi-static morphological evolution of an initially isotropic spheroid into uniaxial (head–foot), bicephalic (two-headed), and tentacular configurations. Importantly, in all simulations, growth and deformation are not prescribed a priori. Instead, they emerge naturally from the interplay between internal stresses and progressive tissue relaxation, as determined by the solution of the equilibrium problem at each incremental step. For each of the three morphologies considered, we characterize the mechanical state of the tissue by analyzing the spatial distribution of elastic energy throughout the evolution process, thereby obtaining a direct measure of local stress concentrations. Distortions of the nematic director field generate mechanical stresses, which are subsequently redistributed and partially relaxed through tissue deformation. We also examine the evolution of the stress-dependent volumetric growth factor. 

Our results reveal behavior consistent with experimental observations, showing that tissue growth is explicitly coupled to the local elastic energy density. Consequently, growth is stress-driven, with new material being preferentially incorporated into regions experiencing elevated mechanical stress. Finally, the analysis of the stress field indicates that tissue deformation partially relaxes stresses in the vicinity of the mouths, facilitating enlargement of the oral apertures, as well as along the tentacle shafts. In contrast, the tentacle tips and the basal foot remain persistently stressed throughout the developmental process.

Considerable work remains to be dine to achieve a deeper understanding of morphogenesis. Future developments of the present model should incorporate active stresses and a geometry-induced evolution of the nematic director field, both of which are essential for capturing the intrinsically non-equilibrium dynamics of living tissues. Furthermore, the functional framework proposed here for ABNs requires a thorough mathematical investigation to establish the well-posedness of the differential problems associated with the continuum model. The inclusion of stochastic model parameters \cite{Mihai:2022} could further enhance the inherent variability of the system, thereby improving the representation of real-world uncertainty.

\section*{Data availability statement}
The data supporting the findings of this study will be made openly available upon acceptance of this manuscript for publication.

\section*{Acknowledgments}
The work of S.P. and G.G.G. is partially supported by the European Union–Next Generation EU under the National Recovery and Resilience Plan (NRRP), Mission 4 Component 2 Investment 1.1–Call PRIN 2022 of Italian Ministry of University and Research. Project No. 202249PF73. \\
S.P. gratefully
acknowledge partial financial support provided by the INdAM - GNFM Project, CUP code [E5324001950001].

\section*{Conflict of interest}
The authors have no competing interests to declare that are relevant to the content of this article.

\section*{ORCID IDs}
Silvia Paparini: https://orcid.org/0000-0003-3970-0875

\noindent Giulio G. Giusteri: https://orcid.org/0000-0001-9001-9706

\noindent L. Angela Mihai: https://orcid.org/0000-0003-0863-3729

\appendix

\section{Stereographic construction of a local orthonormal basis on the sphere}\label{sec:stereographic}

To construct a director field $\n_0$ with topological defects at prescribed locations on a spherical surface, we exploit the stereographic projection. This mapping transfers the sphere onto a plane, yielding a simpler domain in which the director field can be explicitly defined using the Cartesian basis of the stereographic coordinates. Once constructed in this planar setting, the field is pulled back onto the sphere and expressed in terms of a suitably defined local orthonormal basis $(\e_1, \e_2)$.

\subsection{Stereographic coordinates and angular parametrization}

Consider a point on a spherical surface of radius $R$, described in Cartesian coordinates by $(X, Y, Z)$ and parameterized by the standard spherical angles: the polar angle $\vartheta \in [0, \pi]$ and the azimuthal angle $\varphi \in [0, 2\pi)$.

Under stereographic projection, the sphere is mapped onto a plane. On this plane, the point is represented by polar coordinates $(\rho_s, \varphi_s)$, where the azimuthal angle is preserved and the radial coordinate depends on the polar angle according to
\begin{equation}
    \varphi_s = \varphi, \qquad \rho_s = \tan\frac{\vartheta}{2}.
\end{equation}
The corresponding Cartesian coordinates on the plane are therefore
\begin{equation}
    X_s = \tan\frac{\vartheta}{2}\cos\varphi, 
    \qquad 
    Y_s = \tan\frac{\vartheta}{2}\sin\varphi.
\end{equation}
Rewriting these expressions in terms of the spatial coordinates $(X, Y, Z)$ yields the projection formulas reported in \eqref{eq:projector_def} in the main text.

On the stereographic plane, we define the planar director field $\n_{0,s}$ as
\begin{equation}
\label{eq:n_0s_def}
    \n_{0,s} = \cos\big({w}(X_s,Y_s)\big)\,\e_{X_s} + \sin\big({w}(X_s,Y_s)\big)\,\e_{Y_s},
\end{equation}
where $\e_{X_s}$ and $\e_{Y_s}$ denote the Cartesian basis vectors of the plane. The orientation angle ${w}$, introduced in \eqref{eq:omega_def}, is constructed so that $\n_{0,s}$ exhibits topological defects precisely at the stereographic projections of the prescribed defect locations on the sphere.

\subsection{Inverse mapping and tangent basis construction}

 Having defined the planar director field $\n_{0,s}$ and its associated field lines on the stereographic plane, we now map this construction back onto the sphere. To this end, we express a generic point on the sphere, $\mathbf{p} = (X, Y, Z)$, entirely in terms of the stereographic coordinates $(X_s, Y_s)$. 

Within this parametrization, the partial derivatives of $\mathbf{p}$ with respect to $X_s$ and $Y_s$ define tangent vectors to the spherical surface. Because stereographic projection is conformal (angle-preserving), the orthogonal Cartesian grid on the plane maps to an orthogonal grid of curves on the sphere. Consequently, these tangent vectors are mutually orthogonal, and their normalization directly yields the desired local orthonormal basis $(\e_1, \e_2)$.

Here, vector components written in array form are understood with respect to the standard Cartesian basis in $\mathbb{R}^3$. To construct the inverse mapping $\mathbf{p}(X_s, Y_s)$, we begin by inverting the projection relations in \eqref{eq:projector_def}, namely, $X = X_s (R + Z)$ and $Y = Y_s (R + Z)$. Introducing the planar radial coordinate $\rho_s^2 = X_s^2 + Y_s^2$ and using the spherical constraint $X^2 + Y^2 = R^2 - Z^2$, we obtain
\begin{equation}
    \rho_s^2 = \frac{R - Z}{R + Z}.
\end{equation}
Solving for $Z$ gives
\begin{equation}
    Z = R \frac{1 - \rho_s^2}{1 + \rho_s^2}.
\end{equation}
Substituting back into the expressions for $X$ and $Y$ yields the inverse stereographic map
\begin{equation}
    \mathbf{p}(X_s, Y_s) = \left( \frac{2R X_s}{1 + \rho_s^2}, \, \frac{2R Y_s}{1 + \rho_s^2}, \, R \frac{1 - \rho_s^2}{1 + \rho_s^2} \right).
\end{equation}
We now construct the local orthonormal basis. The first basis vector $\e_1$ is obtained by differentiating $\mathbf{p}$ with respect to $X_s$ and normalizing. Since $\|\partial \mathbf{p} / \partial X_s\| = 2R / (1 + \rho_s^2)$—the conformal scale factor—we obtain
\begin{equation}
\label{eq:E1s}
    \e_1 = \frac{\partial \mathbf{p} / \partial X_s}{\|\partial \mathbf{p} / \partial X_s\|}
    = \frac{1}{1 + \rho_s^2} \left( 1 - X_s^2 + Y_s^2, \, -2X_s Y_s, \, -2X_s \right).
\end{equation}
Analogously, differentiating with respect to $Y_s$ yields the second basis vector
\begin{equation}
\label{eq:E2s}
    \e_2 = \frac{\partial \mathbf{p} / \partial Y_s}{\|\partial \mathbf{p} / \partial Y_s\|}
    = \frac{1}{1 + \rho_s^2} \left( -2X_s Y_s, \, 1 + X_s^2 - Y_s^2, \, -2Y_s \right).
\end{equation}
These expressions define an orthonormal basis at every point on the sphere, except at the South Pole where the stereographic projection is singular. Substituting the relations \eqref{eq:projector_def} into \eqref{eq:E1s}–\eqref{eq:E2s} recovers the expressions \eqref{eq:E_12_def} reported in the main text.

\section{Gradient identity for the nematic tensor field}
\label{sec:identity_grads}

In this appendix, we derive the identity relating the squared norm of the spatial gradient of the director field, $|\nabla_{\bm{x}} \n|^2$, to that of the associated tensor field $\mathbf{M} = \n \otimes \n$.

Denoting the components of $\mathbf{M}$ by $M_{ij} = n_i n_j$, $i,j=1,2,3$, and taking the spatial gradient and applying the product rule with respect to $x_k$ gives
\[
\partial_k M_{ij} = \partial_k n_i)n_j + n_i(\partial_k n_j.
\]
The squared norm of the tensor gradient is obtained by summing over all components as follows,
\[
|\nabla_{\bm{x}} \mathbf{M}|^2 = \sum_{i,j,k} (\partial_k M_{ij})^2.
\]
Then expanding the squared bracket yields
\[
|\nabla_{\bm{x}} \mathbf{M}|^2 
= \sum_{i,j,k} \left[ (\partial_k n_i)^2 n_j^2 + n_i^2 (\partial_k n_j)^2 + 2 n_i (\partial_k n_i)\, n_j (\partial_k n_j) \right].
\]
This expression simplifies by exploiting two fundamental properties of the director field. First, $\n$ is a unit vector, so $\sum_i n_i^2 = 1$. Second, differentiating this constraint implies orthogonality between $\n$ and its spatial derivatives, $\sum_i n_i (\partial_k n_i) = 0$.

We obtain
\begin{align}
&\sum_{i,j,k} (\partial_k n_i)^2 n_j^2 
= \sum_{i,k} (\partial_k n_i)^2 \sum_j n_j^2
= \sum_{i,k} (\partial_k n_i)^2
= |\nabla_{\bm{x}} \n|^2, \\[6pt]
&\sum_{i,j,k} n_i^2 (\partial_k n_j)^2 
= \sum_i n_i^2\sum_{j,k} (\partial_k n_j)^2
= \sum_{j,k} (\partial_k n_j)^2
= |\nabla_{\bm{x}} \n|^2, \\[6pt]
&2 \sum_{i,j,k} n_i (\partial_k n_i)\, n_j (\partial_k n_j) 
= 2 \sum_k \left(\sum_i n_i \partial_k n_i\right)\left(\sum_j n_j \partial_k n_j\right)
= 0.
\end{align}
Since the cross-term vanishes identically, combining the remaining contributions gives
\[
|\nabla_{\bm{x}} \mathbf{M}|^2 = 2\,|\nabla_{\bm{x}} \n|^2.
\]
We therefore recover equation~\eqref{eq:identity_grads} from the main text.



\begin{thebibliography}{30}

\bibitem{bladon:deformation} Bladon P, Terentjev EM, Warner M. 1994. Deformation-induced orientational transitions in liquid crystals elastomer, Journal of Physics II France 4(1), 75–91 (doi:10.1051/jp2:1994100). 

\bibitem{Bailles:2025:etal} Bailles A, Serafini G, Andreas H, Zechner C, Modes CD, Tomancak P. 2025. Anisotropic stretch biases the self-organization of actin fibers in multicellular Hydra aggregates, Proceedings of the National Academy of Sciences 122(32), e2423437122 (doi: 10.1073/pnas.2423437122).

\bibitem{blanch:turbulent} Blanch-Mercader C, Yashunsky V, Garcia S, Duclos G, Giomi L, Silberzan P. 2018. Turbulent dynamics of epithelial cell cultures, Physical Review Letters 120, 208101 (doi: 10.1103/PhysRevLett.120.208101).

\bibitem{doostmohammadi2018active} Doostmohammadi A, Ign\'{e}s-Mullol J, Yeomans JM, Sagu'{e}s F. 2018. Active nematics, Nature Communication 9, 3246 (doi: 10.1038/s41467-018-05666-8).

\bibitem{ericksen:liquid} Ericksen JL. 1991. Liquid crystals with variable degree of orientations, Archive for Rational Mechanics and Analysis 113, 97-120 (doi:10.1007/BF00380413).

\bibitem{Goodbrake:2021:GGY} Goodbrake C, Goriely A, Yavari A. 2021. The mathematical foundations of anelasticity: existence of smooth global intermediate configurations, Proceedings of the Royal Society A 477, 20200462 (doi: 10.1098/rspa.2020.0462).

\bibitem{goriely2017:growth} Goriely A. 2017. The Mathematics and Mechanics of Biological Growth, Springer-Verlag, New York.

\bibitem{Guillamat:2026:etal} Guillamat P, Mirza W, Bal PK, G\'{o}mez-Gomz\'{a}les M, Roca-Cusachs P, Arroyo M, Trepat X. 2026. Guidance of cellular nematic elastomers into shape-programmable living surfaces, Science 392, 317-323 (doi: 10.1126/science.adz9174).

\bibitem{hernandez2026:mechanics} Hernandez A, Zhu C, Giomi L. 2026. Mechanics of axis formation in Hydra, arXiv a2601.05220 (doi:arxiv.org/abs/2601.05220).

\bibitem{hoffmann2022:theory} Hoffmann LA, Carenza LN, Eckert J, Giomi L. 2022. Theory of defect-mediated morphogenesis, Science Advances 8, eabk2712 (doi:10.1126/sciadv.abk2712).

\bibitem{hoffmann2023:tuneable} Hoffmann LA, Carenza LN, Giomi L. 2023. Tuneable defect-curvature coupling and topological transitions in active shells, Soft Matter 19, 3423–3435 (doi:10.1039/D3SM00071K).

\bibitem{Hopf1983:differential} Hopf H. 1983. Differential Geometry in the Large, Springer-Verlag, Berlin, Heidelberg.

\bibitem{huang2025:modelling} Huang R, Ogden RW, Penta R. 2025. Modelling volumetric growth in soft solids via residual stress, Journal of Elasticity 157(4), 73 (doi: 10.1007/s10659-025-10164-2).

\bibitem{kawaguchi:topological} Kawaguchi K, Kageyama R, Sano M. 2017. Topological defects control collective dynamics in neural progenitor cell cultures, Nature 545, 212 (doi: 10.1038/nature22321).

\bibitem{copenhagen:topological} Copenhagen K, Alert R, Wingreen N, Shaevitz J. 2021. Topological defects promote layer formation in myxococcusxanthus colonies, Nature Physics 17, 211 (doi: 10.1038/s41567-020-01056-4).

\bibitem{keber:topology} Keber FC,  Loiseau E, Sanchez T, DeCamp SJ, Giomi L, Bowick MJ, Marchetti MC, Dogic Z, Bausch AR. 2014. Topology and dynamics of active nematic vesicles, Science 345 (6201), 1135-1139 (doi: 10.1126/science.1254784).

\bibitem{kumar:tunable} Kumar N, Zhang R, de Pablo JJ, Gardel ML. 2018. Tunable structure and dynamics of active liquid crystals, Science Advances 4, eaat7779 (doi: 10.1126/sciadv.aat7779).

\bibitem{lee:crawling} Lee P, Wolgemuth CW. 2011. Crawling cells can close wounds without purse strings or signaling, PLOS Computational Biology 7(3), e1002007 (doi: 10.1371/journal.pcbi.1002007).

\bibitem{livshits:structural} Livshits A, Shani-Zerbib L, Maroudas-Sacks Y. 2017. Structural inheritance of the actin cytoskeletal organization determines the body axis in regenerating Hydra, Cell Reports 18, 1410-1421 (doi: 10.1016/j.celrep.2017.01.036).

\bibitem{Lubarda:2004} Lubarda VA. 2004. Constitutive theories based on the multiplicative decomposition of deformation gradient: thermoelasticity, elastoplasticity and biomechanics, Applied Mechanics Reviews 57(2), 95-108 (doi: 10.1115/1.1591000).

\bibitem{marchetti2013hydrodynamics} Marchetti MC, Joanny JF, Ramaswamy S, Liverpool TB, Prost J, Rao M, Simha RA. 2013. Hydrodynamics of soft active matter, Reviews of Modern Physics 85, 1143 (doi: 10.1103/RevModPhys.85.1143).

\bibitem{maroudas:topological} Maroudas-Sacks Y, Garion L, Shani-Zerbib L, Livshits A, Braun E, Keren K. 2020. Topological defects in the nematic order of actin fibers as organization centers of Hydra morphogenesis, Nature Physics 17(2), 1-9 (doi: 10.1101/2020.03.02.972539).

\bibitem{maroudas2025:mechanical} Maroudas-Sacks Y, Suganthan S, Garion L, Ascoli-Abbina Y, Westfried A, Dori N, Pasvinter I, Popović M, Keren K. 2025. Mechanical strain focusing at topological defect sites in regenerating Hydra, Development 152(4) (doi: 10.1242/dev.204514).

\bibitem{Mihai:2022} Mihai LA. 2022. Stochastic Elasticity: A Nondeterministic Approach to the Nonlinear Field Theory, Springer Cham, Switzerland (doi: 10.1007/978-3-031-06692-4).

\bibitem{napoli2026:nematicbubbles} Napoli G, Paparini S. 2016. Nematic bubbles and the breaking of spherical symmetry, arXiv 2603.07082 (doi: arxiv.org/abs/2603.07082).

\bibitem{narayan:long} Narayan V, Ramaswamy S, Menon N. 2007. Long-lived giant number fluctuations in a swarming granular nematic, Science 317, 105 (doi: 10.1126/science.1140414).

\bibitem{Paparini:2025:PGM} Paparini S, Giusteri GG, Mihai LA. 2025. Shape instabilities driven by topological defects in nematic polymer networks, Journal of Elasticity 157, 69 (doi: 10.1007/s10659-025-10160-6).

\bibitem{pearce:passive}{Pearce DJG, Thibault C, Chaboche Q, Blanch-Mercader C. 2025. Passive defect driven morphogenesis in nematic membranes, Physical Review Letters 134(1), 018402 (doi: 10.1103/PhysRevLett.134.018402)}

\bibitem{perez:activewetting} Pérez-González C, Alert R, Blanch-Mercader C, Gómez-González M, Kolodziej T, Bazellieres E, Casademunt J, Trepat X. 2019. Active wetting of epithelial tissues, Nature Physics 15(1), 79-88 (doi: 10.1038/s41567-018-0279-5).

\bibitem{Perros:2024:etal} Perros T, Biquet-Bisquert A, Ben Meriem Z, Delarue M, Joseph P, Marcq P, Cochet-Escartin O. 2025. Mechanical characterization of regenerating Hydra tissue spheres, Biophysical Journal 123, 1792-1803 (doi: 10.1016/j.bpj.2024.05.022).

\bibitem{poincare1886:courbes} Poincar\'e, H. 1886. Sur les courbes d\'efinies par les \'equations diff\'erentielles (quatri\`eme partie), Journal de Math\'ematiques Pures et Appliqu\'ees, 4e s{\'e}rie, 2, 151--217.

\bibitem{Ravichandran:2025:etal} Ravichandran Y, Vogg M, Kruse K, Pearce DJG, Roux A. 2025. Topology changes of Hydra define actin orientation defects as organizers of morphogenesis, Science Advances 11(3), eadr9855 (doi: 10.1126/sciadv.adr98).

\bibitem{rosso:parallel} Rosso R, Virga EG, Kralj S. 2012. Parallel transport and defects on nematic shells, Continuum Mechanics and Thermodynamics 24, 643-664 (doi:10.1007/s00161-012-0259-4).

\bibitem{Sadik:2017:SY} Sadik S, Yavari A. 2017. On the origins of the idea of the multiplicative decomposition of the deformation gradient, Mathematics and Mechanics of Solids 22(4), 771-772 (doi: 10.1177/1081286515612280).

\bibitem{saw:topological} Saw TB, Doostmohammadi A, Nier V, Kocgozlu L, Thampi S, Toyama Y, Marcq P, Lim C, Yeomans J, Ladoux B. 2017. Topological defects in epithelia govern cell death and extrusion, Nature 544, 212 (doi: 10.1038/nature21718).

\bibitem{stoker1989differential} Stoker JJ. 1989. Differential Geometry, Wiley, Wiley Classics Library.

\bibitem{wang:patterning} Wang Z, Marchetti MC, Brauns F. 2023. Patterning of morphogenetic anisotropy fields. Proceedings
of the National Academy of Sciences 120, e2220167120 (doi: 10.1073/pnas.2220167120).

\bibitem{warner1988:theory} Warner M, Gelling KP, Vilgis TA. 1988. Theory of nematic networks, The Journal of Chemical Physics 88(6), 4008–4013 (doi:10.1063/1.453852).

\bibitem{Warner:2007:WT} Warner M, Terentjev EM. 2007. Liquid Crystal Elastomers, paper back, Oxford University Press, Oxford, UK.

\bibitem{warner1991:elasticity} Warner M, Wang X. 1991. Elasticity and phase behavior of nematic elastomers, Macromolecules 24, 4932-4941 (doi: 10.1021/ma0 0 017a033).

\end{thebibliography}
\end{document}